\documentclass[aps,prb,twocolumn,showpacs,superscriptaddress,bibliography,notitlepage]{revtex4-1}
\usepackage[colorlinks=true,citecolor=blue,linkcolor=blue,breaklinks=true]{hyperref}
\usepackage{amssymb}
\usepackage{graphicx}
\usepackage{amsmath}
\usepackage{mathrsfs}
\usepackage{mathtools}
\usepackage[export]{adjustbox}
\usepackage{epsfig}
\usepackage{times}
\usepackage{color}
\usepackage{setspace}
\usepackage{calc}
\usepackage{natbib}
\usepackage{bm}
\DeclareMathAlphabet{\mathpzc}{OT1}{pzc}{m}{it}
\usepackage{array}
\usepackage{multirow}
\usepackage{comment}
\usepackage{float}

\begin{document}

\newcommand {\beq} {\begin{equation}}
\newcommand {\eeq} {\end{equation}}
\newcommand {\bqa} {\begin{eqnarray}}
\newcommand {\eqa} {\end{eqnarray}}
\newcommand{\dhat}{\ensuremath{\hat{D}}}
\newcommand{\ehat}{\ensuremath{\hat{E}}}
\newcommand{\lhat}{\ensuremath{\hat{\Lambda}}}
\newcommand{\zbar}{\ensuremath{\bar{\zeta}}}
\newcommand{\ebar}{\ensuremath{\bar{\eta}}}
\newcommand {\ba} {\ensuremath{b^\dagger}}
\newcommand {\Ma} {\ensuremath{M^\dagger}}
\newcommand {\psia} {\ensuremath{\psi^\dagger}}
\newcommand {\psita} {\ensuremath{\tilde{\psi}^\dagger}}
\newcommand{\lp} {\ensuremath{{\lambda '}}}
\newcommand{\A} {\ensuremath{{\bf A}}}
\newcommand{\Q} {\ensuremath{{\bf Q}}}
\newcommand{\kk} {\ensuremath{{\bf k}}}
\newcommand{\qq} {\ensuremath{{\bf q}}}
\newcommand{\kp} {\ensuremath{{\bf k'}}}
\newcommand{\rr} {\ensuremath{{\bf r}}}
\newcommand{\rp} {\ensuremath{{\bf r'}}}
\newcommand {\ep} {\ensuremath{\epsilon}}
\newcommand{\nbr} {\ensuremath{\langle ij \rangle}}
\newcommand {\no} {\nonumber}
\newcommand{\up} {\ensuremath{\uparrow}}
\newcommand{\dn} {\ensuremath{\downarrow}}
\newcommand{\rcol} {\textcolor{red}}
\newcommand{\bcol} {\textcolor{blue}}
\newcommand{\bu} {\bold{u}}
\newcommand{\tr}[1]{\mathrm{Tr}\left[#1\right]}
\newcommand{\ve}[1]{\boldsymbol{#1}}
\newcommand{\args}[1]{\ve{#1},\ve{\bar{#1}}}
\newcommand{\mes}[1]{\mathcal{D}\hspace{-2pt}\left[\args{#1}\right]}
\newcommand{\ii}{\iota}
\newcommand{\mnm} {\ensuremath{\mathbb{M}}}
\newcommand{\ncr}[2]{\begin{pmatrix}#1\\#2\end{pmatrix}}

\begin{abstract}
We propose a new method to study non-equilibrium dynamics of
scalar fields starting from non-Gaussian initial conditions using
Keldysh field theory. We use it to study dynamics of phonons coupled
to non-interacting bosonic and fermionic baths, starting from initial Fock states. 
We find that in one dimension long wavelength phonons coupled to
fermionic baths do not thermalize both at low and high bath-temperatures. At low temperature, constraints from energy-momentum
conservation lead to a narrow bandwidth of particle-hole excitations
and the phonons effectively do not see this bath. On
the other hand, the strong band-edge divergence
of the particle-hole density of states leads to an undamped polaritonlike mode
of the dressed phonons above the band edge of the particle-hole
excitations. These undamped modes contribute to the lack of
thermalization of long wavelength phonons at high temperatures. In higher
dimensions, these constraints and the divergence of density of states
are weakened and lead to thermalization at all wavelengths.
\end{abstract}
\title{Non-equilibrium scalar field dynamics starting from Fock states: Absence of thermalization in one dimensional phonons coupled to fermions}
\author{Md Mursalin Islam}\email{mursalin@theory.tifr.res.in}
 \affiliation{Department of Theoretical Physics, Tata Institute of Fundamental
 Research, Mumbai 400005, India.}
\author{ Rajdeep Sensarma} 
\affiliation{Department of Theoretical Physics, Tata Institute of Fundamental
 Research, Mumbai 400005, India.}

\date{\today}

\maketitle
\section{Introduction}
Scalar fields are paradigmatic degrees of freedom in quantum field
theories, with wide ranging applications from particle physics to
astro-particle physics and cosmology to condensed matter physics. In
particle physics they played a big role in the construction of the Higgs
mechanism, which leads to descriptions of massive particles~\cite{englert,higgs1,higgs2}. In
cosmology and astro-particle physics, scalar fields have been used to
describe inflationary expansion of the early universe~\cite{inflationintro,inflationdynamics}; they have also
been projected as possible candidates for dark matter in various
models~\cite{sfdm,uldm,axiondm}. They are also ubiquitous in condensed matter physics, often rearing
their heads as low energy fluctuations around ordered (symmetry
broken) states of matter. From
phonons in solids~\cite{doniach} to magnons in certain types of magnets~\cite{magnon} to phase
fluctuations in a superfluid~\cite{superfluid_anderson,superfluid_book}, there are a large variety of scalar
fields which occur in descriptions of quantum many body systems. More recently, with the advent of technology
to create strong light-matter interactions, photons in optical
cavities~\cite{cvtqedexp1,cvtqedexp2,cvtqedrev1,cvtqedrev2} or dressed excitations like polaritons~\cite{polariton,tolpygo,huang,polaritonrevold,polaritonrevnew} can also be described
by effective scalar fields.

Compared to the well developed equilibrium and ground state theory for
scalar fields (including interacting field theories)~\cite{peskin,srednicki,altlandsimons}, the
non-equilibrium dynamics of scalar fields are relatively less
studied. The primary motivation for earlier studies of non-equilibrium
dynamics of scalar fields was related to non-equilibrium processes in the early universe and in high energy collisions~\cite{neqqf_CalzettaHu,qfielddyn_berges,noneqscalar_anisimov,dynoscscalar}. More
recent advances in creating non-equilibrium states, either by pumping
solid state systems with intense energy sources~\cite{ufpps_expold,ufpps_exp1,ufpps_exp4,ufpps_phdyn_nano_rev}, or by pulse shaping~\cite{pulse1,pulse2,pulse3,pulse4,pulse5}
and cavity engineering~\cite{cvtqedengrev,cvtqwed_sqzlit1,cvtqedeng1,cvtqedeng2}, has led to a renewed focus on this topic. 
In many of these cases, the system evolves starting from an athermal state. A typical example is a cavity where the photons are initiated to a squeezed state~\cite{sqzlightold,sqzlightnew,cvtqwed_sqzlit1,cvtqwed_sqzlit2,cvtqwed_sqzlit3}. Thus it is very important to obtain a description of dynamics of scalar fields starting from athermal initial conditions.

The standard Schwinger-Keldysh field theory used to describe
non-equilibrium dynamics of many body systems, unfortunately requires
a thermal initial condition to work~\cite{schwinger,keldysh,kamenevarticle,kamenevbook,rammer,altlandsimons}. Recently Chakraborty et. al.~\cite{cgs} 
looked at this problem for Schr\"{o}dinger bosons (fields obeying saddle point Schr\"{o}dinger equation with single time derivatives). They
found that arbitrary initial conditions can be incorporated in the
field theory provided an additional source is coupled to the fields at
the initial time. The correlators are calculated within this
theory with the additional source. The physical correlation functions
are then obtained by taking derivatives of these correlators with
respect to the initial sources, with the set of derivatives determined
by the initial conditions. In this paper,
we extend this formalism to describe non-equilibrium dynamics of
scalar fields starting from density matrices diagonal in occupation number (Fock) states. Other than having different symmetries, the key difference between Schr\"{o}dinger bosons and scalar fields is that scalar fields obey a classical equation of motion  which has $2^{nd}$ order time derivatives.
Our formalism also requires additional sources at initial
times. However, these sources
couple to bilinears of both the initial fields and their time
derivatives, reflecting the fact that the classical equations for
these fields are now $2^{nd}$ order in time derivatives and require
knowledge of both initial configurations and their time derivatives
to be solved. The rest of the formalism is similar to Ref.~\onlinecite{cgs}; i.e. we
calculate the correlators in presence of these sources and then
take derivatives with respect to the initial sources to get physical correlation functions.

We use our formalism to study the thermalization of a system of scalar
fields coupled to external baths, and initialized to athermal Fock
states. Although we consider phonons as a concrete example in this
case, similar considerations can be used for other scalar fields
including complex ($O(2)$) and $O(N)$ scalar fields. We first consider coupling
the phonons to a Markovian ohmic bath with a smooth ultra-violet
cut-off~\cite{diss2ss_legget}. Here, as expected, the phonons thermalize at a rate which
becomes momentum independent as we go to the broad bandwidth limits of pure white noise~\cite{quantnoise}. 

We then consider the coupling of phonons with
the particle-hole excitations of a non-interacting fermionic bath. The dynamics of electrons coupled to phonons has been extensively studied in the context of thermalization of ``hot electrons"~\cite{ephrelCu,hotel_metal,edyn_noble,mesothermo,hotdyn_habib,noneqsc_spinsplit,hotel_TiN,ultraspec_hiTsc,hotel_apply}. Relatively less attention has been paid to the dynamics of the phonons coupled to fermions ~\cite{noneq_phdetect,noneqph_gendet,perinati,phdyn_detect,sqzph_ipaul}.
We note that the evolution of one-particle distribution functions have been treated before using Kadanoff-Baym equations~\cite{noneqscalar_anisimov}. In one dimension, at low temperatures, the density of particle-hole excitations with a fixed low momentum transfer is finite only over a narrow region of energy. In the long wavelength limit this almost reduces to a linearly dispersing mode with a width which scales quadratically with the momentum. As a result, the long wavelength phonon modes effectively do not see the fermionic bath as it is not possible to satisfy both energy and momentum conservation during exchange with the bath \cite{matveev1,matveev2}. Hence these modes undergo quantum oscillations and remember their initial occupation numbers even at very long times. These modes thus do not thermalize at low bath temperature.
In principle two phonon scatterings, which occur at higher orders of system-bath couplings, do not suffer from the tight energy momentum conservation requirements. However we find that in presence of a Fermi sea, the decay rate of a phonon with energy $\omega_{ph}$ remains exponentially small ($\sim \omega_{ph}e^{-\frac{\omega_{ph}}{T}}$) for $T\ll\omega_{ph}$. For $T\gg\omega_{ph}$, we obtain a decay rate $\sim T$. This high $T$ classical limit is similar to those considered in Refs.\onlinecite{heavyparticle_castro,spinorgas_kamenev}. The difference in the exponent  between our results and earlier works\cite{heavyparticle_castro,spinorgas_kamenev} comes from considering the fact that the relevant two-phonon process is mediated by a fermion with energy mismatch (i.e. off-shell) $\Delta\epsilon$, which scale with the momentum of the phonon  which is decaying. Note that this is the dominant relaxation process for $\Delta\epsilon\gg T\gg\omega_{ph}$, while the relaxation is dominated by the single phonon process for $T\sim\Delta\epsilon$. 
We note that even higher order (8-th order in system-bath coupling) processes which involve two particle-hole excitations (three phonon processes), result in a decay rate which is polynomial in temperature and hence not exponentially small. But for small values of system-bath coupling these processes are highly suppressed.

At high temperatures, the particle-hole density of states extend from zero energy to a band maximum (due to a finite bandwidth of the fermions).  
An interesting feature of one dimensional fermions  is a divergence in the single particle density of states at the upper edge
of the corresponding band. 
This leads to a strong divergence in the particle-hole density of states near their band edge. This divergence in the bath density of states results in the formation of 
new dressed modes of the phonons just outside the upper band edge for the
particle-hole excitations. This new mode, which we call a ``polarinon''
due to its similarity to polaritons formed by dressing of light with
excitons~\cite{extnplrtn_th,extnplrtn_exp}, remains undamped, leading to the absence of thermalization of long-wavelength phonons, even at high bath temperatures. This is contrary to the general expectation that thermalization should be easier at high temperatures since energy-momentum constraints in scattering processes are relaxed. This new
mode can be studied in one dimensional systems with strong electron-phonon coupling.

To contrast with the singular case of one dimensional particle-hole bath, we also consider a bath of two dimensional non-interacting electrons. In this case, there is no lower bound for the existence of particle-hole excitations and the phonons thermalize unless the Debye frequency of the phonons is larger than the bandwidth of the electrons. Further the singularities in the density of states of the particle-hole bath are weakened, and hence there is no additional mode formation in this case. The scalar fields thermalize unless they are massive, which describes optical phonons in the present context. In that case, coupling to the Fermionic bath fails to thermalize the long wavelength phonons as energy-momentum constraints cannot be satisfied simultaneously.

We now provide a brief route map of the different sections of our paper. In section \ref{formalism}, we set up the Keldysh field theory based formalism to describe non-equilibrium dynamics of scalar fields starting from Fock states. We initially work out the case of a single harmonic oscillator in section \ref{SHO} and extend it to the case of scalar fields in section \ref{Rscalar}. In section \ref{OQS} we first set up the problem of dynamics of phonons coupled to external baths, where the phonons are initialized to a non-thermal Fock state. In section \ref{bosebath}, we focus on the effects of a bosonic ohmic bath, while section \ref{1dfermi} illustrates the example of a one dimensional fermionic bath, where the long wavelength phonons do not thermalize. We finish this section with the case of two dimensional fermionic bath in section \ref{2dfermi}. Finally we have summarized the results in the concluding section \ref{conclusion}.

\section{ Dynamics with initial conditions \label{formalism}}
Several problems in non-equilibrium dynamics relevant to different branches of physics require the description of dynamics of scalar fields starting from athermal initial conditions. Such initial conditions can result from sophisticated pulse-shaping techniques creating squeezed lights in optical cavities~\cite{cvtqwed_sqzlit1,cvtqwed_sqzlit2,cvtqwed_sqzlit3}. They can also be created when a large amount of energy is dumped into  a system with broken symmetry, as is done in the case of pump-probe experiments~\cite{ufpps_expold,ufpps_revold,ufpps_exp1,ufpps_exp2,ufpps_exp3,ufpps_exp4,ufpps_phdyn_nano_rev}. 
This can also be relevant for high energy particles which can be created in non-equilibrium distributions due to particle collision/decay~\cite{heavyioncolrev}.

The standard textbook formalism of Schwinger-Keldysh field theory used to describe non-equilibrium dynamics however works when the initial condition is described by a thermal initial density matrix~\cite{kamenevarticle,kamenevbook,rammer,altlandsimons}. In a recent paper, Chakraborty et. al~\cite{cgs} had constructed a formalism to
treat open system dynamics of bosons and fermions starting from
arbitrary initial conditions. The bosons/fermions were described by
a Schr\"{o}dinger equation, which was suitably modified for open system
dynamics. Here we will extend this formalism to the case of 
scalar fields whose classical equation of motion has $2^{nd}$ order time-derivative which brings additional complications. 
Before we describe the formalism for scalars, 
we first review the structure of the formalism put forward in Ref.~\onlinecite{cgs}.
It will serve as a paradigm for comparing
and contrasting the formalism developed here.

In Ref.~\onlinecite{cgs} it was proposed that the athermal initial density matrix can be exponentiated by using a source term which allows it to be included in the action.  
In this formalism, the Schwinger-Keldysh action is modified by adding a source dependent action $\delta S(u)$ where the source $u$ couples to the bilinears of the fields only at the initial time. The source only couple to the Keldysh $(q-q)$ part of the action which carries information about the initial state. For example, in the case of a single bosonic mode $\delta S(u)=i\frac{1+u}{1-u}\phi^\ast_{q}(0)\phi_{q}(0)$, when the system starts with an arbitrary initial density matrix $\rho_0=\sum_n c_n\vert n\rangle \langle n\vert$ in the Fock basis. The correlation functions are calculated from the action with this extra source dependent part, which makes them dependent on the source $u$ in general. The physical correlation functions $\mathcal{G}_{\rho}$ are then calculated by taking a particular set of derivatives on the source dependent correlation functions $\mathcal{G}(u)$. This set of derivatives depends on the initial condition the system starts with. For the simple case of a single bosonic mode, $\mathcal{G}_{\rho}=\sum_n\frac{c_n}{n!}\partial^n_u\frac{1}{1-u}\mathcal{G}(u)\vert_{u=0}$.
 We will now  extend the Schwinger-Keldysh field theory for scalar fields to include initial density matrices diagonal in Fock basis. We will first consider the case of a single harmonic oscillator in position basis which represents a single mode of scalar fields. Then we will generalize it for the scalar fields containing multiple modes.

\subsection{The Case of a single Harmonic Oscillator\label{SHO}}
It is instructive to first consider the case of a single harmonic
oscillator in the position basis. This will provide us with the basic structure of the
theory, which can then be extended to real scalar fields.
We consider
the dynamics of a closed system of a one dimensional simple harmonic oscillator with
frequency $\Omega$, which is governed by the action
\beq
S = \int dt ~ \frac{m}{2} \dot{x}^2 -\frac{1}{2}m\Omega^2 x^2 .
\label{action}
\eeq
We want to consider the dynamics of this system starting from the
number or Fock states. However, in a harmonic oscillator (and
similarly in scalar field theory), construction of the Fock states
involves an additional energy scale. One can contrast this to Schr\"{o}dinger theories,
where one can define a Fock basis independent of any external scales. To see this, note that creation-annihilation operators defined by $a^\dagger_\omega=[\sqrt{\frac{m \omega}{\hbar}} \hat{x}-i\frac{\hat{p}}{\sqrt{\hbar m \omega}} ]/\sqrt{2}$ and $a_\omega=[\sqrt{\frac{m \omega}{\hbar}} \hat{x}+i
\frac{\hat{p}}{\sqrt{\hbar m \omega}}]/\sqrt{2}$ satisfy the standard bosonic
algebra $[a_\omega,a^\dagger_\omega]=1$ for any value of $\omega$, and can be used
to construct a complete number basis with non-negative integer eigenvalues. Unless $\omega=\Omega$,
these states will not be the stationary states of our harmonic oscillator;
nevertheless they can provide a complete basis to write the initial
density matrix. This will allow us to consider density matrices which
do not commute with the Hamiltonian of the system. In a closed system,
this allows us to consider quench problems with non-trivial dynamics~\cite{quench_rev,quench_das,quench_mandal}. We will consider
an initial density matrix which is diagonal in the number basis
corresponding to $a_\omega$ operators, i.e.
\beq
\hat{\rho}_0 = \sum_{n} c_n |n_\omega\rangle \langle n_\omega | ,
\label{denmat}
\eeq
where $\sum_n c_n=1$ for probability conservation. Shifting now to the Keldysh theory with doubled time contours ($+$ for
forward propagation and $-$ for backward propagation), the Keldysh
partition function can be written as
\begin{widetext}
\beq
Z = \int D[x_+]D[x_-] e^{i \int dt
  \frac{m}{2}\left(\dot{x_+}^2-\dot{x_-}^2\right)-\frac{1}{2}m
  \Omega^2(x_+^2-x_-^2)}\langle x_+(0)|\hat{\rho_0}|x_-(0)\rangle .
\label{partfn0}
\eeq

Using the Harmonic oscillator energy eigenfunctions,
$\psi_n(x)=\langle
x|n_\omega\rangle=\left(\frac{m\omega}{\pi}\right)^{\frac{1}{4}}\frac{1}{\sqrt{2^n
    n!}}e^{-\frac{1}{2}m\omega x^2}H_n(\sqrt{m\omega} x)$, where
$H_n$ are the Hermite polynomials (we have set $\hbar=1$), the
matrix element of the initial density matrix is given by
  \bqa
  \no \langle x_+|\hat{\rho}_0|x_-\rangle
  &=&\sqrt{\frac{m\omega}{\pi}}e^{-\frac{1}{2}m\omega(x_+^2+x_-^2)}\sum_n
  \frac{c_n}{2^nn!}H_n(\sqrt{m\omega} x_+)H_n(\sqrt{m\omega}
  x_-)\\
  &=&\sqrt{\frac{m\omega}{\pi}}\sum_n
\frac{c_n}{n!}\left.\frac{\partial^n}{\partial u^n}\frac{1}{\sqrt{1-u^2}}e^{m\omega
  f(u) x_+x_--\frac{1}{2}m\omega g(u) (x_+^2+x_-^2)} \right\vert_{u=0},
\label{denmatel}
  \eqa
where $f(u)= \frac{2u}{1-u^2}$, $g(u)=\frac{1+u^2}{1-u^2}$, and we
have used the Mehler formula \cite{mehler1} for the generating function of
products of Hermite polynomials to obtain the last expression. 

One can
now write the partition function describing the dynamics starting from
this initial condition as
\beq
Z= \left.\tilde{ {\cal L}}(\partial_u,\rho_0)  Z(u)\right\vert_{u=0},
\label{partfn1}
\eeq
where the differential operator $\tilde{{\cal L}} = \sum_n
\frac{c_n}{n!}(\partial_u)^n \frac{1}{\sqrt{1-u^2}}$ depends on the
initial density matrix. It is useful to write $\tilde{{\cal L}} ={\cal
  L}{\cal M}(u)$, where ${\cal
  L} = \sum_n
\frac{c_n}{n!}(\partial_u)^n$ and ${\cal M}(u)=\frac{1}{\sqrt{1-u^2}}$. Here, the
partition function in presence of source terms at the initial time is 
$Z(u) =\sqrt{m\omega/\pi}\int D[x_+]D[x_-] e^{iS(u)}$, where 
\beq
	S(u)= -\frac{m}{2}\int dt \int dt' (x_+(t),x_-(t)) \left(\begin{array}{cc}%
	\delta(t-t')(\partial_t^2+\Omega^2)& i\omega f(u)\delta(t)\delta(t')\\
	+\delta(t)\delta(t') [\partial_{t'}-i\omega
	g(u)]& \\
	\\
	i\omega f(u)\delta(t)\delta(t')&\delta(t-t')(-\partial_t^2-\Omega^2) \\
	&-\delta(t)\delta(t')
	[\partial_{t'}+i\omega
	g(u)] 
\end{array}\right)
\left(\begin{array}{c}%
	x_+(t')\\
	x_-(t') \end{array}\right).    
 \label{action_sho}                                   
\eeq
\end{widetext}
Note that in addition to the source terms related to imposing the
correct initial condition on the dynamics, there is an additional time derivative term at $t=0$. This boundary term comes from  an
integration by parts, which converts $\dot{x}^2 \rightarrow x\ddot{x}$
in the action and is ignored in theories where the time co-ordinate extends up-to $-\infty$ . As a simple check, one can easily show that without any
additional sources present,
$Z(u)=\sqrt{\frac{1+u}{1-u}}$, and hence
$Z=\sum_nc_n =1$, as one expects for the Keldysh partition function.

The correlation functions in this theory can be calculated by adding
sources coupling linearly to $x$ at arbitrary times and taking derivatives with respect
to these sources. The
derivatives with respect to these linear sources commute with the
derivatives with respect to the bilinear initial source
$u$. Hence one can write 
\beq
D(\rho_0)= \left. {\cal L}(\partial_u,\rho_0) {\cal N}(u) D(u)\right\vert_{u=0}
\label{corfn}
\eeq
where $D(\rho_0)$ is the correlator corresponding to the physical dynamics
with the initial condition, while $D(u)$ is the correlation function
in the theory with the initial sources. The normalization ${\cal
  N}(u)= {\cal M}(u) Z(u)= \frac{1}{1-u}$. Focussing on the single
particle Green's function, we can invert the matrix in
Eq.~(\ref{action_sho}) to write
\begin{widetext}
\bqa
  \no D^{+-}(t,t',u)&=&\frac{1}{2m\Omega} \sin \Omega (t-t')
  -\frac{i}{2m\omega}\frac{1+u}{1-u} \left\{ \cos \Omega t~\cos \Omega t'
    +\frac{\omega^2}{\Omega^2} \sin \Omega t \sin\Omega
    t'\right\} \\
  \no D^{-+}(t,t',u)&=&\frac{-1}{2m\Omega} \sin \Omega (t-t')
  -\frac{i}{2m\omega}\frac{1+u}{1-u} \left\{ \cos \Omega t ~\cos \Omega t'
    +\frac{\omega^2}{\Omega^2} \sin \Omega t \sin\Omega t'\right\} \\
  \no D^{++}(t,t',u)&=&\Theta(t-t') D^{-+}(t,t',u)+\Theta(t'-t)
  D^{+-}(t,t',u)~~ \textrm{and} \\
  D^{--}(t,t',u)&=&\Theta(t-t') D^{+-}(t,t',u)+\Theta(t'-t)
  D^{-+}(t,t',u).
\label{gfnu}
\eqa
\end{widetext} 
The details of this calculation are presented in Appendix~\ref{sho_inv}.

It is useful to work with the Keldysh rotated classical $x_{cl}=\frac{1}{2}(x_++x_-)$ and the
quantum $x_{q}=\frac{1}{2}(x_+-x_-)$ degrees of freedom. In
this basis, the one particle Green's functions take the form
\beq
\hat{D}= \left(\begin{array}{cc}%
                 D^K & D^R\\
                 D^A & 0 \end{array}\right),
\label{gfnstruc_sk}
\eeq
with

\bqa
 D^R(t,t',u) &=&-\Theta(t-t') \frac{\sin \Omega (t-t')}{2m\Omega}=D^A(t',t,u)
~~\textrm{and}
\label{gfnu_sk}\\
\no D^K(t,t',u) &=&-\frac{i}{2m\omega}\frac{1+u}{1-u} \left[\cos \Omega t ~
  \cos \Omega t' +\frac{\omega^2}{\Omega^2} \sin \Omega t ~\sin \Omega
  t' \right]\\
\no &=&\left.-2im\omega
  \frac{1+u}{1-u}D^R(t,\epsilon)\left[1+\frac{1}{\omega^2}
    \overleftarrow{\partial_\epsilon}\overrightarrow{\partial_\epsilon}\right]D^A(\epsilon,t')\right\vert_{\epsilon=0}. 
\eqa
\begin{widetext}
We find that the retarded (and advanced) Green's functions are
independent of the initial conditions/ initial sources, while the
Keldysh Green's function depends on them. It is then easy to show that
the retarded Green's function for the actual dynamics starting from
the initial condition has the same form as in Eq.~(\ref{gfnu_sk}), while
the physical Keldysh Green's function is given by

\beq
D^K(t,t',\rho_0)= -2im\omega\left.\sum_n c_n(1+2n) D^R(t,\epsilon)\left[1+\frac{1}{\omega^2}
    \overleftarrow{\partial_\epsilon}\overrightarrow{\partial_\epsilon}\right]D^A(\epsilon,t')\right\vert_{\epsilon=0}.
\label{dkph}
\eeq

One can now write the action
in the Keldysh rotated basis including the source terms, which will
reproduce the correlators

  \beq
S= -m\int dt \int dt' (x_{cl}(t),x_q(t))\left(\begin{array}{cc}%
   0 &\delta(t-t')(\partial_t^2+\Omega^2) \\
  \delta(t-t')(\partial_t^2+\Omega^2) & -i\omega \frac{1+u}{1-u} \left[\delta(t)\delta(t')+\frac{1}{\omega^2} \overleftarrow{\partial_t}\delta(t)\delta(t')\overrightarrow{\partial_{t'}}\right]   \end{array}\right)
                                \left(\begin{array}{c}%
                                        x_{cl}(t')\\
                                        x_q(t') \end{array}\right).
 \label{actionu_sho_clq}                                   
\eeq
\end{widetext}
Note that in continuum, it is preferable to do a Keldysh rotation on the correlators in $\pm$ basis to get the correlators in $cl-q$ basis and then write an action in the $cl-q$ basis which reproduces the correct correlators. A direct field rotation on the continuum action Eq.~(\ref{action_sho}) does not reproduce Eq.~(\ref{actionu_sho_clq}). This is a well-known problem in standard Keldysh field theory in continuum, which does not appear in discrete time versions~\cite{kamenevbook}. The steps outlined here correctly reproduces all correlators in the system.

We note that the structure of the action is similar to that obtained
for a Schr\"{o}dinger theory of bosons, i.e. we add a bilinear source $u$ which couples
only to the quantum fields at the initial time. This anti-Hermitian term effectively acts as a
Keldysh self energy for the fields. However, there is one key
difference: unlike the Schr\"{o}dinger bosons, where the $u$ terms are only
coupled to bilinears of the fields at $t=0$, here the additional terms
couple both to the position and its time-derivatives at the initial
time. This is a reflection of the fact that the saddle point equation
is second order in time in this case and both
the initial position and the initial velocity of the particle are required to solve this equation.

There is an alternate way of deriving the effective action
Eq.~(\ref{actionu_sho_clq}). One can start with the coherent state
representation of the harmonic oscillator and using the theory derived
in Ref.~\onlinecite{cgs}, derive the effective action in the
Keldysh rotated basis of complex Schr\"{o}dinger bosons. One can then
write the coherent state fields in terms of position and momentum
degrees of freedom in a path integral on the phase space. Integrating
out the momentum degrees of freedom, one can arrive at the effective
action derived above. In this case, it is easy to see that the initial
sources couple to both position and momentum, and hence, on integrating
out the momentum degrees of freedom, they couple to the time derivative of
the position at initial times. We note that this
derivation works when $\omega$ and $\Omega$ are same, while the detailed
derivation provided in this paper also works even when the Fock states of
the initial density matrix are not the eigenstates of the closed system
dynamics.

Before
we move to the case of the scalar fields, we need to discuss how
the dynamics changes if the system is coupled to an external
bath starting at $t=0$, leading to open quantum system dynamics. The addition of a bath ( at least the simplest baths where the
position of the oscillator couples to the bath degrees of freedom),
leads to a quadratic theory where the effect of the bath can be
incorporated through a retarded self-energy $\Sigma^R$ and a Keldysh
self-energy $\Sigma^K$,
corresponding respectively to dissipative and stochastic effects of the bath. The
Keldysh self energy simply adds to the initial self-energy from the
bilinear sources, so the correlators in presence of the external
bath are given by 
\begin{widetext}
  \begin{eqnarray}
    D^R(t-t') &=&D^R_0(t-t') + \int_{t'}^t ~ dt_1 ~\int_{t'}^{t_1} ~dt_2 ~
                  D^R_0(t-t_1) \Sigma^R(t_1-t_2) D^R(t_2-t')
                  \label{gfn_int_sk}\\
 \no   D^K(t,t',\rho_0)&=& \left.-2im\omega\sum_n c_n(1+2n) D^R(t,\epsilon)\left[1+\frac{1}{\omega^2}
    \overleftarrow{\partial_\epsilon}\overrightarrow{\partial_\epsilon}\right]D^A(\epsilon,t')\right\vert_{\epsilon=0} + \int_0^t ~ dt_1 ~\int_0^{t'} ~dt_2 ~
                  D^R(t-t_1) \Sigma^K(t_1,t_2) D^A(t_2-t'),
    \end{eqnarray}
\end{widetext}
where $D^R_0(t-t')$ is the retarded correlator in absence of the bath (given by Eq.~\ref{gfnu_sk}).

Finally, we note that one can introduce effects of interaction by
adding an interaction term in the Keldysh action. In this case, the
dynamics is not exactly solvable, and one needs to make
approximations to treat effects of interaction on the open system
dynamics. It is important to note that even in this case, one can use
standard diagrammatic techniques for calculating $D(u)$, and then
obtain $D(\rho_0)$ by taking appropriate derivatives.

\subsection{Scalar Fields\label{Rscalar}}

We would now like to extend our formalism to the dynamics of 
scalar fields.  In this paper, we will discuss the case of real scalar
fields for the sake of brevity, but the formalism can be extended to
complex scalar fields, or to $O(N)$ fields in a straightforward
way. We consider real scalar fields $\phi(\textbf{x},t)$,
whose dynamics is governed by the action
\begin{widetext}
\beq
S= \int dt ~ \int d^d x ~  (\phi_{cl}(\textbf{x},t) ~,~ \phi_{q}(\textbf{x},t))
\left( \begin{array}{ cc}
         0 & -\partial_t^2 +c^2 \nabla^2 -m^2\\
        -\partial_t^2 +c^2 \nabla^2 -m^2 & 0 \end{array}\right)
    \left(\begin{array}{c}
            \phi_{cl}(\textbf{x},t) \\
            \phi_{q}(\textbf{x},t)\end{array}\right).
 \label{action_rescalar}           
 \eeq
 
 Working with momenta instead of real space, this gives
 \beq
S= \int dt ~ \int d^d k ~  (\phi_{cl}(-\textbf{k},t) ~,~ \phi_{q}(-\textbf{k},t))
\left( \begin{array}{ cc}
         0 & -\partial_t^2 -\Omega_{\textbf{k}}^2\\
        -\partial_t^2 -\Omega_{\textbf{k}}^2& 0 \end{array}\right)
    \left(\begin{array}{c}
            \phi_{cl}(\textbf{k},t) \\
            \phi_{q}(\textbf{k},t)\end{array}\right),
 \label{action_rescalar_mom}           
 \eeq
 where $\Omega_{\textbf{k}} =\sqrt{c^2\textbf{k}^2+m^2}$ is the dispersion of a generic
 massive scalar field. One can of course set $m=0$ to describe
 massless scalar fields.  We note that since the theory would be written
 in terms of $\Omega_{\textbf{k}}$, one can also put the system on a lattice and
 consider lattice dispersions with associated finite range of lattice
 momenta (Brillouin zones).
\end{widetext}
 We assume that the system is initialized to a density matrix which is
 diagonal in the Fock basis of the
 number operators $\hat{n}_\textbf{k}=a^\dagger_\textbf{k}a_\textbf{k}$, where $a_\textbf{k} =
 \sqrt{\frac{\omega_\textbf{k}}{2}} \phi(\textbf{k}) +i
 \sqrt{\frac{1}{2\omega_\textbf{k}}}\dot{\phi}(\textbf{k})$. Here $\omega_\textbf{k}$ is a
 dispersion which can be different from $\Omega_\textbf{k}$, and hence the
 initial state may not be an eigenstate of the Hamiltonian which generates
 the dynamics of the system.

 In this case, we can generalize the answers we obtained for the simple
 harmonic oscillator in a straightforward way. The source $u$ is generalized to a source field $u_{\textbf{k}}$ for each mode $\textbf{k}$ and the source dependent action in the Keldysh basis is 
 \begin{widetext}
 	\beq
 	S= \int dt \int dt'~ \int d^d k ~  (\phi_{cl}(-\textbf{k},t) ~,~ \phi_{q}(-\textbf{k},t))
 	\left( \begin{array}{ cc}
 		0 & \delta(t-t')(-\partial_t^2 -\Omega_\textbf{k}^2)\\
 	\delta(t-t')(-\partial_t^2 -\Omega_{\textbf{k}}^2)& i\omega_{\textbf{k}} \frac{1+u_{\textbf{k}}}{1-u_\textbf{k}} \left[\delta(t)\delta(t')+\frac{1}{\omega_{\textbf{k}}^2} \overleftarrow{\partial_t}\delta(t)\delta(t')\overrightarrow{\partial_{t'}}\right]  \end{array}\right)
 	\left(\begin{array}{c}
 		\phi_{cl}(\textbf{k},t') \\
 		\phi_{q}(\textbf{k},t')\end{array}\right).
 	\label{action_rescalar_field}           
 	\eeq
\end{widetext} 	
 We consider a system starting from $\rho_0=\sum_{\{n\}}c_{\{n\}}\vert \{n\} \rangle \langle \{n\}\vert$. The physical correlation functions for this dynamics is given by 
 \begin{widetext}
 	\begin{eqnarray}
 	D^R(\textbf{k};t-t') &=&D^R_0(\textbf{k};t-t') + \int_{t'}^t ~ dt_1 ~\int_{t'}^{t_1} ~dt_2 ~
 	D^R_0(\textbf{k};t-t_1) \Sigma^R(\textbf{k};t_1-t_2) D^R(\textbf{k};t_2-t')
 	\label{gfn_rsfield}\\
 \no   D^K(\textbf{k};t,t',\rho_0)&=& \left.-2i\omega_\textbf{k}\sum_{\{n\}} c_{\{n\}}(1+2n_\textbf{k}) D^R(\textbf{k};t,\epsilon)\left[1+\frac{1}{\omega_\textbf{k}^2}
 	\overleftarrow{\partial_\epsilon}\overrightarrow{\partial_\epsilon}\right]D^A(\textbf{k};\epsilon,t')\right\vert_{\epsilon=0} \\
 	\no &+& \int_0^t ~ dt_1 ~\int_0^{t'} ~dt_2 ~
 	D^R(\textbf{k};t-t_1) \Sigma^K(\textbf{k};t_1,t_2) D^A(\textbf{k};t_2-t'),
 	\end{eqnarray}
 \end{widetext}
where $\Sigma^R(\textbf{k};t-t')$ and $\Sigma^K(\textbf{k};t,t')$ are retarded self-energy and Keldysh self-energy respectively for each mode. The self energies can arise if the system is coupled to an external bath, or from the interaction between the modes of the scalar fields. 
We will now use this formalism to study the non-equilibrium dynamics of scalar fields coupled to external baths. While this formalism is valid for any scalar fields, we will look into the concrete example of phonons coupled to various kinds of external baths to illustrate the use of this formalism.

\section{Dynamics of Phonons coupled to external bath \label{OQS}}

\begin{figure*}[t]
	\centering
	\includegraphics[width=\textwidth]{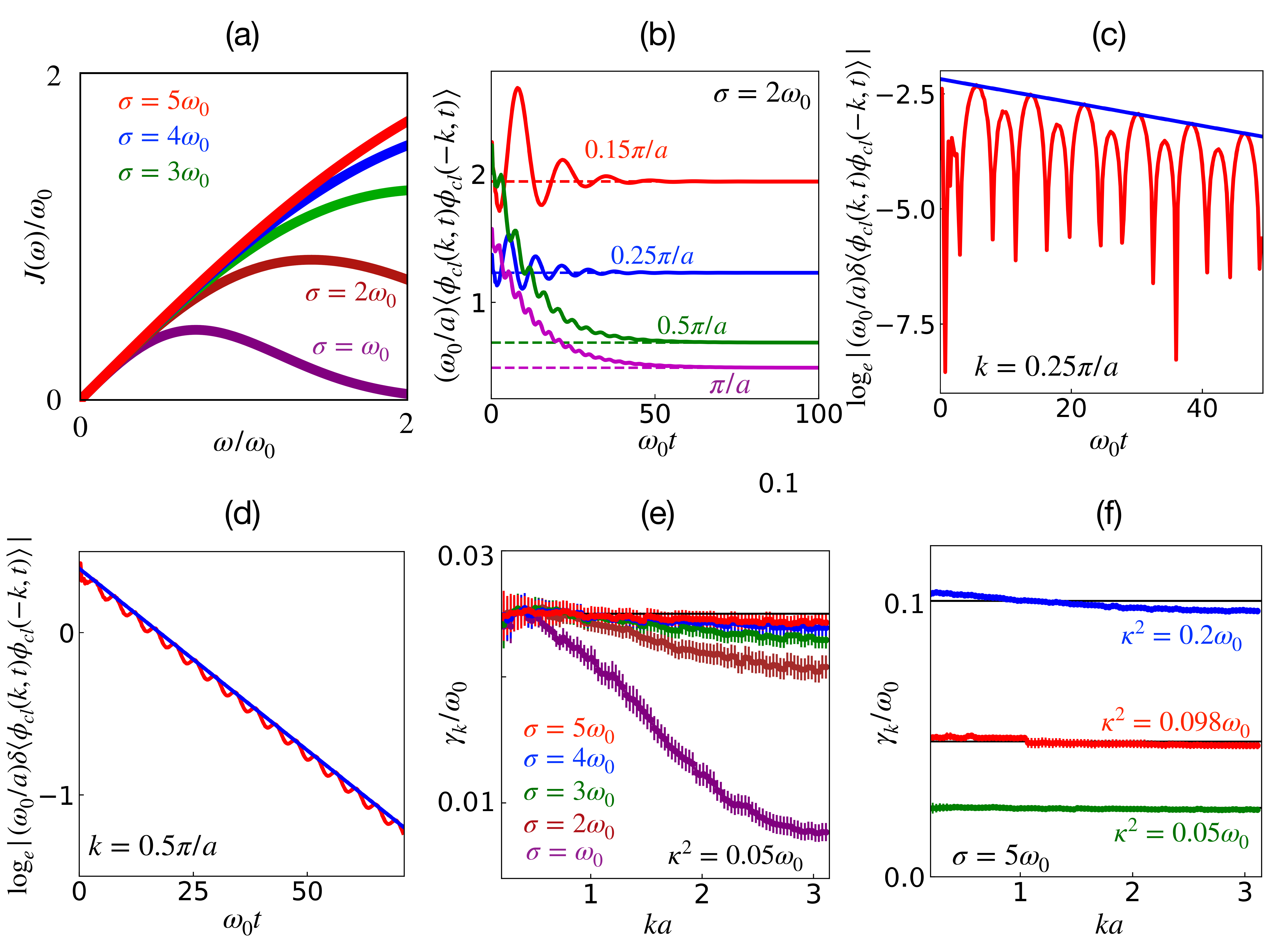}
	\caption{Non-equilibrium dynamics of phonons coupled to an ohmic bath: 
		(a) The bath spectral function $J(\omega)=\omega
		e^{-\frac{\omega^2}{\sigma^2}}$ as a function of
		$\omega$ for different values of $\sigma$. Here $\omega_0$ is the bandwidth of bare phonons. $J(\omega)$ deviates from linear behaviour at progressively smaller energies as $\sigma$ is decreased.
		(b) Thermalizing dynamics of correlation function
		$iD^K(k;t,t)$ for four different momentum modes at a
		low temperature $T=0.02\omega_0$ with system-bath
		coupling strength $\kappa^2=0.05\omega_0$. Here $\sigma=2\omega_0$. The time dependent correlation functions (solid lines) relax to their thermal values (dashed lines) at long times. 
		(c) and (d) Difference of the time dependent
		correlation function and its thermal value as a
		function of time in a log plot for two momentum
		modes, (c) $k=0.25\pi/a$, whose initial occupation is 0 and (d) $k=0.5\pi/a$, whose initial occupation is 1. The decay rate is extracted from the slope of the envelope in the log plot. 
		(e) Decay rate ($\gamma_k$) versus momentum ($k$) for different values of the effective bath bandwidth $\sigma$. $\gamma_k$ approaches the constant value $\kappa^2/2$ independent of k as $\sigma/\omega_0$ is increased and the bath approaches the limit of pure white noise.
		(f) Decay rate ($\gamma_k$) versus momentum ($k$) for different values of system bath coupling. $\sigma$ is fixed at the highest value $5\omega_0$ we considered. We see that $\gamma_k$ goes very close to the constant value $\kappa^2/2$. Note that with increasing system bath coupling, $\gamma_k$ starts to have a weak $k$ dependence.
	}
	\label{bosonbath}
\end{figure*}
 
We now consider the dynamics of a system of phonons (quantized lattice vibrations) initialized to
a non-thermal state and coupled to an external bath.
In one dimension, the bare dispersion of the phonons is given by
\beq
\Omega_k=\sqrt{ \omega_0^2 \sin^2 (ka/2) +m^2},
\label{ph1d}
\eeq
where $a$ is the lattice spacing and $m$ is the mass, which can be set to $0$ to describe
longitudinal phonons. For the massless phonons, the long
wavelength modes have a linear dispersion $\Omega_k \sim c_s k$, with
the sound velocity $c_s= \frac{\omega_0 a}{2}$. Since we are working on a lattice, the Brillouin zone extends from
$k=-\pi/a$ to $k= \pi/a$. The bandwidth of the massless phonons is then
given by $\omega_0$, which controls both the ultraviolet and the
infrared dispersions in this simple model. We also note that this kind of dispersion is not specific to phonons, there exist other systems (for example magnons or quantized spin waves), which can be described by similar scalar field dispersion. The massive fields, which can be used to represent optical phonons, have a low
energy dispersion $\Omega_k \sim \sqrt{c_s^2 k^2+m^2} \sim m +
\frac{k^2}{2m^\ast}$, where the curvature
$m^\ast \sim m/c^2_s$.

In two dimensions, we consider phonons on a square lattice of
lattice spacing $a$, with the bare dispersion given by
\beq
\Omega_\textbf{k}=\sqrt{ \omega_0^2[\sin^2 (k_xa/2)+ \sin^2 (k_ya/2)] +m^2}.
\label{ph2d}
\eeq
Once again, for the 
 massless scalar fields, the long
wavelength modes have a linear dispersion $\Omega_\textbf{k} \sim c_s\vert \textbf{k}\vert$, with speed $c_s= \frac{\omega_0 a}{2}$, while for the massive fields, the low
energy dispersion $\Omega_\textbf{k} \sim \sqrt{c_s^2 \textbf{k}^2+m^2} \sim m +
\frac{\textbf{k}^2}{2m^\ast}$, where $m^\ast$ has the same form as in the one dimensional case. We note that for a square lattice, the Brillouin zone extends from
$k_x=-\pi/a$ to $k_x= \pi/a$ and $k_y=-\pi/a$ to $k_y= \pi/a$. 

We study the dynamics of these phonon modes starting from an athermal
initial condition. Since the coupling to the bath will anyway generate
non-trivial dynamics, we consider the initial state of the system to
be a Fock state corresponding to the bare phonon dispersion
$\Omega(\textbf{k})$  i.e. we consider the annihilation operator $a_\textbf{k}=
\sqrt{\frac{\Omega_\textbf{k}}{2}} \phi(\textbf{k}) + i\sqrt{\frac{1}{2\Omega_\textbf{k}}}
\dot{\phi}(\textbf{k})$ and the occupation number states corresponding to
$\hat{n}_\textbf{k}=a^\dagger_\textbf{k}a_\textbf{k}$. 
In one dimension, we consider a lattice of even number of
sites $2L$ with momenta $k= \pm\pi i/L$ with integer $i$. The initial
Fock state then corresponds to $n_k=1$ for even values of $i$ and $0$
for odd values of $i$.
Similarly in two dimensions we consider a square lattice of $(2L)^2$
sites with momenta $(k_x,k_y)= (\pm\pi i/L,\pm\pi j/L)$ with integer $i$ and $j$. The initial
Fock state then corresponds to $n_\textbf{k}=1$ for even values of $i$ and $n_\textbf{k}=0$
for odd values of $i$, so we have strips of occupation numbers 0 and 1 along the $y$-axis. 
These states are quite different from
a thermal state of the bosons. The thermal state has a large occupancy near $\textbf{k}=(0,0)$
which decreases monotonically as $\textbf{k}$ is increased. 
The main broad features of the dynamics, like presence/absence of thermalization or the rates of thermalization, do not depend on this particular choice of initial state. However, details of phase oscillations in systems that do not thermalize will depend on the initial conditions. The initial conditions are chosen to provide a simple system to distinguish between the modes that thermalize versus the modes that do not thermalize. 
We will now consider different kinds of baths and study their effects on the dynamics of the phonons.

\subsection{Phonons Coupled to an Ohmic Bath \label{bosebath}}
We first consider phonons in one dimension coupled to a bosonic bath. The effect of the bath on the phonons is controlled by the bath spectral function $J(\omega)$.
Here we consider the case where all the phonon modes see the same ohmic bath with $J(\omega)=\omega e^{-\frac{\omega^2}{\sigma^2}}$. This bath spectral function is linear in energy for low $\omega(<<\sigma)$ which is characteristic of the ohmic bath \cite{qds_ohm,diss2ss_legget}. The parameter $\sigma$ ensures that
the bath is well behaved in the ultra-violet limit with the spectral function smoothly decaying to zero at large energy. So $\sigma $ plays the role of an effective bandwidth for the bath without producing non-analyticities which lead to non-Markovian dynamics \cite{nonmarkov_oqs}. 
The bath spectral function $J(\omega)$ is plotted as a function of energy for five different values of $\sigma$ ranging from $\sigma=\omega_0$ to $\sigma=5\omega_0$ in Fig.~\ref{bosonbath}(a). It is clear that the bath spectral function deviates from its linear behaviour at progressively lower energies as $\sigma$ is decreased. We note that we have absorbed certain constants into the system bath coupling here, so that both $\kappa^2$ and $J(k,\omega)$ have dimensions of energy in this problem.

In this case, the momentum independent retarded self energy is given by 
\beq
\Sigma^R(\omega)=-\kappa^2\int_{-\infty}^{\infty}\frac{dz}{2\pi}J(z)\left[\frac{1}{z-\omega-i0^+}+\frac{1}{z+\omega+i0^+}\right].
\label{sigr_ob}
\eeq 
Note that $Im~\Sigma^R(\omega)=-\kappa^2J(\omega)$ and $Re~\Sigma^R(\omega)=-\frac{\kappa^2}{\sqrt{\pi}}\left[\sigma-2\omega  F_D\left(\frac{\omega}{\sigma}\right)\right]
$, where $F_D(x)=e^{-x^2}\int_0^x  dy\ e^{y^2}$ is the Dawson's function \cite{dawson}.

The Keldysh self energy is
\beq
\Sigma^K(\omega)=-i\kappa^2 2\coth\left(\frac{\omega}{2T}\right)J(\omega),
\label{sigk_ob}
\eeq
which follows from fluctuation-dissipation theorem \cite{kamenevbook}.
Here $T$ is the temperature of the bath. 

If the effective bandwidth of the bath $\sigma$ is much larger than the phonon bandwidth $\omega_0$, then
all the phonon modes effectively see a Gaussian white noise and one would expect the modes to thermalize at the same rate $\gamma=\kappa^2/2$. On the other hand, if $\sigma$ is less than or comparable to $\omega_0$ then the relaxation of different phonon modes depends on their energies with the slowest relaxation rate for the highest phonon energy.
To see this, we track the dynamics of the phonons coupled to this ohmic bath 
($T=0.2\omega_0$ and $\sigma=2\omega_0$) with a system-bath coupling strength $\kappa^2=0.05\omega_0$. In Fig.~\ref{bosonbath}(b), we plot the time evolution of the correlation function $\langle\phi_{cl}(k,t)\phi_{cl}(-k,t)\rangle$ for four different values of momentum $k$. The correlation function decays to its thermal value with damped oscillations. While the steady state is independent of the initial condition, the modes which are initially populated approach the steady value from above while the correlation in the unpopulated modes oscillate around the long time value. We plot the absolute value of the deviation of the correlation function from its steady value on a log scale for $k=0.25\pi/a$ ($n_k=0$ at $t=0$) in Fig.~\ref{bosonbath}(c) and for $k=0.5\pi/a$ ($n_k=1$ at $t=0$) in Fig.~\ref{bosonbath}(d). An exponential fit to the envelope of such curves is used to determine the momentum dependent decay rate $\gamma_k$. In Fig.~\ref{bosonbath}(e), we plot $\gamma_k$ as a function of $k$ for the five different bath bandwidths ranging from $\sigma=\omega_0$ to $\sigma=5~\omega_0$. For $\sigma=\omega_0$ we find that the decay rate strongly depends on $k$, decreasing by a factor of 2.5 as we go from the zone centre to the edge of the Brillouin zone. As $\sigma$ is increased, the dependence of $\gamma_k$ on $k$ is weakened with an almost $k$-independent $\gamma$ for $\sigma=5\omega_0$. The large bandwidth bath thus behaves like a source of Gaussian white noise \cite{quantnoise}. In the large bandwidth limit, we consider the effect of the system bath coupling on the decay rate $\gamma$ in Fig.~\ref{bosonbath}(f). As expected, the system decays faster as system bath coupling is increased. We note that with increasing system bath coupling $\gamma$ acquires a weak $k$-dependence even for $\sigma=5\omega_0$.

\subsection{Phonons Coupled to a Fermionic Bath \label{fermibath}}

 We now consider the dynamics a system of scalar fields initialized to
 a Fock state and coupled to a bath of
 non-interacting fermions. This is prototype of a system which can be
 found in varied contexts in nature, like phonons coupled to
 electrons \cite{agd,ephint,ephintobs}, light coupled to metallic systems \cite{phtelint1,phtelint2}, ultracold atoms in cavity \cite{coldatmcvtrev}, descriptions of
 early universe and multi-component dark matter systems \cite{axiontherm}. For concreteness, we will consider phonons coupled to non-interacting fermions. 
 
 We consider a bath of non-interacting spinless fermions with a Hamiltonian given by
 \beq
 H_F= \sum_\textbf{k}\epsilon_{\textbf{k}} \ \psi^\dagger(\textbf{k}) \psi(\textbf{k}), 
 \label{fb}
 \eeq
 where $\psi(\textbf{k})$ are the fermion annihilation operators in mode $\textbf{k}$ and $\epsilon_{\textbf{k}}$ is the corresponding dispersion. 
We consider fermions hopping on a linear chain in one dimension and on
a square lattice in two dimensions with only nearest neighbour hopping,
which results in the dispersions $\epsilon_k=-\epsilon_B \cos ka$ and
$\epsilon_{\textbf{k}}=-\epsilon_B[ \cos k_xa + \cos k_ya]$
respectively, where $a$ is the lattice constant. Here, the bandwidth
of the fermions is given by $2\epsilon_B$ (1-D) and $4\epsilon_B$
(2-D) respectively. The fermionic bath is
 characterized by a temperature $T$ and a chemical potential $\mu$,
 which fixes the particle density in the bath.

 The phonons couple to the fermionic bath through the Hamiltonian
 \beq
 H_{int} = \kappa \sum_{\textbf{k},\textbf{q}} \lambda(\textbf{k})\psi^\dagger(\textbf{k}+\textbf{q})\psi(\textbf{q}) \phi(\textbf{k}), 
 \label{eph}
 \eeq
 where $\kappa$ is the system bath coupling strength and the form factor $\lambda(\textbf{k})=\sqrt{\sum_{j=1}^D\sin^2(k_ja/2)}$ is related to the deformation potential acting between electrons and phonons \cite{mahan}. Here $\phi(q)$
 has the dimension of $[ \sqrt{\textrm{volume}/ \textrm{energy}}]$,
 while $\psi(k)$ has the dimension of $\sqrt{\textrm{volume}}$, hence
 $\kappa^2$ has the dimension of $[\textrm{energy}]^3\times \textrm{volume}$. We would like
 to note that the system bath coupling in this problem has different
 dimensions than in the problem with the ohmic bath.
 
 The non-unitary dynamics of the phonons is governed by the retarded self energy $\Sigma^R(\textbf{k};t-t')$ and the Keldysh self energy $\Sigma^K(\textbf{k};t,t')$. The real part of $\Sigma^R$ leads to the dressing of phonon dispersion while its imaginary part controls the dissipation in the system. The Keldysh self energy $\Sigma^K$ controls the stochastic noise from the external bath. It is clear from Eq.~(\ref{eph}) that the phonons actually couple to the particle-hole excitations of the fermionic system. The retarded self energy  is then given by,
 \begin{equation}
 \Sigma^R(\textbf{k},\omega)=\kappa^2\lambda(\textbf{k})^2\sum_{\textbf{q}}\frac{F(\epsilon_{\textbf{k}+\textbf{q}})-F(\epsilon_\textbf{q})}{\omega+i\eta+\epsilon_\textbf{q}-\epsilon_{\textbf{k}+\textbf{q}}},
 \label{sigr_fb}
 \end{equation}
 where $F(x)=\tanh\left(\frac{x-\mu}{2T}\right)$ (see Appendix~\ref{sigma_fermibath} for a derivation). The imaginary part
 of $\Sigma^R$, which controls the dissipation in the system, is
 related to the spectral function of the effective bath
 $J(\textbf{k},\omega)$ by
 \begin{equation}
   \begin{split}
   &\textrm{Im} [\Sigma^R(\textbf{k},\omega)]=-\kappa^2
   J(\textbf{k},\omega)\\
   &=-\pi\kappa^2\lambda(\textbf{k})^2\sum_{\textbf{q}}\delta(\omega-\epsilon_{\textbf{k}+\textbf{q}}+\epsilon_{\textbf{q}})[F(\epsilon_{\textbf{k}+\textbf{q}})-F(\epsilon_\textbf{q})].
 \end{split}
 \end{equation}
   
 Using fluctuation-dissipation theorem, the Keldysh self energy is given by
 \begin{equation}
 \begin{split}
 &\Sigma^K(\textbf{k},\omega)=2i\coth\left(\frac{\omega}{2T}\right)\textrm{Im} [\Sigma^R(\textbf{k},\omega)]\\
 &=i2\pi\kappa^2\lambda(\textbf{k})^2\sum_{\textbf{q}}\delta(\omega-\epsilon_{\textbf{k}+\textbf{q}}+\epsilon_{\textbf{q}})[F(\epsilon_{\textbf{k}+\textbf{q}})F(\epsilon_\textbf{q})-1].
 \end{split}
 \label{sigk_fb}
 \end{equation}
For the non-equilibrium evolution of the phonon correlator, we find it easier to construct the retarded Green's function in frequency space, $D^R(\textbf{k},\omega)=\frac{1}{2[(\omega+i0^+)^2-\Omega_{\textbf{k}}^2]-\Sigma^R(\textbf{k},\omega)}$ and then Fourier transform $D^R(\textbf{k},\omega)$ and $i\omega D^R(\textbf{k},\omega)$ to get $D^R(\textbf{k},t-t')$ and its derivatives. $\Sigma^K(\textbf{k},\omega)$ is Fourier transformed to get $\Sigma^K(\textbf{k},t-t')$. The integrals for the Dyson equation for $D^K(\textbf{k},t,t)$ [Eq.~(\ref{gfn_rsfield})] are then performed numerically to obtain the physical correlators.

 The particle-hole excitations, to which the scalar fields couple, have qualitatively different density of states (DOS) in one and two dimensions. This difference is reflected in the dynamics of the scalar fields through the self-energies. So, we will consider the case of one and two dimensional systems separately. 

\subsubsection{One dimensional systems\label{1dfermi}} 
\begin{figure*}[t]
	\centering
	\includegraphics[width=\textwidth]{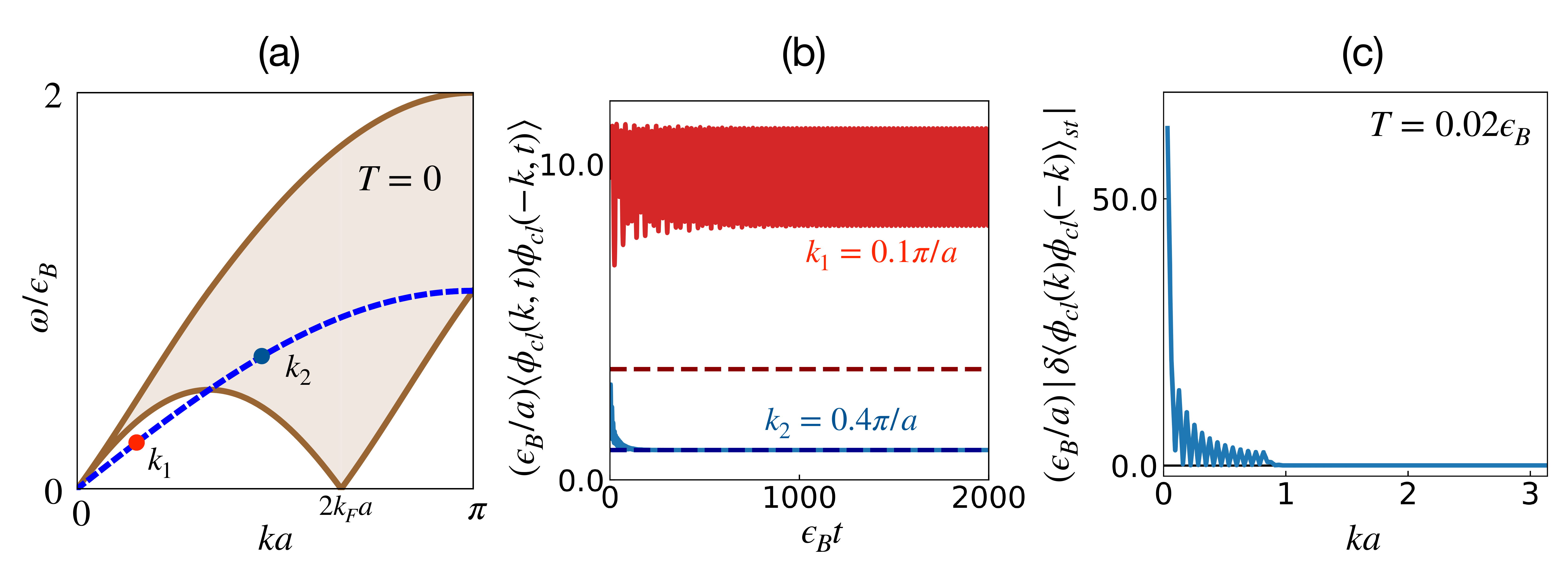}
	\caption{Non-equilibrium dynamics of phonons coupled to a low temperature fermionic bath in one dimension.
		(a) The region of finite density of particle-hole excitations in the fermionic bath at $T=0$ is shown
		in the $k-\omega$ plane as a shaded region. The solid brown lines define the upper and lower threshold energies. Note that the region collapses to a line (with slope $v_F$) at low momenta. The phonon dispersion for $c_s=\frac{1}{\sqrt{3}}v_F$ is also shown as a dashed line. The two momenta $k_1$ and $k_2$, shown by the dots are chosen such that for $k_1$, the phonon energy lies below the particle-hole bath while for $k_2$, the energy lies within the bath. 
		(b) Time evolution of the correlation function $iD^K(k;t,t)=\langle\phi_{cl}(k,t)\phi_{cl}(-k,t)\rangle$ for $k_1=0.1\pi/a$, where the phonons do not see a bath (red solid line) and for $k_2=0.4\pi/a$, where the phonons see an effective bath (solid blue line). $k_1$ and $k_2$ are are shown in (a). The correlator at $k_1$ oscillates about its initial value and never approaches its thermal value shown by the dashed red line. On the other hand, the mode at $k_2$ is damped and the correlator approaches its thermal value (dashed blue line) at long times. We have used a bath with $T=0.02\epsilon_B$ and $\mu=0.5\epsilon_B$ and a system-bath coupling strength $\kappa^2=0.5\epsilon_B^3a$. 
		(c) The deviation of the long time value of the
                correlator from its thermal value plotted as a
                function of momenta. The long wavelength modes do not
                thermalize while the higher momentum correlators
                approach their thermal value. The value of the
                correlator for modes that do not thermalize depend on
                the initial conditions. The oscillations corresponds
                to the initial (1,0,1,0,1,0...) pattern of occupation
                of the momentum modes.
	}
		\label{fermibath1dp1}
	\end{figure*}

\begin{figure*}
	\centering
	\includegraphics[width=\textwidth]{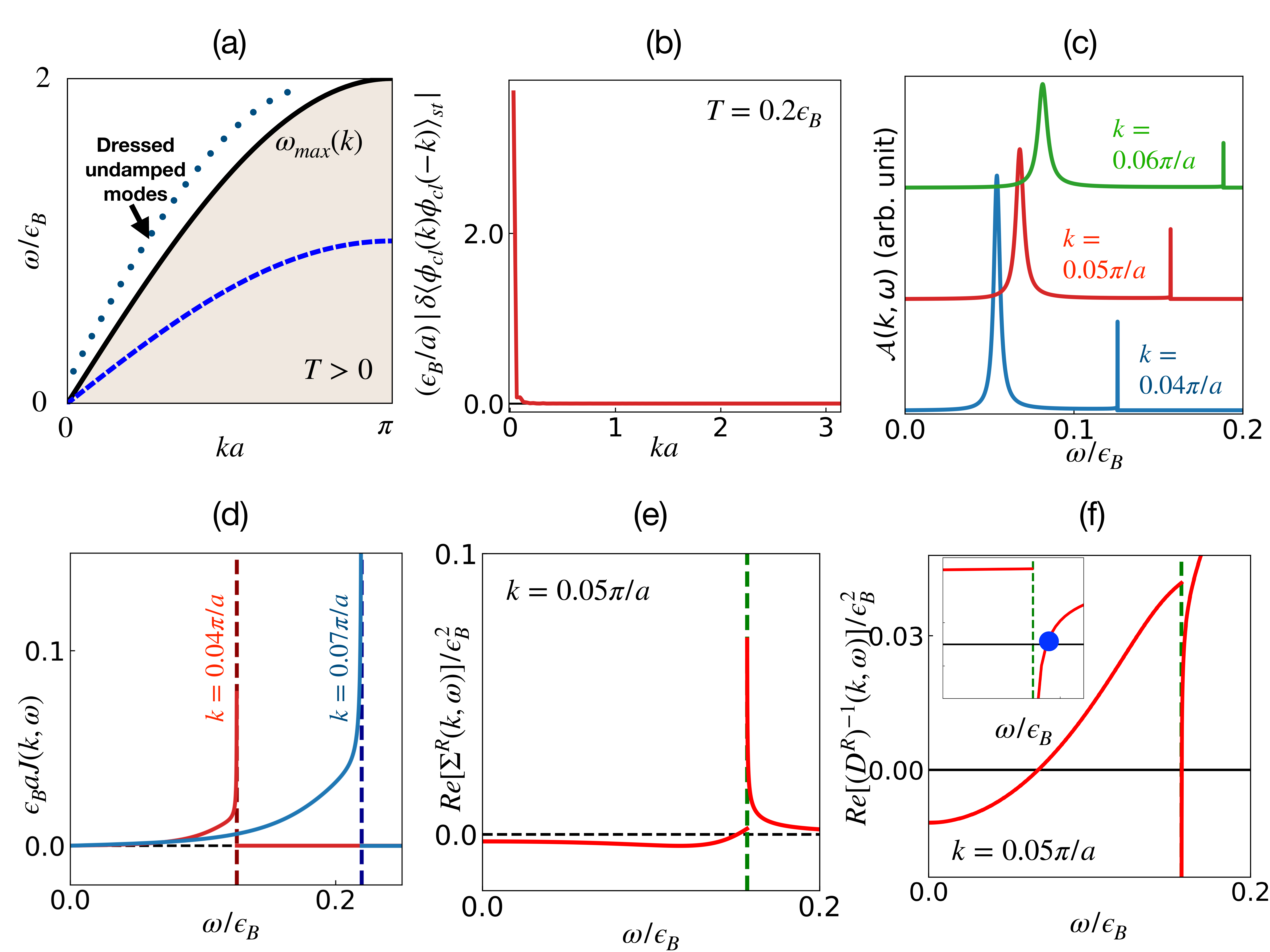}
	\caption{Dynamics of phonons coupled to high temperature fermionic bath and emergence of dressed undamped modes in one dimension. 
	(a) Particle-hole density of states (DOS) of the fermionic
        bath is finite only in the shaded region in the $k-\omega$
        plane at finite temperature. It is bounded above by the solid
        black line given by $\omega_{max}(k)=2\epsilon_B\sin(ka/2)$
        . The phonon dispersion with $c_s=0.5\epsilon_Ba$ is also
        plotted as a dashed blue line. The dots above $\omega_{max}(k)$ correspond to the undamped ``polarinon" modes formed due to strong fermion-phonon coupling.
	(b) The deviation of the correlation function at long times from its thermal value at a moderate temperature $T=0.2\epsilon_B$ and system-bath coupling strength $\kappa^2=0.5\epsilon_B^3a$. A few long wavelength phonon modes do not thermalize. 
	(c) Phonon spectral function for a few low momentum modes. The broader peak represents the original dressed damped modes of the phonons. The sharp peaks indicate undamped modes.
	(d) Bath spectral functions for $k=0.04\pi/a$ and $k=0.07\pi/a$ are plotted in solid red and blue lines respectively. They have sharp cut-offs at $\omega=\omega_{max}(k)$, indicated by dashed vertical lines. The spectral function has a $\frac{1}{\sqrt{\omega_{max}-\omega}}$ singularity near this edge.  
	(e) $Re\Sigma^R(k,\omega)$ as function of $\omega$ for $k=0.05\pi$. It diverges at $\omega=\omega_{max}(k)$ (vertical green dashed line) from above and goes to a constant value inside.
	(f) The real part of the inverse propagator of phonon modes dressed by the fermionic bath. The inverse propagator has two zero crossings, one near the bare phonon frequency, corresponding to damped dressed phonons, and one above the bath edge ($\omega_{max}(k)$, shown as dashed green line) corresponding to undamped polarinons [inset shows the region near $\omega_{max}(k)$ clearly showing the second zero crossing].
	}
	\label{fermibath1dp2}
\end{figure*}

\begin{figure}[h]
	\includegraphics[scale=0.13]{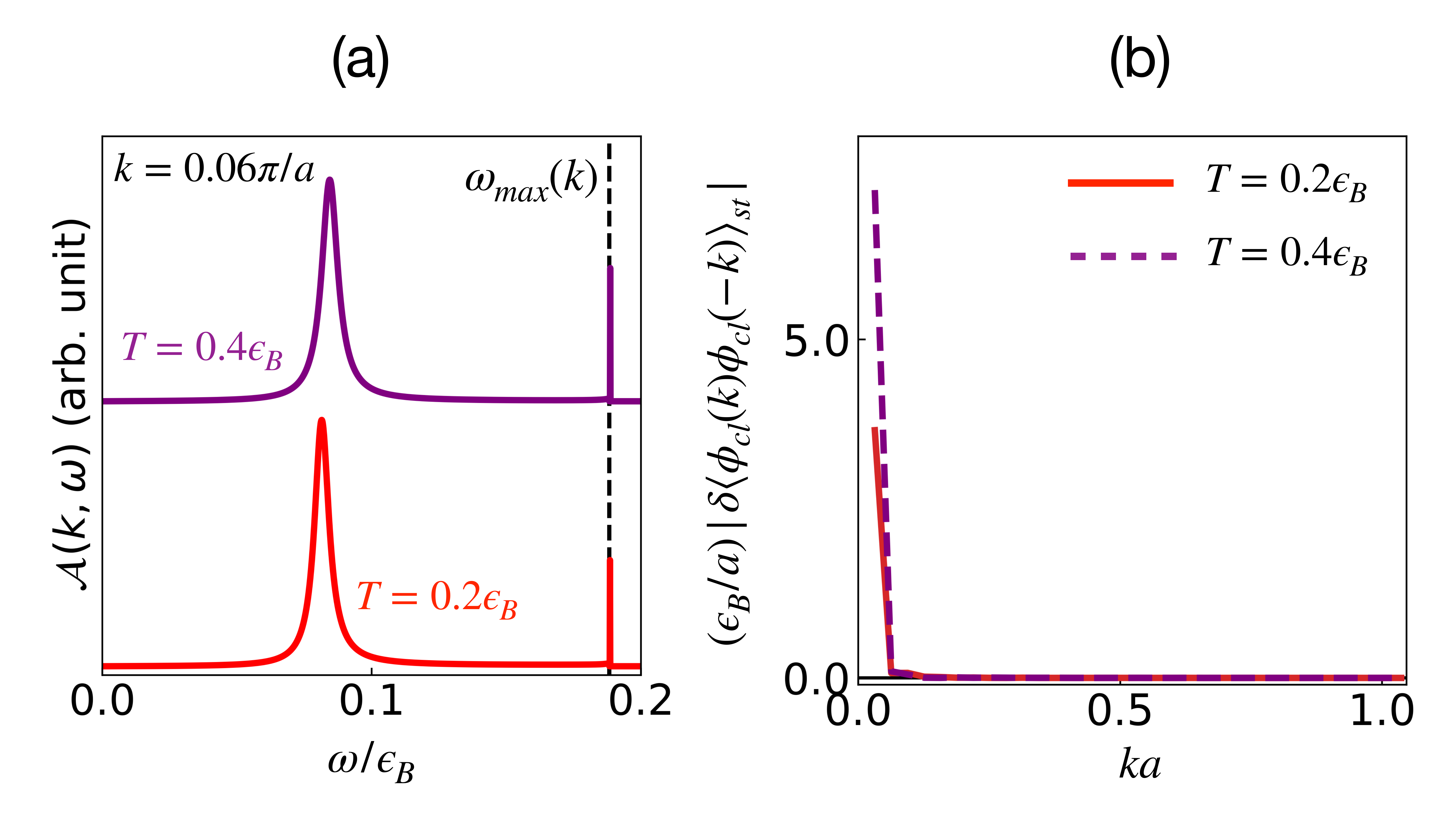}
	\caption{Effect of temperature on non-thermalizing behaviour of phonon modes.
	(a) Phonon spectral function at $T=0.2\epsilon_B$ and $T=0.4\epsilon_B$ for the mode $k=0.06\pi/a$ which does not thermalize. As temperature is increased, the polarinon mode remains sharp.
	(b) Absolute difference of the long time correlation function
        from its thermal value at long time at a high temperatures
        $T=0.2\epsilon_B$ (solid red line) and $T=0.4\epsilon_B$
        (dashed purple line). We notice finite deviation at low momenta at both the temperatures. The system-bath coupling strength is $\kappa^2=0.5\epsilon_B^3a$.
    }   
	\label{fermibath1dp3}
\end{figure}

There are two reasons why one dimensional systems show atypical behaviour: (a) energy-momentum conservation relations impose tight constraints on possible processes in one dimension and (b) the density of states of excitations have strong singular behaviour in one dimension. As we will see in this section, both these factors play an important role in the absence of thermalization for long wavelength phonons in one dimension.  

To understand the novel behaviour of the system, we focus on the
polarization function of a one dimensional free Fermi gas.
We first consider $T=0$. In
Fig.~\ref{fermibath1dp1}(a), the region in the $k-\omega$
plane where $J(k,\omega)$ is finite at $T=0$ is shown as a shaded
region bounded by solid lines. The upper limit is given by the curves
$\omega=(v_F/a)\sin~ka+\mu(1-\cos~ka)$ for $k \leq \vert 2k_F-\pi\vert$
and $\omega=2\epsilon_B\sin~(ka/2)$ for $k>\vert 2k_F-\pi\vert$, while the lower limit is given by $\omega=(v_F/a)\sin~ka-\mu(1-\cos~ka)$ for $k \leq 2k_F$
and $\omega=-(v_F/a)\sin~ka+\mu(1-\cos~ka)$ for $k>2k_F$, where $k_F$ is the Fermi wave-vector of the
Fermions and $v_F$ is the Fermi velocity. At low momenta, both the upper and lower limits $\sim v_F k$,
and thus the region actually merges into a single line with a slope of
$v_F$. For the particle-hole symmetric point at half-filling $(\mu=0)$, the collapse of upper and lower limits is exact for low $k$. As we move away from half-filling, the width of the region at low $k\sim(\mu a^2/2)k^2$. This is a
consequence of the strong constraints of energy and momentum
conservation in one dimension. In this
case, the long-wavelength phonons will not see any bath and undergo
unitary quantum dynamics unless the phonon velocity $c_s$ is exactly
equal to $v_F$. Note that, in an interacting
system, where one would expect a Luttinger liquid like behaviour~\cite{giamarchi}, one
would once again get a linearly dispersing mode with the velocity
tuned by the Luttinger parameter. Further, at low temperatures, the particle-hole spectral weight
outside this region is exponentially small. Since this fine tuning $(v_F=c_s)$ is impossible in real systems,
the long wavelength phonons will not thermalize at low temperature in one
dimension.
More precisely, for $T \ll c_s q$, the thermalization timescale will be exponentially large in inverse temperature ($\tau^{-1}\sim e^{-\frac{(v_F-c_s)k}{T}}$).
One might think that this statement is only true for $\mathcal{O}(\kappa^2)$ and when one includes two-phonon processes $[\sim\mathcal{O}(\kappa^4)]$, where the energy-momentum constraints can be relaxed, the decay rate of phonons will be a power law. In fact for scattering of a single massive particle by scalar fields, this is true, as shown in Refs.~\onlinecite{heavyparticle_castro,spinorgas_kamenev}. However for a finite density of fermions, with a Fermi surface, it can be shown that for $T\ll c_sk$, the decay rate remains exponentially small ($\tau^{-1}\sim ke^{-\frac{c_sk}{T}}$) even when two-phonon processes are included in the description. This difference primarily arises from the fact that while the single particle can be treated classically ($\frac{p^2}{2m}\sim T$), the particle-hole excitations in our problem remain quantum objects as long as the phonons retain their quantum nature ($T\ll c_sk$).
In the limit $T\gg c_sk$, we obtain $\tau^{-1}\sim T$.
The difference in the exponent  between our results and earlier works\cite{heavyparticle_castro,spinorgas_kamenev} comes from considering the fact that the relevant two-phonon process is mediated by a fermion with energy mismatch (i.e. off-shell) $\sim(v_F-c_s)k$. Note that this is the dominant relaxation process for $(v_F-c_s)k\gg T\gg c_sk$, while the relaxation is dominated by the single phonon process for $T\sim(v_F-c_s)k$. 
See Appendix~\ref{multiphonon} for the detailed calculation. Note that two-phonon processes involve effectively one particle-hole excitation. Three-phonon processes, which involve two particle-hole excitations also satisfy energy-momentum conservation constraints. They can result in a decay rate which is polynomial in temperature and hence not exponentially small. But these processes are of $\mathcal{O}(\kappa^8)$ and for small values of system-bath coupling they are highly suppressed.

To see this, we consider the case where the density of bath electrons
gives $\mu=0.5 \epsilon_B$, so that
$v_F=\frac{\sqrt{3}}{2}\epsilon_Ba$. We choose a phonon dispersion
with  $\omega_0=\epsilon_B$ so that $c_s=0.5\epsilon_Ba$, i.e. $c_s = \frac{1}{\sqrt{3}} v_F$.
Here the phonon dispersion lies below the electronic dispersion at
low momenta, as shown by the dashed line in
Fig.~\ref{fermibath1dp1}(a). We consider a low bath temperature of
$T=0.02 \epsilon_B$ to illustrate the low temperature behaviour of
the system. We have set the system-bath coupling strength $\kappa^2=0.5\epsilon_B^3a$. In Fig.~\ref{fermibath1dp1}(b), we plot
the time dependence of the correlation function $iD^K(k;t,t)=\langle
\phi_{cl}(k,t)\phi_{cl}(-k,t)\rangle$ for the setup given above (solid
lines) for two different momenta: a low
momentum $k_1=
0.1 \pi/a$, where the phonon dispersion is outside the energy range of
the bath, and a high momentum $k_2=0.4 \pi/a$, where the phonon
dispersion lies within the bath energy range. The thermal value of the
correlator is also indicated in this plot with dashed lines. It is
clear that the mode at $k_2$ thermalizes, whereas the long wavelength
mode $k_1$ does not thermalize in this case.
The correlator at $k_1$ undergoes unitary quantum dynamics with a dressed phonon energy, oscillating about its initial value and never approaching the thermal value shown with the dashed red line. On the other hand, the damped correlator at $k_2$ approaches the thermal value (dashed blue line) at long times.  
To see which modes are
thermalizing, we plot the difference between the long-time value of
the non-equilibrium correlators and its thermal value, $i\delta
D^K(k;t,t)=\delta\langle \phi_{cl}(k,t)\phi_{cl}(-k,t)\rangle$, as a
function of $k$ in Fig.~\ref{fermibath1dp1}(c). 
We use the value of the correlator at $\epsilon_Bt=10^4$ as the ``long time" value, and average over a timescale of $\epsilon_Bt\sim10^3$ to smooth out effects of the oscillations in modes that do not thermalize. It is clear that the low momentum modes that lie outside the effective bath do not thermalize. For these modes, the long time value of the correlator depends on the initial condition, and the oscillations with k represent the (1,0,1,0,1,0...) pattern of initial occupation.

The energy conservation constraints are relaxed as we increase the
bath temperature. In this case, the bath spectral function gains weight at
low energies starting from $\omega=0$ for all $k$. The region in the $k-\omega$
plane where $J(k,\omega)$ is finite at high temperatures is shown as
shaded region  in
Fig.~\ref{fermibath1dp2}(a). Note that the finite bandwidth of the
fermions still results in an ultraviolet cut-off in the bath; i.e. for
each $k$, there is a maximum energy $\omega_{max}(k)=
  2\epsilon_B\sin (ka/2)$ beyond
which the bath has no spectral weight at any temperature. This is
plotted as a solid line in Fig.~\ref{fermibath1dp2}(a).  One would thus
expect the low momentum modes to thermalize once the bath has
sufficiently high temperature, provided the phonon dispersion 
lies below this maximum energy, i.e. the phonon velocity $c_s <
\epsilon_B a$, which is expected to be valid in generic one
dimensional material systems like nanowires \cite{phnNbSe3nano,bandNbSe3}. In
Fig.~\ref{fermibath1dp2}(b), we plot the difference between the long time value and the thermal value of the correlator, $i\delta
D^K(k;t,t)=\delta\langle \phi_{cl}(k,t)\phi_{cl}(-k,t)\rangle$ for the
phonon modes with $c_s =\frac{v_F}{\sqrt{3}}$, for which we had earlier looked
at thermalization at low bath temperatures. The dispersion of this
mode is shown as a dashed blue line in
Fig.~\ref{fermibath1dp2}(a). We see that the low momentum modes
do not thermalize even at a reasonably high temperature of $T=0.2
\epsilon_B$, although all the modes now see a particle-hole bath
with sufficient spectral weight. We note that this lack of
thermalization persists if the temperature of the bath is further
increased.

To get an insight into the lack of thermalization of low momentum
modes even at high temperatures, in Fig.~\ref{fermibath1dp2}(c), we plot the spectral function of the
dressed phonons in the steady state,
 \beq
 {\cal A}(k,\omega)=
-(1/\pi)\textrm{Im} D^R(k,\omega),
\label{spectral_ph}
\eeq 
as a function of $\omega$ for
several values of $k$. In Fig.~\ref{fermibath1dp2}(c), the spectral functions for different modes are shifted by arbitrary amounts to make them visible. In addition to the original phonon mode, which
is dressed and has a width, we find another sharp mode with no
damping at an energy above the boundary of the particle-hole
bath. This mode appears due to the strong coupling between the phonons
and the electron-hole pairs, similar to polariton modes which occur
due to strong coupling between photons and excitons \cite{extnplrtn_th,extnplrtn_exp}. We will call
these modes the ``polarinon'' modes. A key difference with polaritons
is that unlike excitons, the particle-hole excitations are not
coherent; i.e. a simple mode coupling theory would not work here. Rather the
origin of this mode can be traced back to the strong $\sim
1/\sqrt{\omega_{max}(k)-\omega}$ divergence of the spectral density of
the particle-hole excitations at the band edge in 1-d, which gives a similar divergence in
$\textrm{Im} \Sigma^R(k,\omega)$. This is shown in
Fig.~\ref{fermibath1dp2}(d), where we plot the bath spectral
function for two different values of $k$ at $T=0.2 \epsilon_B$ and
$\mu=0.5\epsilon_B$.  Using Kramers-Kronig relations~\cite{arfken}, one
can easily show that this would also lead to a $\sim
1/\sqrt{\omega-\omega_{max}(k)}$ divergence in $\textrm{Re}
\Sigma^R(k,\omega)$ for frequencies just above the band edge. This is shown in
Fig.~\ref{fermibath1dp2}(e), where it is evident that $\textrm{Re}
\Sigma^R(k,\omega)\rightarrow +\infty$ as we approach the band edge
from above. In contrast, the real part of the self energy approaches a
constant as the frequency comes close to the band edge from below. The real part of
the retarded self energy dresses the spectrum, and its divergence
ensures that $D^{R-1}(k,\omega)= 2(\omega^2 -\Omega_k^2)
-\Sigma^R(k,\omega)$ has zeroes just above the upper edge of the bath;
i.e. close to $\omega \sim \omega_{max}(k)$. This is shown in
Fig.~\ref{fermibath1dp2}(f), where $\textrm{Re}~D^{R-1}(k,\omega)$
is plotted as a function of $\omega$, revealing the two zero
crossings. The lower energy zero crossing is related to the phonon mode (now dressed
by the bath), while the higher frequency zero crossing (shown more
clearly in the inset) is
related to divergence of the bath spectral function. Thus the
strong divergence of the bath density of states leads to an
additional pole in the dressed phonon Green's function above the upper edge of the bath. 
This undamped ``polarinon" mode, which shows up as a sharp
feature in Fig.~\ref{fermibath1dp2}(c), does not thermalize even at
high temperatures. This explains the lack of thermalization seen in
Fig.~\ref{fermibath1dp2}(b) even for high temperatures where the bare phonon
sees a substantial bath spectral density.

The dispersion of the polarinon mode $\omega^{pl}_k$ is shown schematically as a dotted line in
Fig.~\ref{fermibath1dp2}(a).
We note that the quasiparticle residue for the polarinon is given by (see Appendix~\ref{polarinon} for details)
\beq
Z^{pl}(k) =\left[ 4\omega^{pl}_k-\left.\frac{\partial
  \textrm{Re}\Sigma^R(k,\omega)}{\partial
  \omega}\right\vert_{\omega^{pl}_k}\right]^{-1}\sim \frac{\left[\omega^{pl}_k-\omega_{max}(k)\right]^{\frac{3}{2}}}{[\lambda(k)]^2}
\label{qpres}
\eeq
As $k$ increases, $\omega^{pl}_k-\omega_{max}(k)$ increases, while $1/[\lambda(k)]^2$ decreases. The suppression from the form factor dominates over the increase in separation between the polarinon energy and the band edge, and the quasiparticle residue of this mode decreases with increasing $k$. Thus the polarinon dominated lack of thermalization is not seen at large momenta. Further, since the polarinon occurs outside the band edge it is not smeared out by thermal fluctuations. This is shown in
Fig.~\ref{fermibath1dp3}(a), where we plot the spectral function of
a phonon mode with $k =0.06 \pi/a$ at two different temperatures,
$T_1=0.2\epsilon_B$ and $T_2=0.4 \epsilon_B$. While the original phonon mode broadens with increasing temperature, the polarinon mode remains sharp with almost constant spectral weight at these two temperatures. With increasing
temperature, the deviations of the phonon correlators from their
thermal values increases slightly, as seen in Fig.~\ref{fermibath1dp3}(b).

To summarize, in one dimension, phonons coupled to fermionic baths do
not thermalize, especially the modes at low momenta. At low
temperature, this is primarily due to strong energy momentum
constraints, which leads to a very narrow energy band of particle-hole
excitations at low momenta. However, the lack of thermalization at
high temperatures is dominated by the formation of undamped
polarinon modes with energies above the bath band edge due to strong
divergence of the bath density of states near the band edge. The
effect of these modes increases with
electron-phonon coupling in the system. The effect also increases if
the phonons are stiffer, so that the gap between the phonon dispersion
and the band edge is reduced.

\begin{figure*}[t]
	\centering
	\includegraphics[width=\textwidth]{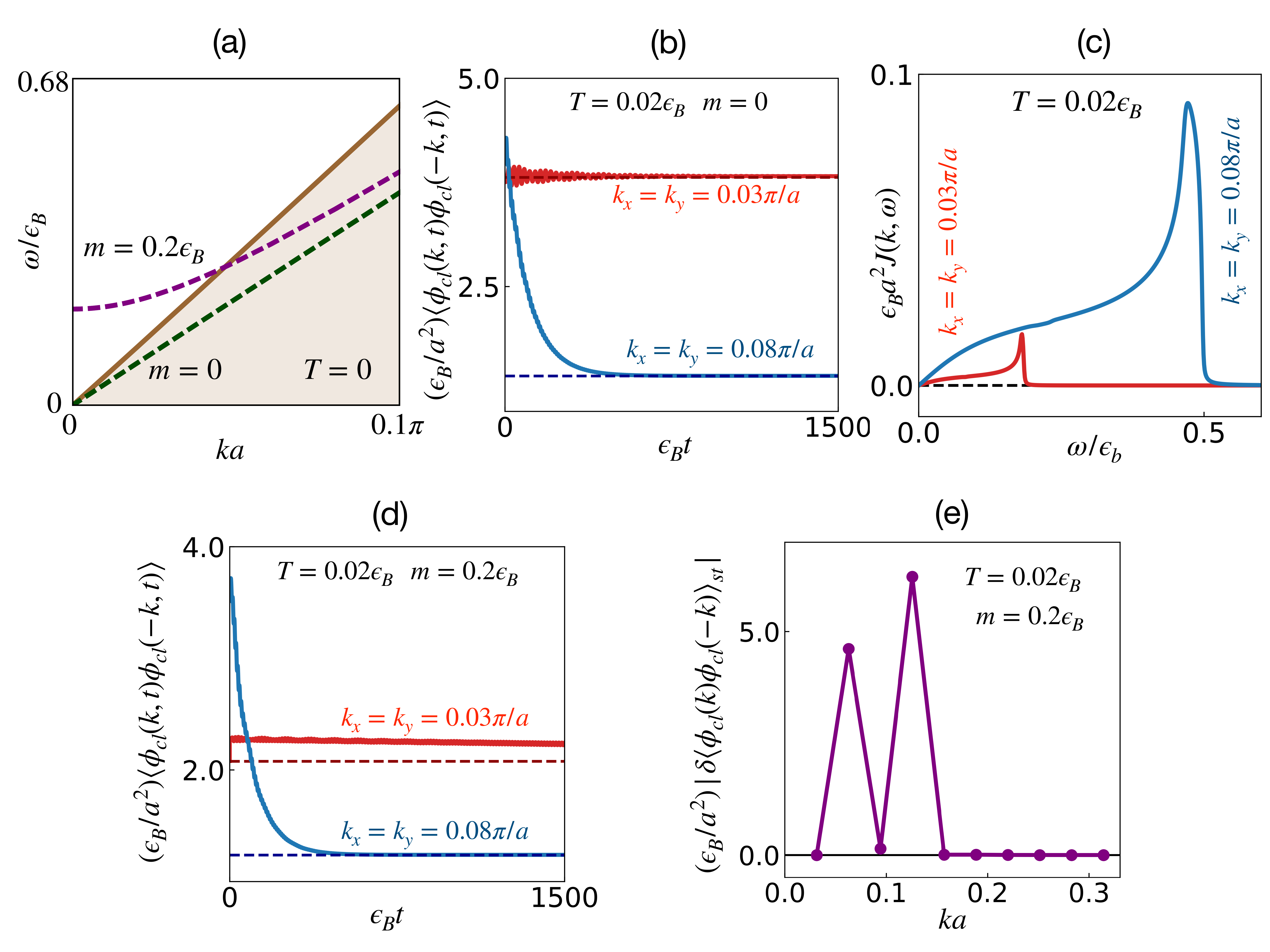}
	\caption{Non-equilibrium dynamics of phonons coupled to fermionic bath in two dimensions.
		(a) A schematic depiction of particle-hole density of states (DOS) of the fermionic bath on $\textbf{k}-\omega$ plane along $(1,1)$ direction where $\textbf{k}=(k,k)$ at $T=0$. The shaded region, from $\omega=0$ to $\omega_{max}(k)=\frac{\sqrt{2}v_F}{a}\sin~ka +\mu(1-\cos~ka)$ (shown in brown solid line) has finite DOS. The dispersion of acoustic (massless) phonons ($m=0$) with $c_s=\epsilon_Ba ~=0.73v_F$ is given by the dashed green line. The dashed purple line is the dispersion of optical (massive) phonons with $m=0.2\epsilon_B$ and $c_s=\epsilon_Ba~=0.73v_F$.
		(b) Time evolution of the correlation function $iD^K(\textbf{k};t,t)=\langle\phi_{cl}(\textbf{k},t)\phi_{cl}(-\textbf{k},t)\rangle$ at $\textbf{k}=(0.03\pi/a,0.03\pi/a)$ (red solid line) and $\textbf{k}=(0.08\pi/a,0.08\pi/a)$ (blue solid line) for massless phonons are plotted. Both the correlators relax to their thermal values (shown with dashed lines) at long times. 
		(c) Bath spectral functions for  $\textbf{k}=(0.03\pi/a,0.03\pi/a)$ and  $\textbf{k}=(0.08\pi/a,0.08\pi/a)$ are plotted in solid red and blue lines respectively. Their finite value at low $\omega$ indicates the absence of a lower bound in two dimensions. Unlike in one dimension, there is no divergence in the bath spectral function in two dimensions. 
		(d) Dynamics of the correlation function $iD^K(\textbf{k};t,t)$ at $\textbf{k}=(0.03\pi,0.03\pi)$ (red line) and  $\textbf{k}=(0.08\pi,0.08\pi)$ (blue line) for massive phonons with $m=0.2\epsilon_B$ and $c_s=\epsilon_Ba=0.73v_F$. The dispersion of the mode at $\textbf{k}=(0.03\pi/a,0.03\pi/a)$ lies above the band threshold and hence the correlator does not relax to its thermal value (shown by dashed red line). On the other hand, the energy of the mode at $\textbf{k}=(0.08\pi/a,0.08\pi/a)$ lies within the region of finite bath spectral function. So the corresponding correlator relaxes to its thermal value at long times.
		(e) The deviation of the long time value of the correlation function $iD^K(\textbf{k};t,t)$ from its thermal value as a function of momenta. The low momentum modes, with dispersion above the band threshold do not thermalize. All data in this figure are obtained for a bath temperature $T=0.02\epsilon_B$ and chemical potential $\mu=0.5\epsilon_B$ and system-bath coupling strength $\kappa^2=0.5\epsilon_B^3a^2$.
	}
	\label{fermibath2d}
\end{figure*}
\subsubsection{Two dimensional systems\label{2dfermi}}
In this section, we will consider the non-equilibrium dynamics of
phonons coupled to fermions in two dimensions and see how the dynamics
differs from that in one dimension. As we have seen before, the
dynamics is governed by the density of states of particle-hole pairs
of fermions. Contrary to one dimension, the energy momentum
conservation relations can be satisfied much more easily in two
dimensions. Hence even at $T=0$, the density of states of
particle-hole excitations with a given momentum transfer ${\bf k}$ remains
finite at arbitrary low energies. The finite bandwidth of the lattice
fermions leads to an upper threshold energy
$\omega_{max}({\textbf{k}})=\frac{\sqrt{2}v_F}{a}\sin~ka+\mu(1-\cos~ka)$ along the $(1,1)$ direction where  $\textbf{k}=(k,k)$.
It disperses linearly at low $|{\bf k}|$, i.e.
$\omega_{max}({\textbf{k}})\sim v_Fk$, where $v_F$ is the Fermi
velocity of the fermions in the bath. 
Beyond this threshold energy the density of states of particle-hole excitations vanish.
In Fig.~\ref{fermibath2d}(a) the particle-hole DOS is sketched schematically in the $\textbf{k}-\omega$ plane along the $(1,1)$ direction in $\textbf{k}$-space up-to $\textbf{k}=(0.1\pi/a,0.1\pi/a)$. In the figure, the upper bound $\omega_{max}(\textbf{k})$ is represented by a solid brown line. In the shaded region below this line we have finite particle-hole DOS.

Linearly dispersing long wavelength phonon modes with $c_s <
v_F$ would see a
bath with substantial spectral density even at very low temperatures
and would thermalize. In Fig.~\ref{fermibath2d}(a), we also plot the bare
dispersion of a longitudinal phonon mode with $c_s = 0.73 v_F$ as a green dashed line to
illustrate this point. Note that the criterion $c_s <v_F$ is satisfied
for most material systems other than compensated semimetals \cite{semimetal1,semimetal2,semimetal3} or bilayer
graphene \cite{graphene_band} near its charge neutrality point. In these
systems, by tuning the carrier density, one can possibly see a lack of
thermalization of the phonons, although at these very low densities,
disorder and interaction would play a very important role \cite{graphene_transport,graphene_int} and the
simple picture of a non-interacting bath has to be suitably
modified. In Fig.~\ref{fermibath2d}(b) we have plotted the dynamics of
the correlation function $\langle
\phi_{cl}(\textbf{k},t)\phi_{cl}(-\textbf{k},t)\rangle$ as a function of time for
massless phonon modes ($c_s = 0.73v_F$) with two different values
of momenta, ${\bf k}_1=(0.03\pi,0.03\pi)$ (red line)  and ${\bf
  k}_2=(0.08\pi,0.08\pi)$ (blue line) at a low temperature
$T=0.02\epsilon_B$ and system-bath coupling strength $\kappa^2=0.5\epsilon_B^3a^2$. The thermal values of the correlators are
indicated by the dashed lines. It is clear in this case that both
these modes thermalize at long times.

We would like to note that the bath density of states in two
dimensions do not have the strong divergence at the band edges that
was present in the one dimensional case. This is shown in
Fig.~\ref{fermibath2d}(c), where we plot the spectral density of the
particle-hole excitations for two different momentum transfers. The
softening of the band edge non-analyticity means that there is no
corresponding divergence in the real part of self energy and hence
there are no additional undamped dressed states (poles in the Green's
function) of the system. Thus the system continues to thermalize as
temperature is increased, unlike the one dimensional phonons.

The thermalization of the scalar fields can be prohibited if the
fields are massive. In the context of phonons, this would correspond
to optical phonons in the system \cite{ashcroft}. The low momentum dispersion of
the massive fields, with $m=0.2\epsilon_B$ is plotted in
Fig.~\ref{fermibath2d}(a) as a dashed purple line. As $k\rightarrow
0$, the dispersion of the phonons lies above the bandwidth of the
particle-hole excitations and hence one would expect that the long
wavelength modes would not thermalize. As the momentum is increased,
the dispersion enters the region of finite bath spectral density and
these higher momentum modes thermalize. This is
clearly seen in Fig.~\ref{fermibath2d}(d) where we plot the time
evolution of the correlator $\langle
\phi_{cl}(\textbf{k},t)\phi_{cl}(-\textbf{k},t)\rangle$ (solid lines) for two different values
of ${\bf k}$ along with the thermal values of the correlators (dashed lines). The mode with the low momentum $(0.03 \pi/a, 0.03\pi/a)$
(shown by red line) does not thermalize as its energy lies above the
bath band. It oscillates about its initial value sufficiently far from the thermal value (dashed red line). On the other hand, the correlator of the mode at $(0.08 \pi/a,
0.08\pi/a)$, which lies within the bath energy ranges, approaches its
thermal value in the long time limit. Fig.~\ref{fermibath2d}(e) plots
the deviation of the long time correlation functions from their
thermal value as a function of $|\textbf{k}|/\sqrt{2}$ along the $(1,1)$
direction. It is clear that there is a sharp cut-off in momentum, below
which the modes do not thermalize for the massive scalar field. This corresponds to the lowest momentum
for which the dressed dispersion lies below $\omega_{max}({\bf k})$.

\section{Conclusion \label{conclusion}}

In this paper, we have proposed a new method to study non-equilibrium
dynamics of scalar fields starting from non-thermal initial
conditions. This extends earlier work on Schr\"{o}dinger bosons \cite{cgs} to the
case of scalar fields. The method works by adding a source to the
bilinears of quantum fields at the initial time in a Schwinger-Keldysh field theoretic
formalism. The correlation functions are calculated in presence of
this source. One then takes a set of derivatives of this correlation
function with respect to this
source (with the set determined by the initial conditions) to obtain
the physical correlators. The key difference between the earlier and present formalism is the coupling of the sources to both
the fields and their time derivatives, reflecting the nature of the
classical equations of motion.

We use this method to study non-equilibrium dynamics of massless and
massive scalar fields, initialized to athermal states, and coupled to
external baths. For concreteness, we consider a system of phonons with
the massive fields corresponding to optical phonons and the massless
fields corresponding to longitudinal phonons. We first consider
coupling the system to an ohmic bath with a smooth ultraviolet
cut-off. In this case, the system thermalizes with the one particle
distributions relaxing to their thermal values. The relaxation rate is
momentum dependent when the ultraviolet cut-off is small, and
approaches a momentum independent Gaussian white noise limit as the
cut-off is increased.

We then consider the dynamics of these phonon modes coupled to a
system of non-interacting fermions, where the phonons couple to the
particle-hole excitations of the system. In one dimension we find that
the long wavelength phonons fail to thermalize at all temperatures. At
low temperatures, this is due to the effectively zero bandwidth of the
particle-hole excitations with small momentum transfer, as a result of
which the long wavelength phonon modes do not see any bath and
undergoes unitary quantum motion. At high temperatures, the lack of
thermalization is dominated by the formation of undamped dressed modes
just above the upper threshold of the bath. These modes are neither
phonon, nor fermion modes; rather these ``polarinon'' modes arise due
to strong fermion phonon coupling in the system together with
divergences in the density of states of one dimensional particle-hole
excitations. These modes remain sharp with increasing
temperature; as a result the long wavelength modes fail to thermalize
at any temperature.

We finally consider phonons coupled to fermionic baths in two
dimensions. Here simultaneous energy momentum conservation does not
lead to stringent criteria and the bath bandwidth is finite at low
momenta at all temperatures. Further the strong divergence of density
of states is also absent. As a result, we recover the typical
thermalizing behaviour of the phonon modes in two dimensions.

We note that the method we have constructed is much more widely
applicable than the models we have considered in this paper, including
the study of non equilibrium dynamics of interacting scalar field
theories. We believe this method will find much wider applications in
the future.

\begin{acknowledgments}
 The authors are grateful
 to Ahana Chakraborty for useful discussions and suggestions. The authors acknowledge
  the use of computational facilities at the Department of Theoretical
  Physics, Tata Institute of Fundamental Research, Mumbai for this paper. The authors acknowledge support of the Department of Atomic Energy,
  Government of India, under Project Identi cation No. RTI 4002.

\end{acknowledgments}

\appendix
\begin{widetext}
\section{Green's functions for single Harmonic
  Oscillator \label{sho_inv}}

In developing the extended Keldysh formalism for the harmonic
oscillators in the position basis, a crucial step was the inversion of
the matrix that appears in the $u$ dependent action,
Eq.~(\ref{action_sho}). This leads to the $u$ dependent one-particle
correlator $D(u)$, which finally leads to the physical correlation
functions in the system. In this appendix, we present the details of
this non-trivial inversion.

The matrix in Eq.~(\ref{action_sho}) can be separated two parts, one
of which is independent of the source $u$. This decomposition can be
written as
\beq
D^{-1}(t,t';u)=D^{-1}(t,t';0)-\Delta(t,t';u),
\eeq
where

\beq
D^{-1}(t,t';0)=
\delta(t-t')\left[
\begin{matrix}
	(-\partial^2_t-\Omega^2)+\delta_{t0}(-\partial_t+i\omega) & 0\\
	0 & (\partial^2_t+\Omega^2)+\delta_{t0}(\partial_t+i\omega)
\end{matrix}
\right],
\eeq
and 
\beq
\Delta(t,t';u)
=i\omega\frac{2u}{1-u^2}[-u\mathbb{I}_2+\sigma^x]\delta_{t0}\delta_{t'0}.
\eeq
Here $\mathbb{I}_2$ is $2\times2$ identity matrix. Inverting $D^{-1}(t,t';0)$ we get the components of $D(t,t';0)$
 \bqa
 \no D^{+-}(t,t',0)&=&\frac{1}{2\Omega} \sin \Omega (t-t')
 -\frac{i}{2\omega}\left\{ \cos \Omega t~\cos \Omega t'
 +\frac{\omega^2}{\Omega^2} \sin \Omega t \sin\Omega
 t'\right\} \\
 \no D^{-+}(t,t',0)&=&-\frac{1}{2\Omega} \sin \Omega (t-t')
 -\frac{i}{2\omega} \left\{ \cos \Omega t ~\cos \Omega t'
 +\frac{\omega^2}{\Omega^2} \sin \Omega t \sin\Omega t'\right\} \\
 \no D^{++}(t,t',0)&=&\Theta(t-t') D^{-+}(t,t',0)+\Theta(t'-t)
 D^{+-}(t,t',0)~~ \text{and} \\
 D^{--}(t,t',0)&=&\Theta(t-t') D^{+-}(t,t',0)+\Theta(t'-t)
 D^{-+}(t,t',0).
 \label{gfn0}
 \eqa
One can easily check that
$D^{-1}(t,t'';0)D(t'',t';0)=\delta(t-t')\mathbb{I}_2$. Now $D(t,t';u)$
can be found using the Dyson series
\beq
D(t,t';u)=D(t,t';0)+D(t,0;0)[\Delta(0,0;u)+\Delta(0,0;u)D(0,0;0)\Delta(0,0;u)+...]D(0,t';0)
\label{Dttpu}
\eeq
Using the fact that $D(0,0;0)=\frac{i}{2\omega}[\mathbb{I}+\sigma^x]$,
the series inside the brackets of Eq.~(\ref{Dttpu}) can be evaluated
using standard Pauli matrix identities to be
$i\omega\left(\frac{2u}{1-u}\right)\sigma^x$. Using this and
Eq.~(\ref{gfn0}) we get the source dependent Green's functions in
Eq.~(\ref{gfnu}). One then takes the required $u$ derivatives to get
the physical correlation function. 

\section{Phonon self energy from the fermionic bath \label{sigma_fermibath}}
\begin{figure*}
	\centering
	\includegraphics[width=\textwidth]{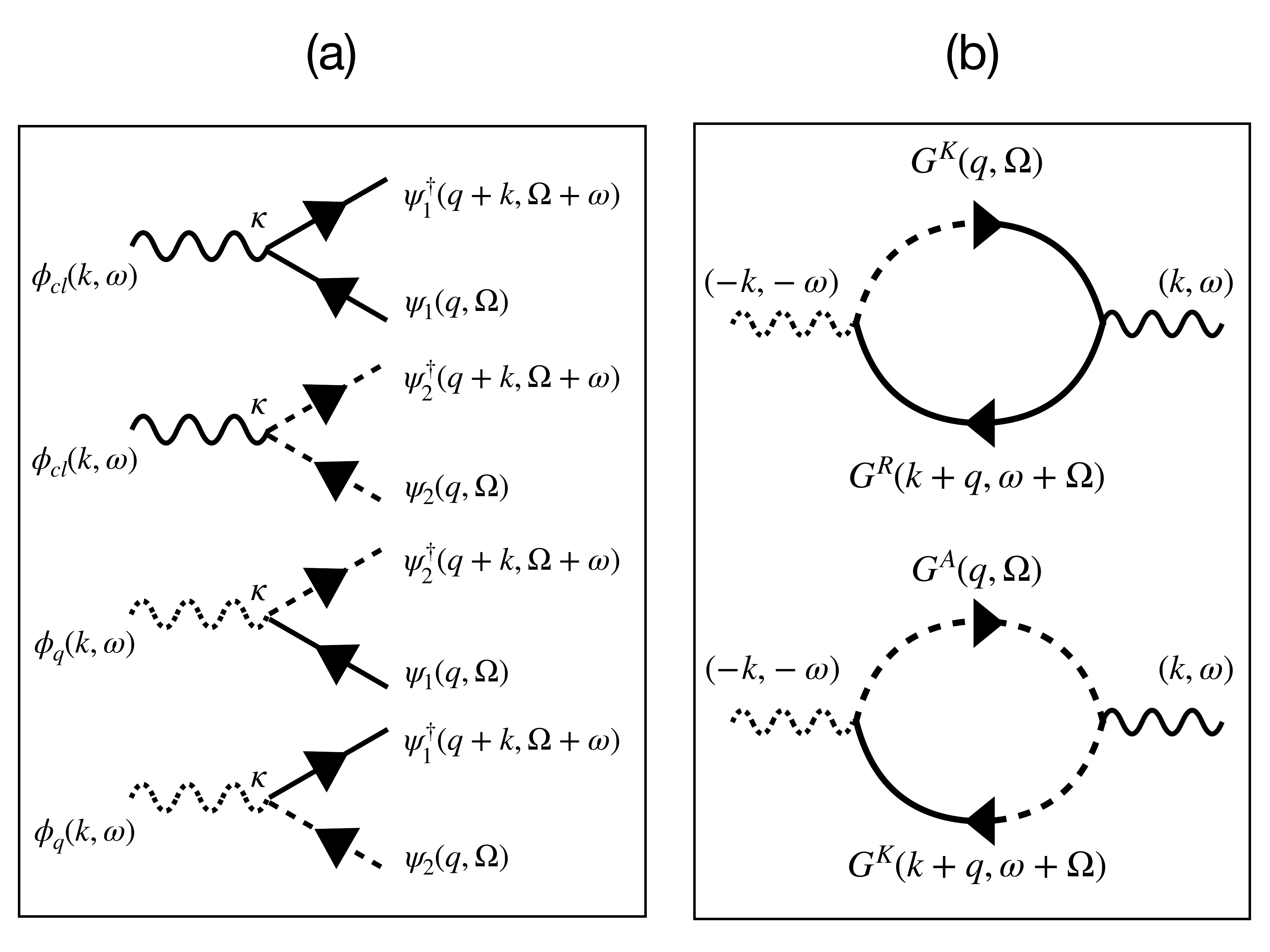}
	\caption{Dressing of the phonons coupled to a fermionic bath. 
		(a) Allowed interaction vertices in terms of the Keldysh rotated fields. The solid line represents the fermionic field $\psi_{1}(\psi^{\dagger}_1)$ and the dashed line represents $\psi_{2}(\psi^{\dagger}_2)$. Solid wavy line represents the scalar field $\phi_{cl}$ and the dotted wavy line stands for the scalar field $\phi_{q}$. $\kappa$ is the coupling strength between the scalar field and the fermions.   
		(b) Diagrams for retarded self energy $\Sigma^R(\textbf{k},\omega)$ of phonons coupled to a fermionic bath.  
		 Note that the retarded self energy of phonons is related to the polarization function of the fermions. Note that the diagrammatic representations for $G^R,G^A,G^K$ look different for bosons and fermions, since the Keldysh rotation is different, e.g. for fermions $G^K=i\langle \psi_1\psi_2^{\dagger}\rangle$ is represented by a dashed-solid line, while for bosons $G^K=i\langle \phi_{cl} \phi_{cl}\rangle$ would be represented by a solid-solid line. 
	}
	\label{fermibathint}
\end{figure*}
In this appendix we present the details of the derivation of the
phonon self-energy when the phonons are coupled to a bath of free
fermions through the Hamiltonian
\beq
 H_{int} = \kappa \sum_{\textbf{k},\textbf{q}}\lambda(\textbf{k})\psi^\dagger(\textbf{k}+\textbf{q},t)\psi(\textbf{q},t) \phi(\textbf{k},t), 
\label{ephapp}
\eeq
where $\kappa$ is the system bath coupling strength and $\lambda(\textbf{k})=\sqrt{\sum_{j=1}^D\sin^2(k_ja/2)}$ is related to the deformation potential acting between electrons and phonons \cite{mahan}.
On the Keldysh contour\cite{kamenevbook} (+/- basis) its contribution to the action is given by
\beq
S_{int}=-\kappa\sum_{\textbf{k},\textbf{q}}\lambda(\textbf{k})\int_{-\infty}^{+\infty}~dt \left[ \psi_{+}^\dagger(\textbf{k}+\textbf{q},t)\psi_{+}(\textbf{q},t) \phi_{+}(\textbf{k},t)-\psi_{-}^\dagger(\textbf{k}+\textbf{q},t)\psi_{-}(\textbf{q},t) \phi_{-}(\textbf{k},t)\right]
\eeq
It is useful to work in the Keldysh rotated basis \cite{kamenevbook}. The Keldysh rotated basis for scalar fields are given by
\begin{equation}
\begin{split}
\phi_{cl}(\textbf{k},t)&=\frac{1}{2}[\phi_+(\textbf{k},t)+\phi_-(\textbf{k},t)]\\
\phi_{q}(\textbf{k},t)&=\frac{1}{2}[\phi_+(\textbf{k},t)-\phi_-(\textbf{k},t)],
\end{split}
\label{clq}
\end{equation} 

while Keldysh rotated basis for fermions is given by
\begin{equation}
\begin{split}
&\psi_1(\textbf{k},t)=\frac{1}{\sqrt{2}}[\psi_+(\textbf{k},t)+\psi_-(\textbf{k},t)] \ \ \ \ \ \ \ \ \ \ \ \psi_2(\textbf{k},t)=\frac{1}{\sqrt{2}}[\psi_+(\textbf{k},t)-\psi_-(\textbf{k},t)] \\
&\psi^{\dagger}_1(\textbf{k},t)=\frac{1}{\sqrt{2}}[\psi^{\dagger}_+(\textbf{k},t)-\psi^{\dagger}_-(\textbf{k},t)] \ \ \ \ \ \ \ \ \ \ \ \psi^{\dagger}_2(\textbf{k},t)=\frac{1}{\sqrt{2}}[\psi^{\dagger}_+(\textbf{k},t)+\psi^{\dagger}_-(\textbf{k},t)] .
\end{split}
\label{fermi12}
\end{equation}
In the Keldysh rotated basis the fermion-phonon coupling can be written as 
\beq
\begin{split}
S_{int}=-\kappa\sum_{\textbf{k},\textbf{q}}\lambda(\textbf{k})\int_{-\infty}^{+\infty}~dt&[ \psi_{1}^\dagger(\textbf{k}+\textbf{q},t)\psi_{1}(\textbf{q},t) \phi_{cl}(\textbf{k},t)+\psi_{2}^\dagger(\textbf{k}+\textbf{q},t)\psi_{2}(\textbf{q},t) \phi_{cl}(\textbf{k},t)\\
&+\psi_{1}^\dagger(\textbf{k}+\textbf{q},t)\psi_{2}(\textbf{q},t) \phi_{q}(\textbf{k},t)+\psi_{2}^\dagger(\textbf{k}+\textbf{q},t)\psi_{1}(\textbf{q},t) \phi_{q}(\textbf{k},t)].
\end{split}
\label{actionkeldysh}
\eeq

In frequency space it turns into
\beq
\begin{split}
	S_{int}=-\kappa\sum_{\textbf{k},\textbf{q}}\lambda(\textbf{k})\int_{-\infty}^{+\infty}~\frac{d\omega}{2\pi}\int_{-\infty}^{+\infty}\frac{d\Omega}{2\pi}&[ \psi_{1}^\dagger(\textbf{k}+\textbf{q},\omega+\Omega)\psi_{1}(\textbf{q},\Omega) \phi_{cl}(\textbf{k},\omega)+\psi_{2}^\dagger(\textbf{k}+\textbf{q},\omega+\Omega)\psi_{2}(\textbf{q},\Omega) \phi_{cl}(\textbf{k},\omega)\\
	&+\psi_{1}^\dagger(\textbf{k}+\textbf{q},\omega+\Omega)\psi_{2}(\textbf{q},\Omega) \phi_{q}(\textbf{k},\omega)+\psi_{2}^\dagger(\textbf{k}+\textbf{q},\omega+\Omega)\psi_{1}(\textbf{q},\Omega) \phi_{q}(\textbf{k},\omega)].
\end{split}
\label{actionSKfreq}
\eeq
The terms in Eq.~(\ref{actionSKfreq}) gives allowed vertices for system-bath coupling. They are sketched in Fig.~\ref{fermibathint}(a). Diagrams for retarded self energy of the phonons are sketched in Fig.~\ref{fermibathint}(b), where solid straight lines represent $\psi_{1}(\psi^{\dagger}_1)$, dashed straight lines represent $\psi_{2}(\psi^{\dagger}_2)$, solid wavy lines indicate $\phi_{cl}$ and dashed wavy lines represent $\phi_q$. Note that due to different Keldysh rotations for fermionic and bosonic fields (we follow the convention in Ref.~\onlinecite{kamenevbook}), the diagrammatic representations of the bosonic and fermionic propagators look different, e.g. while $G^K=i\langle \phi_{cl} \phi_{cl}\rangle$ for bosons is represented by a fully solid line, for fermions $G^K=i\langle\psi_1\psi_2^{\dagger}\rangle$ is represented by a dashed-solid line. Similar adjustments occur for $G^R$ and $G^A$ as well. The diagrams for self-energy are given in terms of the vertices and free fermionic Green's functions.
The retarded self energy is 
\begin{equation}
\begin{split}
\Sigma^R(\textbf{k},\omega)&=-i\kappa^2\lambda(\textbf{k})^2\sum_{\textbf{q}}\int^{\infty}_{-\infty}\frac{d\Omega}{2\pi}[G^R(\textbf{k}+\textbf{q},\omega+\Omega)G^K(\textbf{q},\Omega)+G^K(\textbf{k}+\textbf{q},\omega+\Omega)G^A(\textbf{q},\Omega)].\\
\end{split}
\label{sigrder1}
\end{equation}

Here Green's functions of the free fermions at thermal equilibrium are
\bqa
&G^R_0(\textbf{k},\omega)=[G^A(\textbf{k},\omega)]^*=\frac{1}{\omega-\epsilon_\textbf{k}+i0^{+}}~\text{and}\\
&G^K_0(\textbf{k},\omega)=-2\pi i F(\omega)\delta(\omega-\epsilon_\textbf{k}),
\eqa
where $F(\omega)=\tanh\left(\frac{\omega-\mu}{2T}\right)$. $T$ is the temperature of the bath and $\mu$ is the chemical potential. 

It is then easy to see (see Fig.~\ref{fermibathint}(b) for the corresponding Feynman diagrams) that the retarded self energy for phonons is just the polarization function of free Fermi gas multiplied by factors of system-bath coupling strength. Upon simplifying we get, 

\begin{equation}
\begin{split}
\Sigma^R(\textbf{k},\omega)&=\kappa^2\lambda(\textbf{k})^2\sum_{\textbf{q}}\frac{F(\epsilon_{\textbf{k}+\textbf{q}})-F(\epsilon_\textbf{q})}{\omega+i0^++\epsilon_\textbf{q}-\epsilon_{\textbf{k}+\textbf{q}}}\\
&=-\kappa^2\lambda(\textbf{k})^2\int^{\infty}_{-\infty}\frac{d\omega'}{2\pi}\frac{J(\textbf{k},\omega')}{\omega'-\omega-i0^+},
\end{split}
\label{sigrder2}
\end{equation}
where bath spectral function $J(k,\omega)$ is given by 
\beq
J(\textbf{k},\omega)=\pi\lambda(\textbf{k})^2\sum_\textbf{q}[F(\epsilon_{\textbf{k}+\textbf{q}})-F(\epsilon_\textbf{q})]\delta(\omega-\epsilon_{\textbf{k}+\textbf{q}}+\epsilon_\textbf{q}).
\label{bathspecfn}
\eeq
The Keldysh self energy is given by (using fluctuation-dissipation theorem)
\beq
\begin{split}
	\Sigma^K(\textbf{k},\omega)&=2i\coth\left(\frac{\omega}{2T}\right)\text{Im}[\Sigma^R(\textbf{k},\omega)]\\
	&=-i2\pi\kappa^2\lambda(\textbf{k})^2\sum_{\textbf{q}} \coth\left(\frac{\omega}{2T}\right) [F(\epsilon_{\textbf{k}+\textbf{q}})-F(\epsilon_\textbf{q})]\delta(\omega-\epsilon_{\textbf{k}+\textbf{q}}+\epsilon_\textbf{q})\\
	&=-i2\pi\kappa^2\lambda(\textbf{k})^2\sum_{\textbf{q}} \coth\left(\frac{\epsilon_{\textbf{k}+\textbf{q}}-\mu}{2T}-\frac{\epsilon_{\textbf{q}}-\mu}{2T}\right)\left [\tanh \left(\frac{\epsilon_{\textbf{k}+\textbf{q}}-\mu}{2T}\right)-\tanh\left(\frac{\epsilon_\textbf{q}-\mu}{2T}\right)\right]\delta(\omega-\epsilon_{\textbf{k}+\textbf{q}}+\epsilon_\textbf{q})\\
	&=-i2\pi\kappa^2\lambda(\textbf{k})^2\sum_{\textbf{q}} \left [1-\tanh \left(\frac{\epsilon_{\textbf{k}+\textbf{q}}-\mu}{2T}\right)\tanh\left(\frac{\epsilon_\textbf{q}-\mu}{2T}\right)\right]\delta(\omega-\epsilon_{\textbf{k}+\textbf{q}}+\epsilon_\textbf{q})\\
	&=i2\pi\kappa^2\lambda(\textbf{k})^2\sum_{\textbf{q}} \left [F \left(\epsilon_{\textbf{k}+\textbf{q}}\right)F\left(\epsilon_\textbf{q}\right)-1\right]\delta(\omega-\epsilon_{\textbf{k}+\textbf{q}}+\epsilon_\textbf{q}).
\end{split}
\label{sigkder}
\eeq

\section{Multi-phonon processes and phonon relaxation \label{multiphonon}}
\begin{figure}[h]
	\includegraphics[scale=0.4]{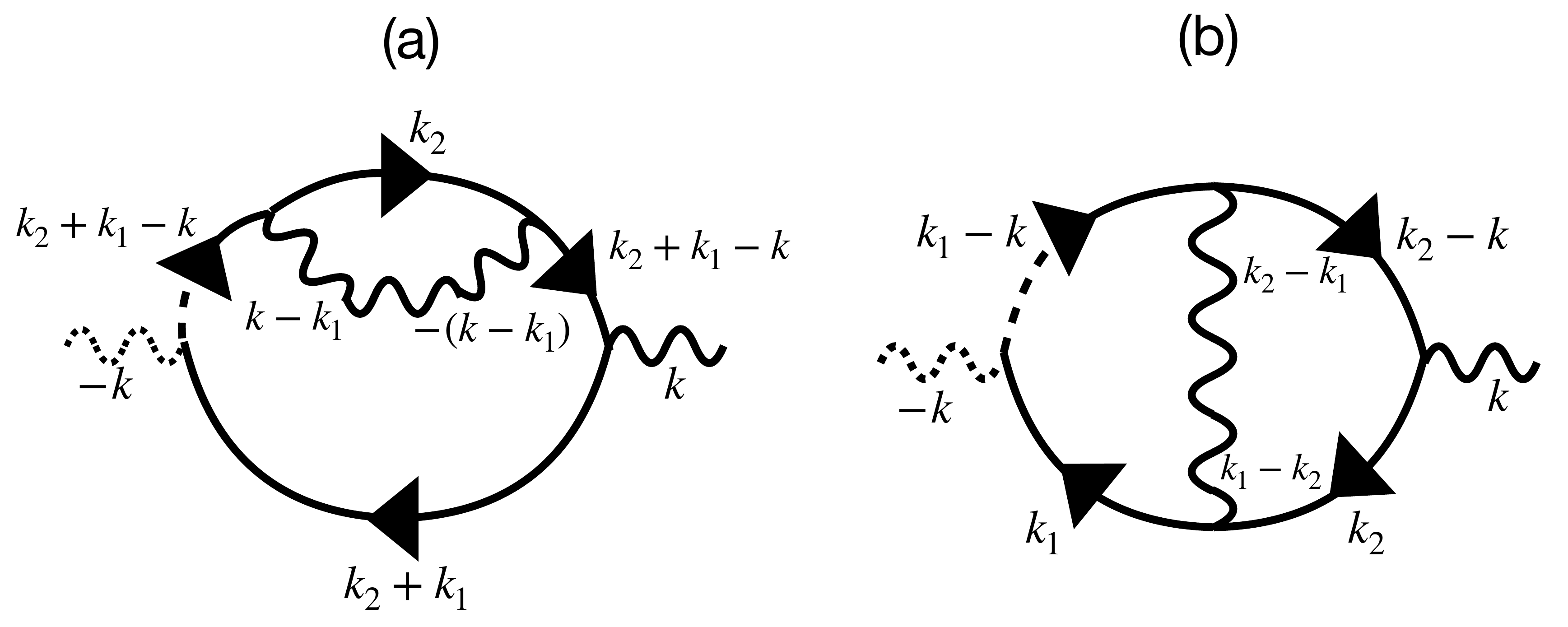}
	\caption{Multi-phonon processes giving rise to $\mathcal{O}(\kappa^4)$ corrections to the phonon self energy: (a) A representative diagram where the fermion correlators in the polarization function are dressed by self energy corrections due to emission/absorption of an additional phonon. Note that the particle-hole pair with momenta $(k_1+k_2,k_2)$ together with the phonon at $k-k_1$ can satisfy the energy-momentum conservation criterion. The intermediate fermion lines at $k_2+k_1-k$ are off-shell propagators. There are many such diagrams for the retarded self energy. We have shown one to indicate the structure of the diagram. (b) A representative diagram where the polarization function of the fermions is dressed by a vertex. These vertex corrections do not play leading role in relaxing the energy momentum constraints and are neglected in our calculations. Once again we have shown one of the many vertex diagrams to illustrate their structure.
	}   
	\label{multiph_fig1}
\end{figure}
\begin{figure}[h]
	\includegraphics[scale=0.5]{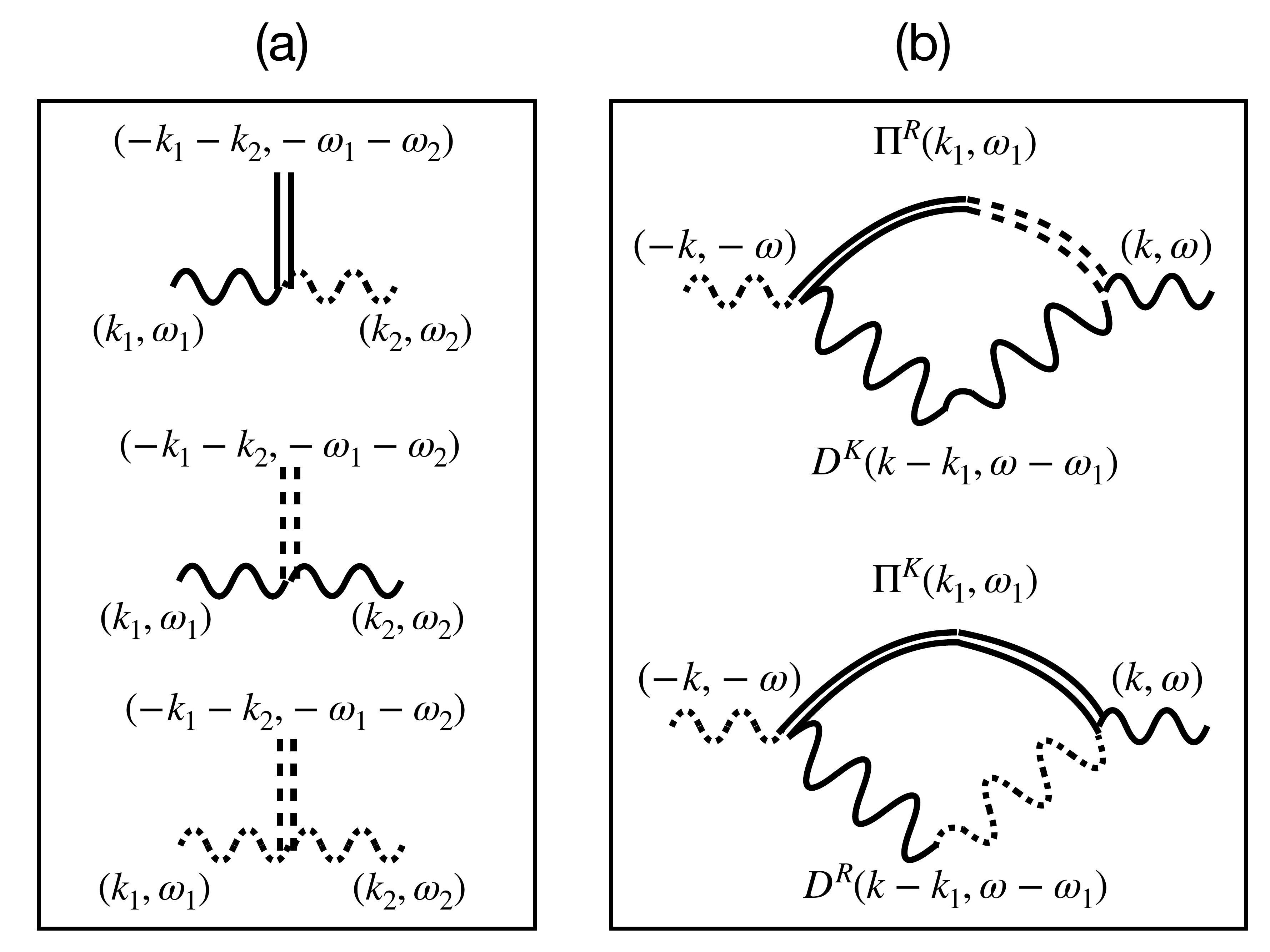}
	\caption{Phonon relaxation due to two phonons scattering with fermions: (a) Allowed vertices  for effective interaction between 2 phonons and particle-hole excitations. Solid (dashed) wavy lines are classical (quantum) phonon fields and solid (dashed) double lines are classical (quantum) particle-hole excitation fields. (b) Retarded self-energy diagrams for phonons: $D^{R/K}$ are phonon correlators (retarded/Keldysh components) and  $\Pi^{R/K}$ are polarization functions (retarded/Keldysh components).
	}   
	\label{multiph_fig2}
\end{figure}
\begin{figure}[h]
	\includegraphics[scale=0.25]{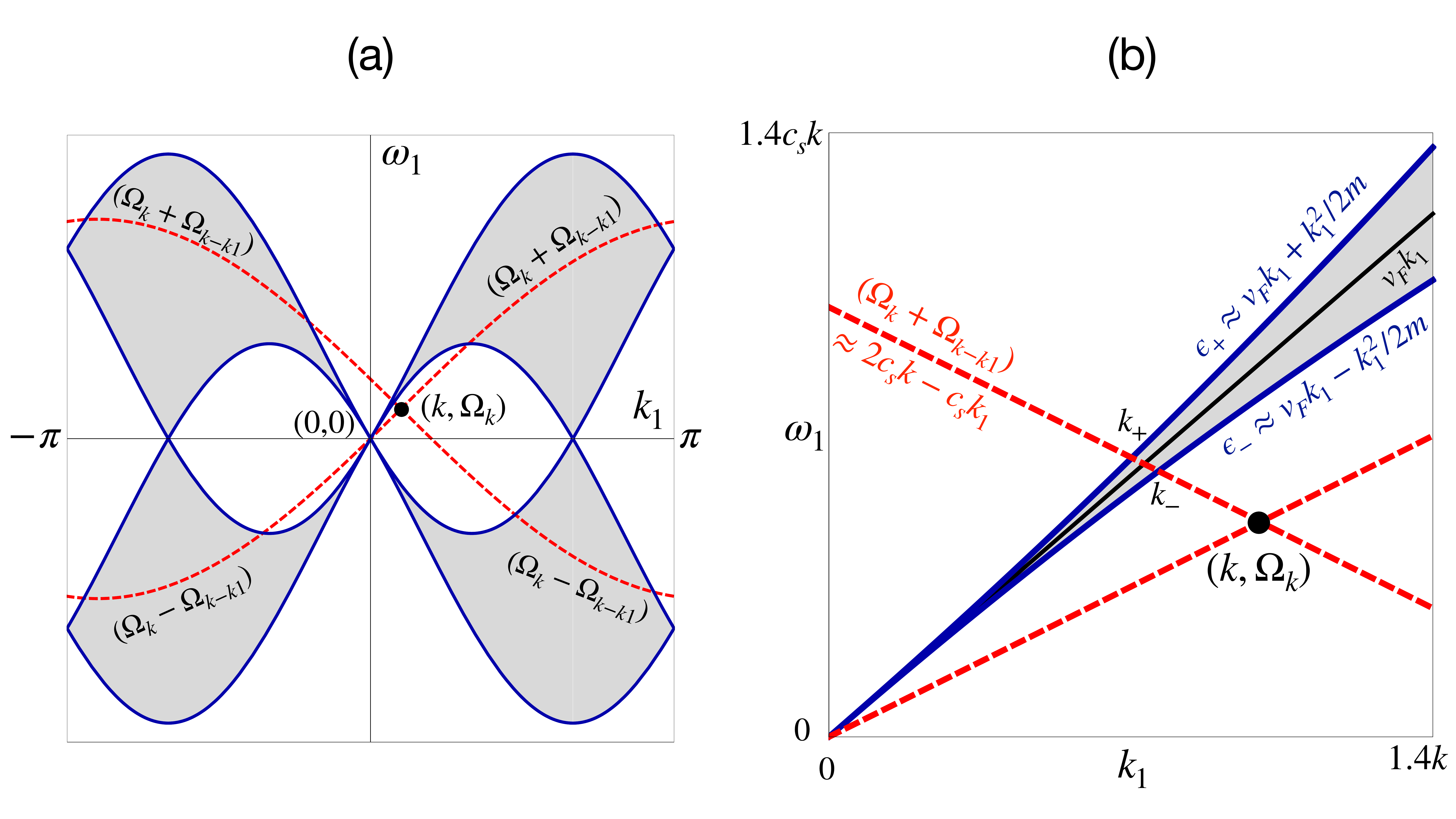}
	\caption{ Particle-hole continuum and contribution of 2-phonon processes to the decay rate (at momenta $k$): (a) The 1D particle-hole excitations have a finite spectral weight in the shaded region in $k_1,\omega_1$ plane, bounded by solid blue lines. The dashed red lines correspond to the sum and difference of phonon energies, $\Omega_k+\Omega_{k-k_1}$ and $\Omega_k-\Omega_{k-k_1}$. The location ($k,\Omega_k$) is shown as a solid black circle. The crossing of the red lines with the shaded regions indicate the values of $(k_1,\omega_1)$ which contributes to the self energy integral. For small $k$ at low temperature, the contribution of the regions at large $k_1$, where the phonon lines enter the particle hole continuum, is exponentially suppressed by thermal factors.
	(b)The small region at low values of $k_1$, where the energy-momentum conditions are satisfied, is shown in detail. The shaded region, where particle-hole excitations have finite spectral weight, is around $v_Fk$ and is bounded by $\epsilon_+=v_Fk+k_1^2/2m$ and $\epsilon_-=v_Fk-k_1^2/2m$ (solid blue lines). The region of interest, where the phonon line crosses the shaded region, lies between $k_+$ and $k_-$, where $k_\pm\approx\frac{2c_s}{v_F+c_s}k\mp \frac{2c_s^2}{m(v_F+c_s)^3}k^2$. 
	}   
	\label{multiph_fig3}
\end{figure}
We have shown in the main text that the energy momentum constraints in 1D lead to an exponentially small relaxation rate for the phonons at low temperatures. We have shown this to ${\cal O} (\kappa^2)$, where the phonon self energy is proportional to the polarization function of the fermionic bath. A natural question arises: Do multi-phonon processes, where other phonons can carry away energy and momentum, relax the constraints and lead to a power law behaviour at low $T$? Earlier works, which modelled relaxation of a single particle in a Luttinger liquid ~\cite{heavyparticle_castro} or in spinor gas~\cite{spinorgas_kamenev}, predicted a scattering rate $\Gamma\sim T^4$, coming from two-phonon processes. In this appendix we show that for our problem of phonons coupled to a Fermi sea, the scattering rate remains exponentially small at low $T << c_s k$; i.e. $\Gamma_k\sim k e^{-c_s k/T}$. For $T \gg c_s k$, we recover the expected thermal broadening $\Gamma_k \sim T$. The key difference between our paper and earlier works stems from
the fact that we are considering a finite density of particles with a Fermi sea whereas the earlier works~\cite{heavyparticle_castro,spinorgas_kamenev} considered a single particle.

We would like to note that since we are interested in the decay rate for phonons, we focus on the imaginary part of the retarded self-energy of the phonons on shell, i.e. 
\beq\label{decayrate}
\Gamma_k \sim -\frac{1}{\Omega_k}\textrm{Im}~ \Sigma^R_{(4)}(k,\Omega_k),
\eeq
calculated to ${\cal  O}(\kappa^4)$. There are two types of diagrams which contribute at ${\cal  O}(\kappa^4)$: (i) diagrams where the fermion lines in Fig.~\ref{fermibathint}(b) are dressed by emission/absorption of phonons [an example is shown in Fig.~\ref{multiph_fig1} (a)] and (ii) diagrams where the fermionic polarization is dressed by vertex functions [an example is shown in Fig.~\ref{multiph_fig1} (b)]. The self-energy corrections of the fermion lines help in relaxing the energy momentum constraints and are equivalent to the diagrams which lead to power laws in the earlier works~\cite{heavyparticle_castro}; hence we will focus on them in this appendix and ignore the vertex corrections.

Let us focus on the diagram shown in Fig.~\ref{multiph_fig1}(a). As we argued in the section on phonon self energy, the particle-hole pair with momenta $(k_2+k_1,k_2+k_1-k)$ cannot satisfy energy-momentum conservation. However, the particle hole pair with momenta $(k_2+k_1,k_2)$, together with the phonon at $(k-k_1)$ can satisfy the on-shell condition and give rise to a finite contribution to $\textrm{Im}~ \Sigma^R_{(4)}(k,\Omega_k)$. Note that the intermediate fermion at $k_2+k_1-k$ is necessarily off-shell. To make initial headway, and to compare with the earlier works, we replace these off-shell propagators by a constant (which we take to be $1$, since we are interested in scaling of the decay rate). This effectively gives a 2-phonon, 2-fermion scattering vertex ${\cal O}(\kappa^4)$, similar to Ref. ~\onlinecite{heavyparticle_castro}.
We can also think of this as a vertex between two phonons and a particle-hole excitation. These effective vertices are shown in Fig.~\ref{multiph_fig2}(a), where the double line represents a particle-hole propagator. The retarded self energy corrections for the phonons due to these effective vertices are shown in Fig.~\ref{multiph_fig2}(b). The self energy is given by
\beq
\Sigma^R_{(4)}(k,\omega)=i\kappa^4\lambda(k)^2\int\frac{d\omega_1}{2\pi}\sum_{k_1}\lambda(k-k_1)^2\left[D^R(k-k_1,\omega-\omega_1)\Pi^K(k_1,\omega_1)+D^K(k-k_1,\omega-\omega_1)\Pi^R(k_1,\omega_1)\right],
\eeq
where $\lambda$ is the form factor from the deformation potential, and $\Pi^{R/K}$ are polarization functions of the fermions which are also the particle-hole propagators.  We note that solving the full non-equilibrium problem at this order is beyond the scope of this paper; however the calculation can be simplified by assuming thermal equilibrium, i.e.  $D(\Pi)^K(p,\omega) = 2i \coth \left[\frac{\omega}{2T}\right] Im D(\Pi)^R(p,\omega)$, which is simply a statement of fluctuation dissipation theorem. The imaginary part of $\Sigma^R_{(4)}(k,\omega)$ is then given by
\beq
\textrm{Im}~[\Sigma^R_{(4)}(k,\omega)]=-\kappa^4\lambda(k)^2\int\frac{d\omega_1}{\pi}\sum_{k_1}\lambda(k-k_1)^2\left[\coth\left(\frac{\omega_1}{2T}\right)+\coth\left(\frac{\omega-\omega_1}{2T}\right)\right]ImD^R(k-k_1,\omega-\omega_1)Im\Pi^R(k_1,\omega_1).
\eeq
Now, we are interested in the behaviour of this function at small $k$ and $\omega=\Omega_k\sim c_sk$ (we assume $k,\Omega_k >0$). In Fig.~\ref{multiph_fig3}(a) we plot the region in $k_1,\omega_1$ plane where the 1D particle-hole excitations have a finite spectral weight. This corresponds to the shaded area between the solid lines. The dashed lines correspond to the sum and difference of phonon energies, $\Omega_k+\Omega_{k-k_1}$ and $\Omega_k-\Omega_{k-k_1}$. The location ($k,\Omega_k$) is shown as a solid circle in this figure. The crossing of the red lines with the shaded regions indicate the values of $(k_1,\omega_1)$ which contributes to the self energy integral. For small $k$ at low temperature, the contribution of the regions at large $k_1$, where the phonon lines enter the particle hole continuum, is exponentially suppressed by thermal factors. However there is a small region at low values of $k_1$, where the energy-momentum conditions are satisfied, and this region contributes to the decay rate in the leading order. This region is shown in detail in  Fig.~\ref{multiph_fig3}(b), where the region of interest lies between $k_+$ and $k_-$, where $k_\pm\approx\frac{2c_s}{v_F+c_s}k\mp \frac{2c_s^2}{m(v_F+c_s)^3}k^2$. In this case, one can show that $\textrm{Im}~\Pi^R(k_1,\omega_1)\sim\frac{-1}{|k_1|}\Theta[(\omega_1-\epsilon_-)(\epsilon_+-\omega_1)]$, where $\epsilon_+=v_Fk_1+k_1^2/2m$ and $\epsilon_-=v_Fk_1-k_1^2/2m$. Here $v_F$ is the Fermi velocity of the Fermi gas and $m$ is the mass of the Fermions in the low energy long wavelength continuum description. Note that the width of this region $\sim k_1^2$ and is small at low $k_1$. Further, the retarded phonon propagator has two poles at $\omega_1=\Omega_k \pm \Omega_{k-k_1}$. From Fig.~\ref{multiph_fig3}(b), we see that $\omega_1 > \Omega_k$ and hence we only consider the residue of the propagator at $\omega_1=\Omega_k + \Omega_{k-k_1}$. Putting all these together, we get
\beq
\textrm{Im}~[\Sigma^R_{(4)}(k,c_sk)]\sim\frac{\kappa^4}{4}\lambda(k)^2\int_{k_+}^{k_-}dk_1\frac{\lambda(k-k_1)^2}{\Omega_{k-k_1}}\frac{1}{|k_1|}\left[\coth\left(\frac{c_sk+\Omega_{k-k_1}}{2T}\right)-\coth\left(\frac{\Omega_{k-k_1}}{2T}\right)\right],
\eeq
For small $k$, we can replace the integration by the function value at $k_0=\frac{1}{2}(k_++k_-)=\frac{2c_s}{v_F+c_s}k$ multiplied by the width of the region $\Delta k=\frac{4c_s^2}{m(v_F+c_s)^3}k^2$ to get the leading order estimate of  $\textrm{Im}~[\Sigma^R_{(4)}(k,c_sk)]$,
\beq
\begin{split}
\textrm{Im}~[\Sigma^R_{(4)}(k,c_sk)]&\sim\frac{\kappa^4}{4c_s}k^2\frac{|k-k_0|}{|k_0|}\left[\coth\left(\frac{c_sk+\Omega_{k-k_0}}{2T}\right)-\coth\left(\frac{\Omega_{k-k0}}{2T}\right)\right]\Delta k\\
&\sim-\kappa^4\left(\frac{v_F-c_s}{2m(v_F+c_s)^3}\right)k^4\left[\coth\left(\frac{v_F-c_s}{v_F+c_s}\frac{c_sk}{2T}\right)-\coth\left(\frac{2v_F}{v_F+c_s}\frac{c_sk}{2T}\right)\right].
\end{split}
\eeq
We assume that $v_F$ and $c_s$ are of similar magnitude but $c_s< v_F$, so that $2v_F/(v_F+c_s) \sim 1$ and   $(v_F-c_S)/(v_F+c_S) \sim 1$. In this case, for $T\ll c_sk$, the arguments of both the $\coth$ functions in the above expression will be large and the value of each thermal factor will be exponentially (in inverse temperature) close to 1. So their difference will result in a decay rate which is exponentially small in inverse temperature, i.e. $\Gamma_k \sim k^3 e^{-c_sk/T}$.  Note that if $v_F \gg c_s$, this argument is bolstered even more. For $T \gg c_s k $, the argument of the $\coth$ functions are small, and
we get $\Gamma_k \sim Tk^2$.

The above calculations have neglected the energy momentum dependence of the off-shell fermionic propagators, which will give additional momentum dependence to the vertex between two phonons and particle-hole excitations.  Since the energy mismatch (which determines how far off shell the fermionic propagators are) $\sim (v_F-c_s)k$, one would expect the fermion propagators to be $\sim \frac{1}{ (v_F-c_s)k}$. This would reduce a factor of $k^2$ from the scaling. In this case, one would obtain $\Gamma_k \sim k e^{-c_s k/T}$ for $T \ll c_s k$ and $ \Gamma_k \sim  T$ for $T \gg c_s k$. Note that the high temperature result is equivalent to a standard thermal broadening, while the low $T$ result is exponentially small rather than being a power law. This high $T$ limit is similar to the case considered in Ref.~\onlinecite{heavyparticle_castro} (where $\frac{k^2}{2m}\sim T$ for the single particle). In addition, the vertex in Ref.~\onlinecite{heavyparticle_castro} does not have the extra $1/k$ factor from the off-shell fermion propagator.

We finally comment on why our calculations at the lowest temperatures show a different scaling of the decay rate from the earlier works~\cite{heavyparticle_castro,spinorgas_kamenev}. In the earlier works, which looks at scattering of single particles, the relevant momenta of fermions are close to $0$ ( $\sim \sqrt{T}$), while we are considering a Fermi sea at a finite density of fermions; so the fermions relevant for phonon relaxation in our case have momenta $\sim k_F$.
Refs.\onlinecite{heavyparticle_castro,spinorgas_kamenev} considered the scattering of a massive particle which can be treated classically ($\frac{p^2}{2m}\sim T$). Since we have $c_s<v_F$, we do not have a temperature regime where the phonons are quantum but the relevant particle-hole excitations can be treated classically. Thus both degrees of freedom have quantum nature in our calculations. For low $T\ll c_sk$, the thermal factor in both the loss and gain rates in the collision integral approaches 1 to exponential accuracy. The leading order 1 is cancelled, leaving us with an exponentially small remaining term. 
 This is the key reason the relaxation rate of phonons coupled to a Fermi sea in 1D remains exponential even when two phonon processes are considered, in contrast to the case of scattering of a single massive particle. Further the additional $1/k$ factors from the off-shell fermion lines change the power of $k$  sitting in the prefactor to the exponential.

We also note that three-phonon processes involve two particle-hole excitations and satisfy energy-momentum conservation constraints. They can result in a decay rate which is polynomial in temperature and hence not exponentially small. However, these processes are of $\mathcal{O}(\kappa^8)$ and for small system-bath coupling they are suppressed.

\section{Characterization of polarinon mode in 1D \label{polarinon}}
In this appendix we will discuss some details about the undamped polarinon mode situated just outside particle-hole band edge. We noticed (as discussed in Section~\ref{1dfermi}) that particle-hole DOS has inverse square root divergence near band edge in 1D. Using Kramers-Kronig relations~\cite{arfken} one can show that the real part of the retarded self energy diverges in the same way just outside the particle-hole band edge [see Fig.~\ref{fermibath1dp2}(e)]. 
\begin{figure}[h]
	\includegraphics[scale=0.25]{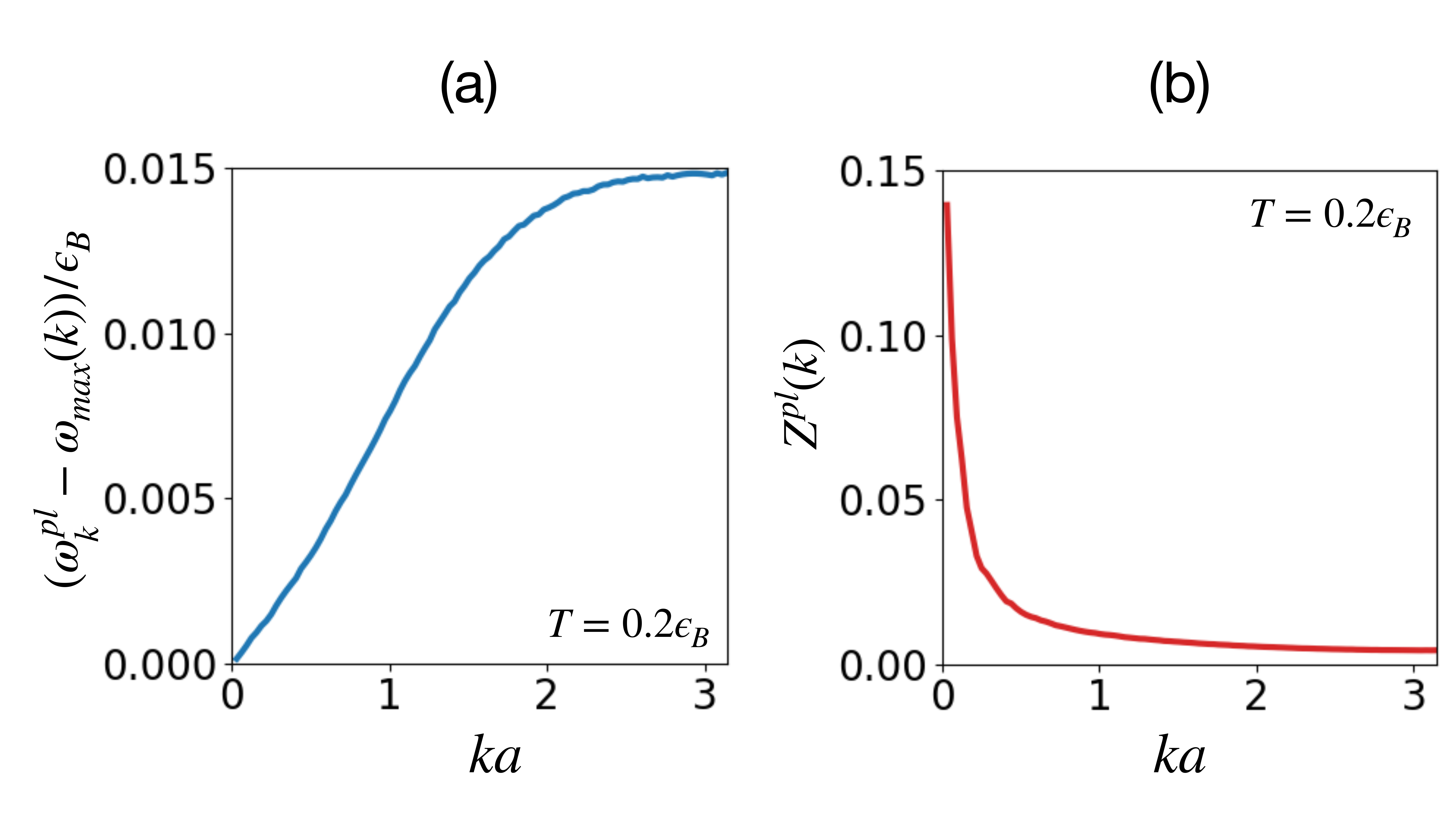}
	\caption{Specifications of ``polarinon" modes.
		(a) Distance of the polarinon modes from particle-hole band edge is plotted as a function of momenta. It increases smoothly with momenta. 
		(b) Spectral weight of the polarinon modes is plotted as a function of momenta. The weight falls sharply with momenta. Hence the polarinon modes will not affect the dynamics of the phonons at higher momenta. The bath is set at $T=0.2\epsilon_B$ and $\mu=0.5\epsilon_B$. The system-bath coupling strength is $\kappa^2=0.9\epsilon_B^3a$.
	}   
	\label{polfig}
\end{figure}

Now, for the pole of the inverse Green's function outside the particle-hole band edge we have the equation 
\beq\label{outpole}
\omega^2-\Omega^2_k-\frac{g(k,\omega)}{\sqrt{\omega-\omega_{max}(k)}}=0,
\eeq      
where we have assumed $\Sigma^R(k,\omega)\sim\frac{2g(k,\omega)}{\sqrt{\omega-\omega_{max}(k)}}$and $g(k,\omega)$ is a function of $k$ and $\omega$ (also depends on the bath parameters) but varies slowly with $\omega$ as we have separated out the divergent piece. We also considered a phonon mode below the particle-hole band edge so that $\Omega_k<\omega_{max}(k)$.

Now we define $x=\omega-\omega_{max}(k)$. For small positive $x$, the solution of Eq.~(\ref{outpole}) becomes
\beq
x(k)=\frac{[g(k,\omega_{max}(k))]^2}{(\omega_{max}(k)^2-\Omega^2_k)^2}.
\eeq
The spectral weight of this mode is given by
\beq
Z^{pl}(k)\approx\frac{[x(k)]^{3/2}}{g(k,\omega_{max}(k))}=\frac{[g(k,\omega_{max}(k))]^2}{(\omega_{max}(k)^2-\Omega^2_k)^3}.
\eeq

We have $\omega_{max}=2\epsilon_B|\sin(ka/2)|$ and $\Omega_k=\omega_0|\sin(ka/2)|$ where $2\epsilon_B$ is the fermion bandwidth and $\omega_0$ is the phonon bandwidth. For the electron-phonon coupling we have considered, we can separate out leading $k$-dependence of $g$ as $g(k,\omega_{max}(k))=\sin^2(ka/2)f(k)$, where $f(k)$ is a smooth function. With these inputs we have 
\bqa\label{polspecs}
x(k)&=\frac{1}{[4\epsilon^2_B-\omega^2_0]^2}[f(k)]^2\\
Z^{pl}(k)&=\frac{1}{[4\epsilon^2_B-\omega^2_0]^2}[\frac{f(k)}{\sin(k/2)}]^2.
\eqa
We have calculated $x(k)$ and $Z^{pl}(k)$ numerically and plotted them as a function of momenta $k$ at bath temperature $T=0.2\epsilon_B$ and chemical potential $\mu=0.5\epsilon_B$ in Fig.~\ref{polfig}. The system-bath coupling strength is $\kappa^2=0.9\epsilon_B^3a$. We note that their behaviour is consistent with the analytical forms [see Eq.~\ref{polspecs}]. While $x(k)$ increases with $k$, $Z^{pl}(k)$ falls sharply with $k$ and hence they have little role to play in the dynamics of the phonons at higher momenta.


\end{widetext}
\bibliography{realscalar}

\begin{thebibliography}{103}%
\makeatletter
\providecommand \@ifxundefined [1]{%
 \@ifx{#1\undefined}
}%
\providecommand \@ifnum [1]{%
 \ifnum #1\expandafter \@firstoftwo
 \else \expandafter \@secondoftwo
 \fi
}%
\providecommand \@ifx [1]{%
 \ifx #1\expandafter \@firstoftwo
 \else \expandafter \@secondoftwo
 \fi
}%
\providecommand \natexlab [1]{#1}%
\providecommand \enquote  [1]{``#1''}%
\providecommand \bibnamefont  [1]{#1}%
\providecommand \bibfnamefont [1]{#1}%
\providecommand \citenamefont [1]{#1}%
\providecommand \href@noop [0]{\@secondoftwo}%
\providecommand \href [0]{\begingroup \@sanitize@url \@href}%
\providecommand \@href[1]{\@@startlink{#1}\@@href}%
\providecommand \@@href[1]{\endgroup#1\@@endlink}%
\providecommand \@sanitize@url [0]{\catcode `\\12\catcode `\$12\catcode
  `\&12\catcode `\#12\catcode `\^12\catcode `\_12\catcode `\%12\relax}%
\providecommand \@@startlink[1]{}%
\providecommand \@@endlink[0]{}%
\providecommand \url  [0]{\begingroup\@sanitize@url \@url }%
\providecommand \@url [1]{\endgroup\@href {#1}{\urlprefix }}%
\providecommand \urlprefix  [0]{URL }%
\providecommand \Eprint [0]{\href }%
\providecommand \doibase [0]{http://dx.doi.org/}%
\providecommand \selectlanguage [0]{\@gobble}%
\providecommand \bibinfo  [0]{\@secondoftwo}%
\providecommand \bibfield  [0]{\@secondoftwo}%
\providecommand \translation [1]{[#1]}%
\providecommand \BibitemOpen [0]{}%
\providecommand \bibitemStop [0]{}%
\providecommand \bibitemNoStop [0]{.\EOS\space}%
\providecommand \EOS [0]{\spacefactor3000\relax}%
\providecommand \BibitemShut  [1]{\csname bibitem#1\endcsname}%
\let\auto@bib@innerbib\@empty
\bibitem [{\citenamefont {Englert}\ and\ \citenamefont
  {Brout}(1964)}]{englert}%
  \BibitemOpen
  \bibfield  {author} {\bibinfo {author} {\bibfnamefont {F.}~\bibnamefont
  {Englert}}\ and\ \bibinfo {author} {\bibfnamefont {R.}~\bibnamefont
  {Brout}},\ }\href {\doibase 10.1103/PhysRevLett.13.321} {\bibfield  {journal}
  {\bibinfo  {journal} {Phys. Rev. Lett.}\ }\textbf {\bibinfo {volume} {13}},\
  \bibinfo {pages} {321} (\bibinfo {year} {1964})}\BibitemShut {NoStop}%
\bibitem [{\citenamefont {Higgs}(1964{\natexlab{a}})}]{higgs1}%
  \BibitemOpen
  \bibfield  {author} {\bibinfo {author} {\bibfnamefont {P.}~\bibnamefont
  {Higgs}},\ }\href {\doibase https://doi.org/10.1016/0031-9163(64)91136-9}
  {\bibfield  {journal} {\bibinfo  {journal} {Physics Letters}\ }\textbf
  {\bibinfo {volume} {12}},\ \bibinfo {pages} {132} (\bibinfo {year}
  {1964}{\natexlab{a}})}\BibitemShut {NoStop}%
\bibitem [{\citenamefont {Higgs}(1964{\natexlab{b}})}]{higgs2}%
  \BibitemOpen
  \bibfield  {author} {\bibinfo {author} {\bibfnamefont {P.~W.}\ \bibnamefont
  {Higgs}},\ }\href {\doibase 10.1103/PhysRevLett.13.508} {\bibfield  {journal}
  {\bibinfo  {journal} {Phys. Rev. Lett.}\ }\textbf {\bibinfo {volume} {13}},\
  \bibinfo {pages} {508} (\bibinfo {year} {1964}{\natexlab{b}})}\BibitemShut
  {NoStop}%
\bibitem [{\citenamefont {Liddle}(1999)}]{inflationintro}%
  \BibitemOpen
  \bibfield  {author} {\bibinfo {author} {\bibfnamefont {A.~R.}\ \bibnamefont
  {Liddle}},\ }in\ \href@noop {} {\emph {\bibinfo {booktitle} {{ICTP Summer
  School in High-Energy Physics and Cosmology}}}}\ (\bibinfo  {publisher}
  {World Scientific},\ \bibinfo {address} {Singapore},\ \bibinfo {year}
  {1999})\ \Eprint {http://arxiv.org/abs/astro-ph/9901124}
  {arXiv:astro-ph/9901124} \BibitemShut {NoStop}%
\bibitem [{\citenamefont {Bassett}\ \emph {et~al.}(2006)\citenamefont
  {Bassett}, \citenamefont {Tsujikawa},\ and\ \citenamefont
  {Wands}}]{inflationdynamics}%
  \BibitemOpen
  \bibfield  {author} {\bibinfo {author} {\bibfnamefont {B.~A.}\ \bibnamefont
  {Bassett}}, \bibinfo {author} {\bibfnamefont {S.}~\bibnamefont {Tsujikawa}},
  \ and\ \bibinfo {author} {\bibfnamefont {D.}~\bibnamefont {Wands}},\ }\href
  {\doibase 10.1103/RevModPhys.78.537} {\bibfield  {journal} {\bibinfo
  {journal} {Rev. Mod. Phys.}\ }\textbf {\bibinfo {volume} {78}},\ \bibinfo
  {pages} {537} (\bibinfo {year} {2006})}\BibitemShut {NoStop}%
\bibitem [{\citenamefont {Maga{\~{n}}a}\ and\ \citenamefont
  {Matos}(2012)}]{sfdm}%
  \BibitemOpen
  \bibfield  {author} {\bibinfo {author} {\bibfnamefont {J.}~\bibnamefont
  {Maga{\~{n}}a}}\ and\ \bibinfo {author} {\bibfnamefont {T.}~\bibnamefont
  {Matos}},\ }\href {\doibase 10.1088/1742-6596/378/1/012012} {\bibfield
  {journal} {\bibinfo  {journal} {Journal of Physics: Conference Series}\
  }\textbf {\bibinfo {volume} {378}},\ \bibinfo {pages} {012012} (\bibinfo
  {year} {2012})}\BibitemShut {NoStop}%
\bibitem [{\citenamefont {Hui}\ \emph {et~al.}(2017)\citenamefont {Hui},
  \citenamefont {Ostriker}, \citenamefont {Tremaine},\ and\ \citenamefont
  {Witten}}]{uldm}%
  \BibitemOpen
  \bibfield  {author} {\bibinfo {author} {\bibfnamefont {L.}~\bibnamefont
  {Hui}}, \bibinfo {author} {\bibfnamefont {J.~P.}\ \bibnamefont {Ostriker}},
  \bibinfo {author} {\bibfnamefont {S.}~\bibnamefont {Tremaine}}, \ and\
  \bibinfo {author} {\bibfnamefont {E.}~\bibnamefont {Witten}},\ }\href
  {\doibase 10.1103/PhysRevD.95.043541} {\bibfield  {journal} {\bibinfo
  {journal} {Phys. Rev. D}\ }\textbf {\bibinfo {volume} {95}},\ \bibinfo
  {pages} {043541} (\bibinfo {year} {2017})}\BibitemShut {NoStop}%
\bibitem [{\citenamefont {Chadha-Day}\ \emph {et~al.}(2021)\citenamefont
  {Chadha-Day}, \citenamefont {Ellis},\ and\ \citenamefont {Marsh}}]{axiondm}%
  \BibitemOpen
  \bibfield  {author} {\bibinfo {author} {\bibfnamefont {F.}~\bibnamefont
  {Chadha-Day}}, \bibinfo {author} {\bibfnamefont {J.}~\bibnamefont {Ellis}}, \
  and\ \bibinfo {author} {\bibfnamefont {D.~J.~E.}\ \bibnamefont {Marsh}},\
  }\href@noop {} {\enquote {\bibinfo {title} {Axion dark matter: What is it and
  why now?}}\ } (\bibinfo {year} {2021}),\ \Eprint
  {http://arxiv.org/abs/2105.01406} {arXiv:2105.01406 [hep-ph]} \BibitemShut
  {NoStop}%
\bibitem [{\citenamefont {Doniach}\ and\ \citenamefont
  {Sondheimer}(1998)}]{doniach}%
  \BibitemOpen
  \bibfield  {author} {\bibinfo {author} {\bibfnamefont {S.}~\bibnamefont
  {Doniach}}\ and\ \bibinfo {author} {\bibfnamefont {E.~H.}\ \bibnamefont
  {Sondheimer}},\ }\href {\doibase 10.1142/p067} {\emph {\bibinfo {title}
  {Green's Functions for Solid State Physicists}}}\ (\bibinfo  {publisher}
  {Imperial College press},\ \bibinfo {address} {London, UK},\ \bibinfo {year}
  {1998})\BibitemShut {NoStop}%
\bibitem [{\citenamefont {Patton}(1984)}]{magnon}%
  \BibitemOpen
  \bibfield  {author} {\bibinfo {author} {\bibfnamefont {C.~E.}\ \bibnamefont
  {Patton}},\ }\href {\doibase https://doi.org/10.1016/0370-1573(84)90023-1}
  {\bibfield  {journal} {\bibinfo  {journal} {Physics Reports}\ }\textbf
  {\bibinfo {volume} {103}},\ \bibinfo {pages} {251} (\bibinfo {year}
  {1984})}\BibitemShut {NoStop}%
\bibitem [{\citenamefont {Andrson}(1966)}]{superfluid_anderson}%
  \BibitemOpen
  \bibfield  {author} {\bibinfo {author} {\bibfnamefont {P.~W.}\ \bibnamefont
  {Andrson}},\ }\href {\doibase 10.1103/RevModPhys.38.298} {\bibfield
  {journal} {\bibinfo  {journal} {Rev. Mod. Phys.}\ }\textbf {\bibinfo {volume}
  {38}},\ \bibinfo {pages} {298} (\bibinfo {year} {1966})}\BibitemShut
  {NoStop}%
\bibitem [{\citenamefont {Schmitt}(2015)}]{superfluid_book}%
  \BibitemOpen
  \bibfield  {author} {\bibinfo {author} {\bibfnamefont {A.}~\bibnamefont
  {Schmitt}},\ }\href {\doibase https://doi.org/10.1007/978-3-319-07947-9}
  {\emph {\bibinfo {title} {Introduction to Superfluidity}}}\ (\bibinfo
  {publisher} {Springer International Publishing},\ \bibinfo {address}
  {Switzerland},\ \bibinfo {year} {2015})\BibitemShut {NoStop}%
\bibitem [{\citenamefont {Rempe}\ \emph {et~al.}(1991)\citenamefont {Rempe},
  \citenamefont {Thompson}, \citenamefont {Brecha}, \citenamefont {Lee},\ and\
  \citenamefont {Kimble}}]{cvtqedexp1}%
  \BibitemOpen
  \bibfield  {author} {\bibinfo {author} {\bibfnamefont {G.}~\bibnamefont
  {Rempe}}, \bibinfo {author} {\bibfnamefont {R.~J.}\ \bibnamefont {Thompson}},
  \bibinfo {author} {\bibfnamefont {R.~J.}\ \bibnamefont {Brecha}}, \bibinfo
  {author} {\bibfnamefont {W.~D.}\ \bibnamefont {Lee}}, \ and\ \bibinfo
  {author} {\bibfnamefont {H.~J.}\ \bibnamefont {Kimble}},\ }\href {\doibase
  10.1103/PhysRevLett.67.1727} {\bibfield  {journal} {\bibinfo  {journal}
  {Phys. Rev. Lett.}\ }\textbf {\bibinfo {volume} {67}},\ \bibinfo {pages}
  {1727} (\bibinfo {year} {1991})}\BibitemShut {NoStop}%
\bibitem [{\citenamefont {Thompson}\ \emph {et~al.}(1992)\citenamefont
  {Thompson}, \citenamefont {Rempe},\ and\ \citenamefont
  {Kimble}}]{cvtqedexp2}%
  \BibitemOpen
  \bibfield  {author} {\bibinfo {author} {\bibfnamefont {R.~J.}\ \bibnamefont
  {Thompson}}, \bibinfo {author} {\bibfnamefont {G.}~\bibnamefont {Rempe}}, \
  and\ \bibinfo {author} {\bibfnamefont {H.~J.}\ \bibnamefont {Kimble}},\
  }\href {\doibase 10.1103/PhysRevLett.68.1132} {\bibfield  {journal} {\bibinfo
   {journal} {Phys. Rev. Lett.}\ }\textbf {\bibinfo {volume} {68}},\ \bibinfo
  {pages} {1132} (\bibinfo {year} {1992})}\BibitemShut {NoStop}%
\bibitem [{\citenamefont {Miller}\ \emph {et~al.}(2005)\citenamefont {Miller},
  \citenamefont {Northup}, \citenamefont {Birnbaum}, \citenamefont {Boca},
  \citenamefont {Boozer},\ and\ \citenamefont {Kimble}}]{cvtqedrev1}%
  \BibitemOpen
  \bibfield  {author} {\bibinfo {author} {\bibfnamefont {R.}~\bibnamefont
  {Miller}}, \bibinfo {author} {\bibfnamefont {T.~E.}\ \bibnamefont {Northup}},
  \bibinfo {author} {\bibfnamefont {K.~M.}\ \bibnamefont {Birnbaum}}, \bibinfo
  {author} {\bibfnamefont {A.}~\bibnamefont {Boca}}, \bibinfo {author}
  {\bibfnamefont {A.~D.}\ \bibnamefont {Boozer}}, \ and\ \bibinfo {author}
  {\bibfnamefont {H.~J.}\ \bibnamefont {Kimble}},\ }\href {\doibase
  10.1088/0953-4075/38/9/007} {\bibfield  {journal} {\bibinfo  {journal}
  {Journal of Physics B: Atomic, Molecular and Optical Physics}\ }\textbf
  {\bibinfo {volume} {38}},\ \bibinfo {pages} {S551} (\bibinfo {year}
  {2005})}\BibitemShut {NoStop}%
\bibitem [{\citenamefont {Walther}\ \emph {et~al.}(2006)\citenamefont
  {Walther}, \citenamefont {Varcoe}, \citenamefont {Englert},\ and\
  \citenamefont {Becker}}]{cvtqedrev2}%
  \BibitemOpen
  \bibfield  {author} {\bibinfo {author} {\bibfnamefont {H.}~\bibnamefont
  {Walther}}, \bibinfo {author} {\bibfnamefont {B.~T.~H.}\ \bibnamefont
  {Varcoe}}, \bibinfo {author} {\bibfnamefont {B.-G.}\ \bibnamefont {Englert}},
  \ and\ \bibinfo {author} {\bibfnamefont {T.}~\bibnamefont {Becker}},\ }\href
  {\doibase 10.1088/0034-4885/69/5/r02} {\bibfield  {journal} {\bibinfo
  {journal} {Reports on Progress in Physics}\ }\textbf {\bibinfo {volume}
  {69}},\ \bibinfo {pages} {1325} (\bibinfo {year} {2006})}\BibitemShut
  {NoStop}%
\bibitem [{\citenamefont {Tonks}\ and\ \citenamefont
  {Langmuir}(1929)}]{polariton}%
  \BibitemOpen
  \bibfield  {author} {\bibinfo {author} {\bibfnamefont {L.}~\bibnamefont
  {Tonks}}\ and\ \bibinfo {author} {\bibfnamefont {I.}~\bibnamefont
  {Langmuir}},\ }\href {\doibase 10.1103/PhysRev.33.195} {\bibfield  {journal}
  {\bibinfo  {journal} {Phys. Rev.}\ }\textbf {\bibinfo {volume} {33}},\
  \bibinfo {pages} {195} (\bibinfo {year} {1929})}\BibitemShut {NoStop}%
\bibitem [{\citenamefont {Tolpygo}(2008)}]{tolpygo}%
  \BibitemOpen
  \bibfield  {author} {\bibinfo {author} {\bibfnamefont {K.}~\bibnamefont
  {Tolpygo}},\ }\href
  {http://archive.ujp.bitp.kiev.ua/files/journals/53/si/53SI21p.pdf} {\bibfield
   {journal} {\bibinfo  {journal} {Ukr. J. Phys.}\ }\textbf {\bibinfo {volume}
  {53}},\ \bibinfo {pages} {93} (\bibinfo {year} {2008})}\BibitemShut {NoStop}%
\bibitem [{\citenamefont {Huang}(1951)}]{huang}%
  \BibitemOpen
  \bibfield  {author} {\bibinfo {author} {\bibfnamefont {K.}~\bibnamefont
  {Huang}},\ }\href {\doibase 10.1038/167779b0} {\bibfield  {journal} {\bibinfo
   {journal} {Nature}\ }\textbf {\bibinfo {volume} {167}},\ \bibinfo {pages}
  {779} (\bibinfo {year} {1951})}\BibitemShut {NoStop}%
\bibitem [{\citenamefont {Mills}\ and\ \citenamefont
  {Burstein}(1974)}]{polaritonrevold}%
  \BibitemOpen
  \bibfield  {author} {\bibinfo {author} {\bibfnamefont {D.~L.}\ \bibnamefont
  {Mills}}\ and\ \bibinfo {author} {\bibfnamefont {E.}~\bibnamefont
  {Burstein}},\ }\href {\doibase 10.1088/0034-4885/37/7/001} {\bibfield
  {journal} {\bibinfo  {journal} {Reports on Progress in Physics}\ }\textbf
  {\bibinfo {volume} {37}},\ \bibinfo {pages} {817} (\bibinfo {year}
  {1974})}\BibitemShut {NoStop}%
\bibitem [{\citenamefont {Basov}\ \emph {et~al.}(2021)\citenamefont {Basov},
  \citenamefont {Asenjo-Garcia}, \citenamefont {Schuck}, \citenamefont {Zhu},\
  and\ \citenamefont {Rubio}}]{polaritonrevnew}%
  \BibitemOpen
  \bibfield  {author} {\bibinfo {author} {\bibfnamefont {D.~N.}\ \bibnamefont
  {Basov}}, \bibinfo {author} {\bibfnamefont {A.}~\bibnamefont
  {Asenjo-Garcia}}, \bibinfo {author} {\bibfnamefont {P.~J.}\ \bibnamefont
  {Schuck}}, \bibinfo {author} {\bibfnamefont {X.}~\bibnamefont {Zhu}}, \ and\
  \bibinfo {author} {\bibfnamefont {A.}~\bibnamefont {Rubio}},\ }\href
  {\doibase doi:10.1515/nanoph-2020-0449} {\bibfield  {journal} {\bibinfo
  {journal} {Nanophotonics}\ }\textbf {\bibinfo {volume} {10}},\ \bibinfo
  {pages} {549} (\bibinfo {year} {2021})}\BibitemShut {NoStop}%
\bibitem [{\citenamefont {Peskin}\ and\ \citenamefont
  {Schroeder}(1995)}]{peskin}%
  \BibitemOpen
  \bibfield  {author} {\bibinfo {author} {\bibfnamefont {M.~E.}\ \bibnamefont
  {Peskin}}\ and\ \bibinfo {author} {\bibfnamefont {D.~V.}\ \bibnamefont
  {Schroeder}},\ }\href@noop {} {\emph {\bibinfo {title} {{An Introduction to
  quantum field theory}}}}\ (\bibinfo  {publisher} {Addison-Wesley},\ \bibinfo
  {address} {Reading, USA},\ \bibinfo {year} {1995})\BibitemShut {NoStop}%
\bibitem [{\citenamefont {Srednicki}(2007)}]{srednicki}%
  \BibitemOpen
  \bibfield  {author} {\bibinfo {author} {\bibfnamefont {M.}~\bibnamefont
  {Srednicki}},\ }\href {\doibase 10.1017/CBO9780511813917} {\emph {\bibinfo
  {title} {Quantum Field Theory}}}\ (\bibinfo  {publisher} {Cambridge
  University Press},\ \bibinfo {address} {Cambridge,UK},\ \bibinfo {year}
  {2007})\BibitemShut {NoStop}%
\bibitem [{\citenamefont {Altland}\ and\ \citenamefont
  {Simons}(2010)}]{altlandsimons}%
  \BibitemOpen
  \bibfield  {author} {\bibinfo {author} {\bibfnamefont {A.}~\bibnamefont
  {Altland}}\ and\ \bibinfo {author} {\bibfnamefont {B.~D.}\ \bibnamefont
  {Simons}},\ }\href {\doibase 10.1017/CBO9780511789984} {\emph {\bibinfo
  {title} {Condensed Matter Field Theory}}},\ \bibinfo {edition} {2nd}\ ed.\
  (\bibinfo  {publisher} {Cambridge University Press},\ \bibinfo {address}
  {Cambridge,UK},\ \bibinfo {year} {2010})\BibitemShut {NoStop}%
\bibitem [{\citenamefont {Calzetta}\ and\ \citenamefont
  {Hu}(1988)}]{neqqf_CalzettaHu}%
  \BibitemOpen
  \bibfield  {author} {\bibinfo {author} {\bibfnamefont {E.}~\bibnamefont
  {Calzetta}}\ and\ \bibinfo {author} {\bibfnamefont {B.~L.}\ \bibnamefont
  {Hu}},\ }\href {\doibase 10.1103/PhysRevD.37.2878} {\bibfield  {journal}
  {\bibinfo  {journal} {Phys. Rev. D}\ }\textbf {\bibinfo {volume} {37}},\
  \bibinfo {pages} {2878} (\bibinfo {year} {1988})}\BibitemShut {NoStop}%
\bibitem [{\citenamefont {Berges}(2002)}]{qfielddyn_berges}%
  \BibitemOpen
  \bibfield  {author} {\bibinfo {author} {\bibfnamefont {J.}~\bibnamefont
  {Berges}},\ }\href {\doibase https://doi.org/10.1016/S0375-9474(01)01295-7}
  {\bibfield  {journal} {\bibinfo  {journal} {Nuclear Physics A}\ }\textbf
  {\bibinfo {volume} {699}},\ \bibinfo {pages} {847} (\bibinfo {year}
  {2002})}\BibitemShut {NoStop}%
\bibitem [{\citenamefont {Anisimov}\ \emph {et~al.}(2009)\citenamefont
  {Anisimov}, \citenamefont {Buchmüller}, \citenamefont {Drewes},\ and\
  \citenamefont {Mendizabal}}]{noneqscalar_anisimov}%
  \BibitemOpen
  \bibfield  {author} {\bibinfo {author} {\bibfnamefont {A.}~\bibnamefont
  {Anisimov}}, \bibinfo {author} {\bibfnamefont {W.}~\bibnamefont
  {Buchmüller}}, \bibinfo {author} {\bibfnamefont {M.}~\bibnamefont {Drewes}},
  \ and\ \bibinfo {author} {\bibfnamefont {S.}~\bibnamefont {Mendizabal}},\
  }\href {\doibase https://doi.org/10.1016/j.aop.2009.01.001} {\bibfield
  {journal} {\bibinfo  {journal} {Annals of Physics}\ }\textbf {\bibinfo
  {volume} {324}},\ \bibinfo {pages} {1234} (\bibinfo {year}
  {2009})}\BibitemShut {NoStop}%
\bibitem [{\citenamefont {Mukaida}\ and\ \citenamefont
  {Nakayama}(2013)}]{dynoscscalar}%
  \BibitemOpen
  \bibfield  {author} {\bibinfo {author} {\bibfnamefont {K.}~\bibnamefont
  {Mukaida}}\ and\ \bibinfo {author} {\bibfnamefont {K.}~\bibnamefont
  {Nakayama}},\ }\href {\doibase 10.1088/1475-7516/2013/01/017} {\bibfield
  {journal} {\bibinfo  {journal} {Journal of Cosmology and Astroparticle
  Physics}\ }\textbf {\bibinfo {volume} {2013}},\ \bibinfo {pages} {017}
  (\bibinfo {year} {2013})}\BibitemShut {NoStop}%
\bibitem [{\citenamefont {Thomsen}\ \emph {et~al.}(1986)\citenamefont
  {Thomsen}, \citenamefont {Grahn}, \citenamefont {Maris},\ and\ \citenamefont
  {Tauc}}]{ufpps_expold}%
  \BibitemOpen
  \bibfield  {author} {\bibinfo {author} {\bibfnamefont {C.}~\bibnamefont
  {Thomsen}}, \bibinfo {author} {\bibfnamefont {H.~T.}\ \bibnamefont {Grahn}},
  \bibinfo {author} {\bibfnamefont {H.~J.}\ \bibnamefont {Maris}}, \ and\
  \bibinfo {author} {\bibfnamefont {J.}~\bibnamefont {Tauc}},\ }\href {\doibase
  10.1103/PhysRevB.34.4129} {\bibfield  {journal} {\bibinfo  {journal} {Phys.
  Rev. B}\ }\textbf {\bibinfo {volume} {34}},\ \bibinfo {pages} {4129}
  (\bibinfo {year} {1986})}\BibitemShut {NoStop}%
\bibitem [{\citenamefont {Qi}\ \emph {et~al.}(2010)\citenamefont {Qi},
  \citenamefont {Chen}, \citenamefont {Yu}, \citenamefont {Cadden-Zimansky},
  \citenamefont {Smirnov}, \citenamefont {Tolk}, \citenamefont {Miotkowski},
  \citenamefont {Cao}, \citenamefont {Chen}, \citenamefont {Wu}, \citenamefont
  {Qiao},\ and\ \citenamefont {Jiang}}]{ufpps_exp1}%
  \BibitemOpen
  \bibfield  {author} {\bibinfo {author} {\bibfnamefont {J.}~\bibnamefont
  {Qi}}, \bibinfo {author} {\bibfnamefont {X.}~\bibnamefont {Chen}}, \bibinfo
  {author} {\bibfnamefont {W.}~\bibnamefont {Yu}}, \bibinfo {author}
  {\bibfnamefont {P.}~\bibnamefont {Cadden-Zimansky}}, \bibinfo {author}
  {\bibfnamefont {D.}~\bibnamefont {Smirnov}}, \bibinfo {author} {\bibfnamefont
  {N.~H.}\ \bibnamefont {Tolk}}, \bibinfo {author} {\bibfnamefont
  {I.}~\bibnamefont {Miotkowski}}, \bibinfo {author} {\bibfnamefont
  {H.}~\bibnamefont {Cao}}, \bibinfo {author} {\bibfnamefont {Y.~P.}\
  \bibnamefont {Chen}}, \bibinfo {author} {\bibfnamefont {Y.}~\bibnamefont
  {Wu}}, \bibinfo {author} {\bibfnamefont {S.}~\bibnamefont {Qiao}}, \ and\
  \bibinfo {author} {\bibfnamefont {Z.}~\bibnamefont {Jiang}},\ }\href
  {\doibase 10.1063/1.3513826} {\bibfield  {journal} {\bibinfo  {journal}
  {Applied Physics Letters}\ }\textbf {\bibinfo {volume} {97}},\ \bibinfo
  {pages} {182102} (\bibinfo {year} {2010})}\BibitemShut {NoStop}%
\bibitem [{\citenamefont {Novko}\ \emph {et~al.}(2020)\citenamefont {Novko},
  \citenamefont {Caruso}, \citenamefont {Draxl},\ and\ \citenamefont
  {Cappelluti}}]{ufpps_exp4}%
  \BibitemOpen
  \bibfield  {author} {\bibinfo {author} {\bibfnamefont {D.}~\bibnamefont
  {Novko}}, \bibinfo {author} {\bibfnamefont {F.}~\bibnamefont {Caruso}},
  \bibinfo {author} {\bibfnamefont {C.}~\bibnamefont {Draxl}}, \ and\ \bibinfo
  {author} {\bibfnamefont {E.}~\bibnamefont {Cappelluti}},\ }\href {\doibase
  10.1103/PhysRevLett.124.077001} {\bibfield  {journal} {\bibinfo  {journal}
  {Phys. Rev. Lett.}\ }\textbf {\bibinfo {volume} {124}},\ \bibinfo {pages}
  {077001} (\bibinfo {year} {2020})}\BibitemShut {NoStop}%
\bibitem [{\citenamefont {Mante}\ \emph {et~al.}(2018)\citenamefont {Mante},
  \citenamefont {Belliard},\ and\ \citenamefont
  {Perrin}}]{ufpps_phdyn_nano_rev}%
  \BibitemOpen
  \bibfield  {author} {\bibinfo {author} {\bibfnamefont {P.-A.}\ \bibnamefont
  {Mante}}, \bibinfo {author} {\bibfnamefont {L.}~\bibnamefont {Belliard}}, \
  and\ \bibinfo {author} {\bibfnamefont {B.}~\bibnamefont {Perrin}},\ }\href
  {\doibase doi:10.1515/nanoph-2018-0069} {\bibfield  {journal} {\bibinfo
  {journal} {Nanophotonics}\ }\textbf {\bibinfo {volume} {7}},\ \bibinfo
  {pages} {1759} (\bibinfo {year} {2018})}\BibitemShut {NoStop}%
\bibitem [{\citenamefont {Kawashima}\ \emph {et~al.}(1995)\citenamefont
  {Kawashima}, \citenamefont {Wefers},\ and\ \citenamefont {Nelson}}]{pulse1}%
  \BibitemOpen
  \bibfield  {author} {\bibinfo {author} {\bibfnamefont {H.}~\bibnamefont
  {Kawashima}}, \bibinfo {author} {\bibfnamefont {M.~M.}\ \bibnamefont
  {Wefers}}, \ and\ \bibinfo {author} {\bibfnamefont {K.~A.}\ \bibnamefont
  {Nelson}},\ }\href {\doibase 10.1146/annurev.pc.46.100195.003211} {\bibfield
  {journal} {\bibinfo  {journal} {Annual Review of Physical Chemistry}\
  }\textbf {\bibinfo {volume} {46}},\ \bibinfo {pages} {627} (\bibinfo {year}
  {1995})},\ \bibinfo {note} {pMID: 24341370}\BibitemShut {NoStop}%
\bibitem [{\citenamefont {Weiner}\ and\ \citenamefont {Kan'an}(1998)}]{pulse2}%
  \BibitemOpen
  \bibfield  {author} {\bibinfo {author} {\bibfnamefont {A.}~\bibnamefont
  {Weiner}}\ and\ \bibinfo {author} {\bibfnamefont {A.}~\bibnamefont
  {Kan'an}},\ }\href {\doibase 10.1109/2944.686738} {\bibfield  {journal}
  {\bibinfo  {journal} {IEEE Journal of Selected Topics in Quantum
  Electronics}\ }\textbf {\bibinfo {volume} {4}},\ \bibinfo {pages} {317}
  (\bibinfo {year} {1998})}\BibitemShut {NoStop}%
\bibitem [{\citenamefont {Fetterman}\ \emph {et~al.}(1998)\citenamefont
  {Fetterman}, \citenamefont {Goswami}, \citenamefont {Keusters}, \citenamefont
  {Yang}, \citenamefont {Rhee},\ and\ \citenamefont {Warren}}]{pulse3}%
  \BibitemOpen
  \bibfield  {author} {\bibinfo {author} {\bibfnamefont {M.}~\bibnamefont
  {Fetterman}}, \bibinfo {author} {\bibfnamefont {D.}~\bibnamefont {Goswami}},
  \bibinfo {author} {\bibfnamefont {D.}~\bibnamefont {Keusters}}, \bibinfo
  {author} {\bibfnamefont {W.}~\bibnamefont {Yang}}, \bibinfo {author}
  {\bibfnamefont {J.-K.}\ \bibnamefont {Rhee}}, \ and\ \bibinfo {author}
  {\bibfnamefont {W.}~\bibnamefont {Warren}},\ }\href {\doibase
  10.1364/OE.3.000366} {\bibfield  {journal} {\bibinfo  {journal} {Opt.
  Express}\ }\textbf {\bibinfo {volume} {3}},\ \bibinfo {pages} {366} (\bibinfo
  {year} {1998})}\BibitemShut {NoStop}%
\bibitem [{\citenamefont {Weiner}(2000)}]{pulse4}%
  \BibitemOpen
  \bibfield  {author} {\bibinfo {author} {\bibfnamefont {A.~M.}\ \bibnamefont
  {Weiner}},\ }\href {\doibase 10.1063/1.1150614} {\bibfield  {journal}
  {\bibinfo  {journal} {Review of Scientific Instruments}\ }\textbf {\bibinfo
  {volume} {71}},\ \bibinfo {pages} {1929} (\bibinfo {year}
  {2000})}\BibitemShut {NoStop}%
\bibitem [{\citenamefont {Weiner}(2011)}]{pulse5}%
  \BibitemOpen
  \bibfield  {author} {\bibinfo {author} {\bibfnamefont {A.~M.}\ \bibnamefont
  {Weiner}},\ }\href {\doibase https://doi.org/10.1016/j.optcom.2011.03.084}
  {\bibfield  {journal} {\bibinfo  {journal} {Optics Communications}\ }\textbf
  {\bibinfo {volume} {284}},\ \bibinfo {pages} {3669} (\bibinfo {year}
  {2011})},\ \bibinfo {note} {special Issue on Optical Pulse Shaping, Arbitrary
  Waveform Generation, and Pulse Characterization}\BibitemShut {NoStop}%
\bibitem [{\citenamefont {Ginzburg}(2016)}]{cvtqedengrev}%
  \BibitemOpen
  \bibfield  {author} {\bibinfo {author} {\bibfnamefont {P.}~\bibnamefont
  {Ginzburg}},\ }\href {\doibase https://doi.org/10.1016/j.revip.2016.07.001}
  {\bibfield  {journal} {\bibinfo  {journal} {Reviews in Physics}\ }\textbf
  {\bibinfo {volume} {1}},\ \bibinfo {pages} {120} (\bibinfo {year}
  {2016})}\BibitemShut {NoStop}%
\bibitem [{\citenamefont {Weiher}\ \emph {et~al.}(2019)\citenamefont {Weiher},
  \citenamefont {Agudelo},\ and\ \citenamefont {Bohmann}}]{cvtqwed_sqzlit1}%
  \BibitemOpen
  \bibfield  {author} {\bibinfo {author} {\bibfnamefont {K.}~\bibnamefont
  {Weiher}}, \bibinfo {author} {\bibfnamefont {E.}~\bibnamefont {Agudelo}}, \
  and\ \bibinfo {author} {\bibfnamefont {M.}~\bibnamefont {Bohmann}},\ }\href
  {\doibase 10.1103/PhysRevA.100.043812} {\bibfield  {journal} {\bibinfo
  {journal} {Phys. Rev. A}\ }\textbf {\bibinfo {volume} {100}},\ \bibinfo
  {pages} {043812} (\bibinfo {year} {2019})}\BibitemShut {NoStop}%
\bibitem [{\citenamefont {Dey}\ and\ \citenamefont
  {Kulkarni}(2020)}]{cvtqedeng1}%
  \BibitemOpen
  \bibfield  {author} {\bibinfo {author} {\bibfnamefont {A.}~\bibnamefont
  {Dey}}\ and\ \bibinfo {author} {\bibfnamefont {M.}~\bibnamefont {Kulkarni}},\
  }\href {\doibase 10.1103/PhysRevA.101.043801} {\bibfield  {journal} {\bibinfo
   {journal} {Phys. Rev. A}\ }\textbf {\bibinfo {volume} {101}},\ \bibinfo
  {pages} {043801} (\bibinfo {year} {2020})}\BibitemShut {NoStop}%
\bibitem [{\citenamefont {\ifmmode~\check{C}\else \v{C}\fi{}ernot\'{\i}k}\
  \emph {et~al.}(2019)\citenamefont {\ifmmode~\check{C}\else
  \v{C}\fi{}ernot\'{\i}k}, \citenamefont {Dantan},\ and\ \citenamefont
  {Genes}}]{cvtqedeng2}%
  \BibitemOpen
  \bibfield  {author} {\bibinfo {author} {\bibfnamefont {O.~c.~v.}\
  \bibnamefont {\ifmmode~\check{C}\else \v{C}\fi{}ernot\'{\i}k}}, \bibinfo
  {author} {\bibfnamefont {A.}~\bibnamefont {Dantan}}, \ and\ \bibinfo {author}
  {\bibfnamefont {C.}~\bibnamefont {Genes}},\ }\href {\doibase
  10.1103/PhysRevLett.122.243601} {\bibfield  {journal} {\bibinfo  {journal}
  {Phys. Rev. Lett.}\ }\textbf {\bibinfo {volume} {122}},\ \bibinfo {pages}
  {243601} (\bibinfo {year} {2019})}\BibitemShut {NoStop}%
\bibitem [{\citenamefont {Loudon}\ and\ \citenamefont
  {Knight}(1987)}]{sqzlightold}%
  \BibitemOpen
  \bibfield  {author} {\bibinfo {author} {\bibfnamefont {R.}~\bibnamefont
  {Loudon}}\ and\ \bibinfo {author} {\bibfnamefont {P.}~\bibnamefont
  {Knight}},\ }\href {\doibase 10.1080/09500348714550721} {\bibfield  {journal}
  {\bibinfo  {journal} {Journal of Modern Optics}\ }\textbf {\bibinfo {volume}
  {34}},\ \bibinfo {pages} {709} (\bibinfo {year} {1987})}\BibitemShut
  {NoStop}%
\bibitem [{\citenamefont {Lvovsky}()}]{sqzlightnew}%
  \BibitemOpen
  \bibfield  {author} {\bibinfo {author} {\bibfnamefont {A.~I.}\ \bibnamefont
  {Lvovsky}},\ }\href@noop {} {\enquote {\bibinfo {title} {Squeezed light},}\
  }\Eprint {http://arxiv.org/abs/1401.4118} {arXiv:1401.4118 [quant-ph]}
  \BibitemShut {NoStop}%
\bibitem [{\citenamefont {Sete}\ \emph {et~al.}(2011)\citenamefont {Sete},
  \citenamefont {Eleuch},\ and\ \citenamefont {Das}}]{cvtqwed_sqzlit2}%
  \BibitemOpen
  \bibfield  {author} {\bibinfo {author} {\bibfnamefont {E.~A.}\ \bibnamefont
  {Sete}}, \bibinfo {author} {\bibfnamefont {H.}~\bibnamefont {Eleuch}}, \ and\
  \bibinfo {author} {\bibfnamefont {S.}~\bibnamefont {Das}},\ }\href {\doibase
  10.1103/PhysRevA.84.053817} {\bibfield  {journal} {\bibinfo  {journal} {Phys.
  Rev. A}\ }\textbf {\bibinfo {volume} {84}},\ \bibinfo {pages} {053817}
  (\bibinfo {year} {2011})}\BibitemShut {NoStop}%
\bibitem [{\citenamefont {Bao}\ \emph {et~al.}(2019)\citenamefont {Bao},
  \citenamefont {Zhu}, \citenamefont {Yang},\ and\ \citenamefont
  {Agarwal}}]{cvtqwed_sqzlit3}%
  \BibitemOpen
  \bibfield  {author} {\bibinfo {author} {\bibfnamefont {D.~Q.}\ \bibnamefont
  {Bao}}, \bibinfo {author} {\bibfnamefont {C.~J.}\ \bibnamefont {Zhu}},
  \bibinfo {author} {\bibfnamefont {Y.~P.}\ \bibnamefont {Yang}}, \ and\
  \bibinfo {author} {\bibfnamefont {G.~S.}\ \bibnamefont {Agarwal}},\ }\href
  {\doibase 10.1364/OE.27.015540} {\bibfield  {journal} {\bibinfo  {journal}
  {Opt. Express}\ }\textbf {\bibinfo {volume} {27}},\ \bibinfo {pages} {15540}
  (\bibinfo {year} {2019})}\BibitemShut {NoStop}%
\bibitem [{\citenamefont {Schwinger}(1961)}]{schwinger}%
  \BibitemOpen
  \bibfield  {author} {\bibinfo {author} {\bibfnamefont {J.}~\bibnamefont
  {Schwinger}},\ }\href {\doibase 10.1063/1.1703727} {\bibfield  {journal}
  {\bibinfo  {journal} {Journal of Mathematical Physics}\ }\textbf {\bibinfo
  {volume} {2}},\ \bibinfo {pages} {407} (\bibinfo {year} {1961})}\BibitemShut
  {NoStop}%
\bibitem [{\citenamefont {Keldysh}(1965)}]{keldysh}%
  \BibitemOpen
  \bibfield  {author} {\bibinfo {author} {\bibfnamefont {L.~V.}\ \bibnamefont
  {Keldysh}},\ }\href
  {http://www.jetp.ac.ru/cgi-bin/e/index/e/20/4/p1018?a=list} {\bibfield
  {journal} {\bibinfo  {journal} {Sov. Phys. JETP}\ }\textbf {\bibinfo {volume}
  {20}},\ \bibinfo {pages} {1018} (\bibinfo {year} {1965})}\BibitemShut
  {NoStop}%
\bibitem [{\citenamefont {Kamenev}\ and\ \citenamefont
  {Levchenko}(2009)}]{kamenevarticle}%
  \BibitemOpen
  \bibfield  {author} {\bibinfo {author} {\bibfnamefont {A.}~\bibnamefont
  {Kamenev}}\ and\ \bibinfo {author} {\bibfnamefont {A.}~\bibnamefont
  {Levchenko}},\ }\href {\doibase 10.1080/00018730902850504} {\bibfield
  {journal} {\bibinfo  {journal} {Advances in Physics}\ }\textbf {\bibinfo
  {volume} {58}},\ \bibinfo {pages} {197} (\bibinfo {year} {2009})}\BibitemShut
  {NoStop}%
\bibitem [{\citenamefont {Kamenev}(2011)}]{kamenevbook}%
  \BibitemOpen
  \bibfield  {author} {\bibinfo {author} {\bibfnamefont {A.}~\bibnamefont
  {Kamenev}},\ }\href {\doibase 10.1017/CBO9781139003667} {\emph {\bibinfo
  {title} {Field Theory of Non-Equilibrium Systems}}}\ (\bibinfo  {publisher}
  {Cambridge University Press},\ \bibinfo {address} {New York},\ \bibinfo
  {year} {2011})\BibitemShut {NoStop}%
\bibitem [{\citenamefont {Rammer}(2007)}]{rammer}%
  \BibitemOpen
  \bibfield  {author} {\bibinfo {author} {\bibfnamefont {J.}~\bibnamefont
  {Rammer}},\ }\href {\doibase 10.1017/CBO9780511618956} {\emph {\bibinfo
  {title} {Quantum Field Theory of Non-equilibrium States}}}\ (\bibinfo
  {publisher} {Cambridge University Press},\ \bibinfo {address} {New York},\
  \bibinfo {year} {2007})\BibitemShut {NoStop}%
\bibitem [{\citenamefont {Chakraborty}\ \emph {et~al.}(2019)\citenamefont
  {Chakraborty}, \citenamefont {Gorantla},\ and\ \citenamefont
  {Sensarma}}]{cgs}%
  \BibitemOpen
  \bibfield  {author} {\bibinfo {author} {\bibfnamefont {A.}~\bibnamefont
  {Chakraborty}}, \bibinfo {author} {\bibfnamefont {P.}~\bibnamefont
  {Gorantla}}, \ and\ \bibinfo {author} {\bibfnamefont {R.}~\bibnamefont
  {Sensarma}},\ }\href {\doibase 10.1103/PhysRevB.99.054306} {\bibfield
  {journal} {\bibinfo  {journal} {Phys. Rev. B}\ }\textbf {\bibinfo {volume}
  {99}},\ \bibinfo {pages} {054306} (\bibinfo {year} {2019})}\BibitemShut
  {NoStop}%
\bibitem [{\citenamefont {Leggett}\ \emph {et~al.}(1987)\citenamefont
  {Leggett}, \citenamefont {Chakravarty}, \citenamefont {Dorsey}, \citenamefont
  {Fisher}, \citenamefont {Garg},\ and\ \citenamefont
  {Zwerger}}]{diss2ss_legget}%
  \BibitemOpen
  \bibfield  {author} {\bibinfo {author} {\bibfnamefont {A.~J.}\ \bibnamefont
  {Leggett}}, \bibinfo {author} {\bibfnamefont {S.}~\bibnamefont
  {Chakravarty}}, \bibinfo {author} {\bibfnamefont {A.~T.}\ \bibnamefont
  {Dorsey}}, \bibinfo {author} {\bibfnamefont {M.~P.~A.}\ \bibnamefont
  {Fisher}}, \bibinfo {author} {\bibfnamefont {A.}~\bibnamefont {Garg}}, \ and\
  \bibinfo {author} {\bibfnamefont {W.}~\bibnamefont {Zwerger}},\ }\href
  {\doibase 10.1103/RevModPhys.59.1} {\bibfield  {journal} {\bibinfo  {journal}
  {Rev. Mod. Phys.}\ }\textbf {\bibinfo {volume} {59}},\ \bibinfo {pages} {1}
  (\bibinfo {year} {1987})}\BibitemShut {NoStop}%
\bibitem [{\citenamefont {Gardiner}\ and\ \citenamefont
  {Zoller}()}]{quantnoise}%
  \BibitemOpen
  \bibfield  {author} {\bibinfo {author} {\bibfnamefont {C.}~\bibnamefont
  {Gardiner}}\ and\ \bibinfo {author} {\bibfnamefont {P.}~\bibnamefont
  {Zoller}},\ }\href@noop {} {\emph {\bibinfo {title} {Quantum Noise}}}\
  (\bibinfo  {publisher} {Springer Berlin},\ \bibinfo {address}
  {Heidelberg})\BibitemShut {NoStop}%
\bibitem [{\citenamefont {Elsayed-Ali}\ \emph {et~al.}(1987)\citenamefont
  {Elsayed-Ali}, \citenamefont {Norris}, \citenamefont {Pessot},\ and\
  \citenamefont {Mourou}}]{ephrelCu}%
  \BibitemOpen
  \bibfield  {author} {\bibinfo {author} {\bibfnamefont {H.~E.}\ \bibnamefont
  {Elsayed-Ali}}, \bibinfo {author} {\bibfnamefont {T.~B.}\ \bibnamefont
  {Norris}}, \bibinfo {author} {\bibfnamefont {M.~A.}\ \bibnamefont {Pessot}},
  \ and\ \bibinfo {author} {\bibfnamefont {G.~A.}\ \bibnamefont {Mourou}},\
  }\href {\doibase 10.1103/PhysRevLett.58.1212} {\bibfield  {journal} {\bibinfo
   {journal} {Phys. Rev. Lett.}\ }\textbf {\bibinfo {volume} {58}},\ \bibinfo
  {pages} {1212} (\bibinfo {year} {1987})}\BibitemShut {NoStop}%
\bibitem [{\citenamefont {Wellstood}\ \emph {et~al.}(1994)\citenamefont
  {Wellstood}, \citenamefont {Urbina},\ and\ \citenamefont
  {Clarke}}]{hotel_metal}%
  \BibitemOpen
  \bibfield  {author} {\bibinfo {author} {\bibfnamefont {F.~C.}\ \bibnamefont
  {Wellstood}}, \bibinfo {author} {\bibfnamefont {C.}~\bibnamefont {Urbina}}, \
  and\ \bibinfo {author} {\bibfnamefont {J.}~\bibnamefont {Clarke}},\ }\href
  {\doibase 10.1103/PhysRevB.49.5942} {\bibfield  {journal} {\bibinfo
  {journal} {Phys. Rev. B}\ }\textbf {\bibinfo {volume} {49}},\ \bibinfo
  {pages} {5942} (\bibinfo {year} {1994})}\BibitemShut {NoStop}%
\bibitem [{\citenamefont {Del~Fatti}\ \emph {et~al.}(2000)\citenamefont
  {Del~Fatti}, \citenamefont {Voisin}, \citenamefont {Achermann}, \citenamefont
  {Tzortzakis}, \citenamefont {Christofilos},\ and\ \citenamefont
  {Vall\'ee}}]{edyn_noble}%
  \BibitemOpen
  \bibfield  {author} {\bibinfo {author} {\bibfnamefont {N.}~\bibnamefont
  {Del~Fatti}}, \bibinfo {author} {\bibfnamefont {C.}~\bibnamefont {Voisin}},
  \bibinfo {author} {\bibfnamefont {M.}~\bibnamefont {Achermann}}, \bibinfo
  {author} {\bibfnamefont {S.}~\bibnamefont {Tzortzakis}}, \bibinfo {author}
  {\bibfnamefont {D.}~\bibnamefont {Christofilos}}, \ and\ \bibinfo {author}
  {\bibfnamefont {F.}~\bibnamefont {Vall\'ee}},\ }\href {\doibase
  10.1103/PhysRevB.61.16956} {\bibfield  {journal} {\bibinfo  {journal} {Phys.
  Rev. B}\ }\textbf {\bibinfo {volume} {61}},\ \bibinfo {pages} {16956}
  (\bibinfo {year} {2000})}\BibitemShut {NoStop}%
\bibitem [{\citenamefont {Giazotto}\ \emph {et~al.}(2006)\citenamefont
  {Giazotto}, \citenamefont {Heikkil\"{a}}, \citenamefont {Luukanen},
  \citenamefont {Savin},\ and\ \citenamefont {Pekola}}]{mesothermo}%
  \BibitemOpen
  \bibfield  {author} {\bibinfo {author} {\bibfnamefont {F.}~\bibnamefont
  {Giazotto}}, \bibinfo {author} {\bibfnamefont {T.~T.}\ \bibnamefont
  {Heikkil\"{a}}}, \bibinfo {author} {\bibfnamefont {A.}~\bibnamefont
  {Luukanen}}, \bibinfo {author} {\bibfnamefont {A.~M.}\ \bibnamefont {Savin}},
  \ and\ \bibinfo {author} {\bibfnamefont {J.~P.}\ \bibnamefont {Pekola}},\
  }\href {\doibase 10.1103/RevModPhys.78.217} {\bibfield  {journal} {\bibinfo
  {journal} {Rev. Mod. Phys.}\ }\textbf {\bibinfo {volume} {78}},\ \bibinfo
  {pages} {217} (\bibinfo {year} {2006})}\BibitemShut {NoStop}%
\bibitem [{\citenamefont {Habib}\ \emph {et~al.}(2018)\citenamefont {Habib},
  \citenamefont {Florio},\ and\ \citenamefont {Sundararaman}}]{hotdyn_habib}%
  \BibitemOpen
  \bibfield  {author} {\bibinfo {author} {\bibfnamefont {A.}~\bibnamefont
  {Habib}}, \bibinfo {author} {\bibfnamefont {F.}~\bibnamefont {Florio}}, \
  and\ \bibinfo {author} {\bibfnamefont {R.}~\bibnamefont {Sundararaman}},\
  }\href {\doibase 10.1088/2040-8986/aac1d8} {\bibfield  {journal} {\bibinfo
  {journal} {Journal of Optics}\ }\textbf {\bibinfo {volume} {20}},\ \bibinfo
  {pages} {064001} (\bibinfo {year} {2018})}\BibitemShut {NoStop}%
\bibitem [{\citenamefont {Bergeret}\ \emph {et~al.}(2018)\citenamefont
  {Bergeret}, \citenamefont {Silaev}, \citenamefont {Virtanen},\ and\
  \citenamefont {Heikkil\"a}}]{noneqsc_spinsplit}%
  \BibitemOpen
  \bibfield  {author} {\bibinfo {author} {\bibfnamefont {F.~S.}\ \bibnamefont
  {Bergeret}}, \bibinfo {author} {\bibfnamefont {M.}~\bibnamefont {Silaev}},
  \bibinfo {author} {\bibfnamefont {P.}~\bibnamefont {Virtanen}}, \ and\
  \bibinfo {author} {\bibfnamefont {T.~T.}\ \bibnamefont {Heikkil\"a}},\ }\href
  {\doibase 10.1103/RevModPhys.90.041001} {\bibfield  {journal} {\bibinfo
  {journal} {Rev. Mod. Phys.}\ }\textbf {\bibinfo {volume} {90}},\ \bibinfo
  {pages} {041001} (\bibinfo {year} {2018})}\BibitemShut {NoStop}%
\bibitem [{\citenamefont {Dal~Forno}\ and\ \citenamefont
  {Lischner}(2019)}]{hotel_TiN}%
  \BibitemOpen
  \bibfield  {author} {\bibinfo {author} {\bibfnamefont {S.}~\bibnamefont
  {Dal~Forno}}\ and\ \bibinfo {author} {\bibfnamefont {J.}~\bibnamefont
  {Lischner}},\ }\href {\doibase 10.1103/PhysRevMaterials.3.115203} {\bibfield
  {journal} {\bibinfo  {journal} {Phys. Rev. Materials}\ }\textbf {\bibinfo
  {volume} {3}},\ \bibinfo {pages} {115203} (\bibinfo {year}
  {2019})}\BibitemShut {NoStop}%
\bibitem [{\citenamefont {Giannetti}\ \emph {et~al.}(2016)\citenamefont
  {Giannetti}, \citenamefont {Capone}, \citenamefont {Fausti}, \citenamefont
  {Fabrizio}, \citenamefont {Parmigiani},\ and\ \citenamefont
  {Mihailovic}}]{ultraspec_hiTsc}%
  \BibitemOpen
  \bibfield  {author} {\bibinfo {author} {\bibfnamefont {C.}~\bibnamefont
  {Giannetti}}, \bibinfo {author} {\bibfnamefont {M.}~\bibnamefont {Capone}},
  \bibinfo {author} {\bibfnamefont {D.}~\bibnamefont {Fausti}}, \bibinfo
  {author} {\bibfnamefont {M.}~\bibnamefont {Fabrizio}}, \bibinfo {author}
  {\bibfnamefont {F.}~\bibnamefont {Parmigiani}}, \ and\ \bibinfo {author}
  {\bibfnamefont {D.}~\bibnamefont {Mihailovic}},\ }\href {\doibase
  10.1080/00018732.2016.1194044} {\bibfield  {journal} {\bibinfo  {journal}
  {Advances in Physics}\ }\textbf {\bibinfo {volume} {65}},\ \bibinfo {pages}
  {58} (\bibinfo {year} {2016})}\BibitemShut {NoStop}%
\bibitem [{\citenamefont {Besteiro}\ \emph {et~al.}(2021)\citenamefont
  {Besteiro}, \citenamefont {Cortés}, \citenamefont {Ishii}, \citenamefont
  {Narang},\ and\ \citenamefont {Oulton}}]{hotel_apply}%
  \BibitemOpen
  \bibfield  {author} {\bibinfo {author} {\bibfnamefont {L.~V.}\ \bibnamefont
  {Besteiro}}, \bibinfo {author} {\bibfnamefont {E.}~\bibnamefont {Cortés}},
  \bibinfo {author} {\bibfnamefont {S.}~\bibnamefont {Ishii}}, \bibinfo
  {author} {\bibfnamefont {P.}~\bibnamefont {Narang}}, \ and\ \bibinfo {author}
  {\bibfnamefont {R.~F.}\ \bibnamefont {Oulton}},\ }\href {\doibase
  10.1063/5.0050796} {\bibfield  {journal} {\bibinfo  {journal} {Journal of
  Applied Physics}\ }\textbf {\bibinfo {volume} {129}},\ \bibinfo {pages}
  {150401} (\bibinfo {year} {2021})}\BibitemShut {NoStop}%
\bibitem [{\citenamefont {Peterreins}\ \emph {et~al.}(1991)\citenamefont
  {Peterreins}, \citenamefont {Jochum}, \citenamefont {Pröbst}, \citenamefont
  {Feilitzsch}, \citenamefont {Kraus},\ and\ \citenamefont
  {Mössbauer}}]{noneq_phdetect}%
  \BibitemOpen
  \bibfield  {author} {\bibinfo {author} {\bibfnamefont {T.}~\bibnamefont
  {Peterreins}}, \bibinfo {author} {\bibfnamefont {J.}~\bibnamefont {Jochum}},
  \bibinfo {author} {\bibfnamefont {F.}~\bibnamefont {Pröbst}}, \bibinfo
  {author} {\bibfnamefont {F.~V.}\ \bibnamefont {Feilitzsch}}, \bibinfo
  {author} {\bibfnamefont {H.}~\bibnamefont {Kraus}}, \ and\ \bibinfo {author}
  {\bibfnamefont {R.~L.}\ \bibnamefont {Mössbauer}},\ }\href {\doibase
  10.1063/1.347233} {\bibfield  {journal} {\bibinfo  {journal} {Journal of
  Applied Physics}\ }\textbf {\bibinfo {volume} {69}},\ \bibinfo {pages} {1791}
  (\bibinfo {year} {1991})}\BibitemShut {NoStop}%
\bibitem [{\citenamefont {Hertzberg}\ \emph {et~al.}(2011)\citenamefont
  {Hertzberg}, \citenamefont {Otelaja}, \citenamefont {Yoshida},\ and\
  \citenamefont {Robinson}}]{noneqph_gendet}%
  \BibitemOpen
  \bibfield  {author} {\bibinfo {author} {\bibfnamefont {J.~B.}\ \bibnamefont
  {Hertzberg}}, \bibinfo {author} {\bibfnamefont {O.~O.}\ \bibnamefont
  {Otelaja}}, \bibinfo {author} {\bibfnamefont {N.~J.}\ \bibnamefont
  {Yoshida}}, \ and\ \bibinfo {author} {\bibfnamefont {R.~D.}\ \bibnamefont
  {Robinson}},\ }\href {\doibase 10.1063/1.3652979} {\bibfield  {journal}
  {\bibinfo  {journal} {Review of Scientific Instruments}\ }\textbf {\bibinfo
  {volume} {82}},\ \bibinfo {pages} {104905} (\bibinfo {year}
  {2011})}\BibitemShut {NoStop}%
\bibitem [{\citenamefont {Perinati}\ \emph {et~al.}(2004)\citenamefont
  {Perinati}, \citenamefont {Barbera}, \citenamefont {Collura}, \citenamefont
  {Serio},\ and\ \citenamefont {Silver}}]{perinati}%
  \BibitemOpen
  \bibfield  {author} {\bibinfo {author} {\bibfnamefont {E.}~\bibnamefont
  {Perinati}}, \bibinfo {author} {\bibfnamefont {M.}~\bibnamefont {Barbera}},
  \bibinfo {author} {\bibfnamefont {A.}~\bibnamefont {Collura}}, \bibinfo
  {author} {\bibfnamefont {S.}~\bibnamefont {Serio}}, \ and\ \bibinfo {author}
  {\bibfnamefont {E.}~\bibnamefont {Silver}},\ }\href {\doibase
  https://doi.org/10.1016/j.nima.2004.05.090} {\bibfield  {journal} {\bibinfo
  {journal} {Nuclear Instruments and Methods in Physics Research Section A:
  Accelerators, Spectrometers, Detectors and Associated Equipment}\ }\textbf
  {\bibinfo {volume} {531}},\ \bibinfo {pages} {459} (\bibinfo {year}
  {2004})}\BibitemShut {NoStop}%
\bibitem [{\citenamefont {Zukerstein}\ \emph {et~al.}(2019)\citenamefont
  {Zukerstein}, \citenamefont {Trojánek}, \citenamefont {Rezek}, \citenamefont
  {Šobáň}, \citenamefont {Kozák},\ and\ \citenamefont
  {Malý}}]{phdyn_detect}%
  \BibitemOpen
  \bibfield  {author} {\bibinfo {author} {\bibfnamefont {M.}~\bibnamefont
  {Zukerstein}}, \bibinfo {author} {\bibfnamefont {F.}~\bibnamefont
  {Trojánek}}, \bibinfo {author} {\bibfnamefont {B.}~\bibnamefont {Rezek}},
  \bibinfo {author} {\bibfnamefont {Z.}~\bibnamefont {Šobáň}}, \bibinfo
  {author} {\bibfnamefont {M.}~\bibnamefont {Kozák}}, \ and\ \bibinfo {author}
  {\bibfnamefont {P.}~\bibnamefont {Malý}},\ }\href {\doibase
  10.1063/1.5119056} {\bibfield  {journal} {\bibinfo  {journal} {Applied
  Physics Letters}\ }\textbf {\bibinfo {volume} {115}},\ \bibinfo {pages}
  {161104} (\bibinfo {year} {2019})}\BibitemShut {NoStop}%
\bibitem [{\citenamefont {Lakehal}\ \emph {et~al.}(2020)\citenamefont
  {Lakehal}, \citenamefont {Schir\'o}, \citenamefont {Eremin},\ and\
  \citenamefont {Paul}}]{sqzph_ipaul}%
  \BibitemOpen
  \bibfield  {author} {\bibinfo {author} {\bibfnamefont {M.}~\bibnamefont
  {Lakehal}}, \bibinfo {author} {\bibfnamefont {M.}~\bibnamefont {Schir\'o}},
  \bibinfo {author} {\bibfnamefont {I.~M.}\ \bibnamefont {Eremin}}, \ and\
  \bibinfo {author} {\bibfnamefont {I.}~\bibnamefont {Paul}},\ }\href {\doibase
  10.1103/PhysRevB.102.174316} {\bibfield  {journal} {\bibinfo  {journal}
  {Phys. Rev. B}\ }\textbf {\bibinfo {volume} {102}},\ \bibinfo {pages}
  {174316} (\bibinfo {year} {2020})}\BibitemShut {NoStop}%
\bibitem [{\citenamefont {Micklitz}\ \emph {et~al.}(2010)\citenamefont
  {Micklitz}, \citenamefont {Rech},\ and\ \citenamefont {Matveev}}]{matveev1}%
  \BibitemOpen
  \bibfield  {author} {\bibinfo {author} {\bibfnamefont {T.}~\bibnamefont
  {Micklitz}}, \bibinfo {author} {\bibfnamefont {J.}~\bibnamefont {Rech}}, \
  and\ \bibinfo {author} {\bibfnamefont {K.~A.}\ \bibnamefont {Matveev}},\
  }\href {\doibase 10.1103/PhysRevB.81.115313} {\bibfield  {journal} {\bibinfo
  {journal} {Phys. Rev. B}\ }\textbf {\bibinfo {volume} {81}},\ \bibinfo
  {pages} {115313} (\bibinfo {year} {2010})}\BibitemShut {NoStop}%
\bibitem [{\citenamefont {Matveev}\ \emph {et~al.}(2010)\citenamefont
  {Matveev}, \citenamefont {Andreev},\ and\ \citenamefont
  {Pustilnik}}]{matveev2}%
  \BibitemOpen
  \bibfield  {author} {\bibinfo {author} {\bibfnamefont {K.~A.}\ \bibnamefont
  {Matveev}}, \bibinfo {author} {\bibfnamefont {A.~V.}\ \bibnamefont
  {Andreev}}, \ and\ \bibinfo {author} {\bibfnamefont {M.}~\bibnamefont
  {Pustilnik}},\ }\href {\doibase 10.1103/PhysRevLett.105.046401} {\bibfield
  {journal} {\bibinfo  {journal} {Phys. Rev. Lett.}\ }\textbf {\bibinfo
  {volume} {105}},\ \bibinfo {pages} {046401} (\bibinfo {year}
  {2010})}\BibitemShut {NoStop}%
\bibitem [{\citenamefont {Castro~Neto}\ and\ \citenamefont
  {Fisher}(1996)}]{heavyparticle_castro}%
  \BibitemOpen
  \bibfield  {author} {\bibinfo {author} {\bibfnamefont {A.~H.}\ \bibnamefont
  {Castro~Neto}}\ and\ \bibinfo {author} {\bibfnamefont {M.~P.~A.}\
  \bibnamefont {Fisher}},\ }\href {\doibase 10.1103/PhysRevB.53.9713}
  {\bibfield  {journal} {\bibinfo  {journal} {Phys. Rev. B}\ }\textbf {\bibinfo
  {volume} {53}},\ \bibinfo {pages} {9713} (\bibinfo {year}
  {1996})}\BibitemShut {NoStop}%
\bibitem [{\citenamefont {Gangardt}\ and\ \citenamefont
  {Kamenev}(2009)}]{spinorgas_kamenev}%
  \BibitemOpen
  \bibfield  {author} {\bibinfo {author} {\bibfnamefont {D.~M.}\ \bibnamefont
  {Gangardt}}\ and\ \bibinfo {author} {\bibfnamefont {A.}~\bibnamefont
  {Kamenev}},\ }\href {\doibase 10.1103/PhysRevLett.102.070402} {\bibfield
  {journal} {\bibinfo  {journal} {Phys. Rev. Lett.}\ }\textbf {\bibinfo
  {volume} {102}},\ \bibinfo {pages} {070402} (\bibinfo {year}
  {2009})}\BibitemShut {NoStop}%
\bibitem [{\citenamefont {Pekar}(1958)}]{extnplrtn_th}%
  \BibitemOpen
  \bibfield  {author} {\bibinfo {author} {\bibfnamefont {S.}~\bibnamefont
  {Pekar}},\ }\href {\doibase https://doi.org/10.1016/0022-3697(58)90127-6}
  {\bibfield  {journal} {\bibinfo  {journal} {Journal of Physics and Chemistry
  of Solids}\ }\textbf {\bibinfo {volume} {5}},\ \bibinfo {pages} {11}
  (\bibinfo {year} {1958})}\BibitemShut {NoStop}%
\bibitem [{\citenamefont {Weisbuch}\ \emph {et~al.}(1992)\citenamefont
  {Weisbuch}, \citenamefont {Nishioka}, \citenamefont {Ishikawa},\ and\
  \citenamefont {Arakawa}}]{extnplrtn_exp}%
  \BibitemOpen
  \bibfield  {author} {\bibinfo {author} {\bibfnamefont {C.}~\bibnamefont
  {Weisbuch}}, \bibinfo {author} {\bibfnamefont {M.}~\bibnamefont {Nishioka}},
  \bibinfo {author} {\bibfnamefont {A.}~\bibnamefont {Ishikawa}}, \ and\
  \bibinfo {author} {\bibfnamefont {Y.}~\bibnamefont {Arakawa}},\ }\href
  {\doibase 10.1103/PhysRevLett.69.3314} {\bibfield  {journal} {\bibinfo
  {journal} {Phys. Rev. Lett.}\ }\textbf {\bibinfo {volume} {69}},\ \bibinfo
  {pages} {3314} (\bibinfo {year} {1992})}\BibitemShut {NoStop}%
\bibitem [{\citenamefont {OCKMAN}\ \emph {et~al.}(1991)\citenamefont {OCKMAN},
  \citenamefont {WANG},\ and\ \citenamefont {ALFANO}}]{ufpps_revold}%
  \BibitemOpen
  \bibfield  {author} {\bibinfo {author} {\bibfnamefont {N.}~\bibnamefont
  {OCKMAN}}, \bibinfo {author} {\bibfnamefont {W.}~\bibnamefont {WANG}}, \ and\
  \bibinfo {author} {\bibfnamefont {R.}~\bibnamefont {ALFANO}},\ }\href
  {\doibase 10.1142/S0217979291001255} {\bibfield  {journal} {\bibinfo
  {journal} {International Journal of Modern Physics B}\ }\textbf {\bibinfo
  {volume} {05}},\ \bibinfo {pages} {3165} (\bibinfo {year}
  {1991})}\BibitemShut {NoStop}%
\bibitem [{\citenamefont {Flock}\ \emph {et~al.}(2014)\citenamefont {Flock},
  \citenamefont {Dekorsy},\ and\ \citenamefont {Misochko}}]{ufpps_exp2}%
  \BibitemOpen
  \bibfield  {author} {\bibinfo {author} {\bibfnamefont {J.}~\bibnamefont
  {Flock}}, \bibinfo {author} {\bibfnamefont {T.}~\bibnamefont {Dekorsy}}, \
  and\ \bibinfo {author} {\bibfnamefont {O.~V.}\ \bibnamefont {Misochko}},\
  }\href {\doibase 10.1063/1.4887483} {\bibfield  {journal} {\bibinfo
  {journal} {Applied Physics Letters}\ }\textbf {\bibinfo {volume} {105}},\
  \bibinfo {pages} {011902} (\bibinfo {year} {2014})}\BibitemShut {NoStop}%
\bibitem [{\citenamefont {Hu}\ \emph {et~al.}(2018)\citenamefont {Hu},
  \citenamefont {Igarashi}, \citenamefont {Sasagawa}, \citenamefont
  {Nakamura},\ and\ \citenamefont {Misochko}}]{ufpps_exp3}%
  \BibitemOpen
  \bibfield  {author} {\bibinfo {author} {\bibfnamefont {J.}~\bibnamefont
  {Hu}}, \bibinfo {author} {\bibfnamefont {K.}~\bibnamefont {Igarashi}},
  \bibinfo {author} {\bibfnamefont {T.}~\bibnamefont {Sasagawa}}, \bibinfo
  {author} {\bibfnamefont {K.~G.}\ \bibnamefont {Nakamura}}, \ and\ \bibinfo
  {author} {\bibfnamefont {O.~V.}\ \bibnamefont {Misochko}},\ }\href {\doibase
  10.1063/1.5016941} {\bibfield  {journal} {\bibinfo  {journal} {Applied
  Physics Letters}\ }\textbf {\bibinfo {volume} {112}},\ \bibinfo {pages}
  {031901} (\bibinfo {year} {2018})}\BibitemShut {NoStop}%
\bibitem [{\citenamefont {Busza}\ \emph {et~al.}(2018)\citenamefont {Busza},
  \citenamefont {Rajagopal},\ and\ \citenamefont {van~der
  Schee}}]{heavyioncolrev}%
  \BibitemOpen
  \bibfield  {author} {\bibinfo {author} {\bibfnamefont {W.}~\bibnamefont
  {Busza}}, \bibinfo {author} {\bibfnamefont {K.}~\bibnamefont {Rajagopal}}, \
  and\ \bibinfo {author} {\bibfnamefont {W.}~\bibnamefont {van~der Schee}},\
  }\href {\doibase 10.1146/annurev-nucl-101917-020852} {\bibfield  {journal}
  {\bibinfo  {journal} {Annual Review of Nuclear and Particle Science}\
  }\textbf {\bibinfo {volume} {68}},\ \bibinfo {pages} {339} (\bibinfo {year}
  {2018})}\BibitemShut {NoStop}%
\bibitem [{\citenamefont {Mitra}(2018)}]{quench_rev}%
  \BibitemOpen
  \bibfield  {author} {\bibinfo {author} {\bibfnamefont {A.}~\bibnamefont
  {Mitra}},\ }\href {\doibase 10.1146/annurev-conmatphys-031016-025451}
  {\bibfield  {journal} {\bibinfo  {journal} {Annual Review of Condensed Matter
  Physics}\ }\textbf {\bibinfo {volume} {9}},\ \bibinfo {pages} {245} (\bibinfo
  {year} {2018})}\BibitemShut {NoStop}%
\bibitem [{\citenamefont {Das}\ \emph {et~al.}(2016)\citenamefont {Das},
  \citenamefont {Galante},\ and\ \citenamefont {Myers}}]{quench_das}%
  \BibitemOpen
  \bibfield  {author} {\bibinfo {author} {\bibfnamefont {S.~R.}\ \bibnamefont
  {Das}}, \bibinfo {author} {\bibfnamefont {D.~A.}\ \bibnamefont {Galante}}, \
  and\ \bibinfo {author} {\bibfnamefont {R.~C.}\ \bibnamefont {Myers}},\ }\href
  {\doibase 10.1007/jhep05(2016)164} {\bibfield  {journal} {\bibinfo  {journal}
  {Journal of High Energy Physics}\ }\textbf {\bibinfo {volume} {2016}}
  (\bibinfo {year} {2016}),\ 10.1007/jhep05(2016)164}\BibitemShut {NoStop}%
\bibitem [{\citenamefont {Mandal}\ \emph {et~al.}(2020)\citenamefont {Mandal},
  \citenamefont {Paranjape},\ and\ \citenamefont
  {Sorokhaibam}}]{quench_mandal}%
  \BibitemOpen
  \bibfield  {author} {\bibinfo {author} {\bibfnamefont {G.}~\bibnamefont
  {Mandal}}, \bibinfo {author} {\bibfnamefont {S.}~\bibnamefont {Paranjape}}, \
  and\ \bibinfo {author} {\bibfnamefont {N.}~\bibnamefont {Sorokhaibam}},\
  }\href@noop {} {\enquote {\bibinfo {title} {Thermalization in 2d critical
  quench and uv/ir mixing},}\ } (\bibinfo {year} {2020}),\ \Eprint
  {http://arxiv.org/abs/1512.02187} {arXiv:1512.02187 [hep-th]} \BibitemShut
  {NoStop}%
\bibitem [{\citenamefont {Watson}(1933)}]{mehler1}%
  \BibitemOpen
  \bibfield  {author} {\bibinfo {author} {\bibfnamefont {G.~N.}\ \bibnamefont
  {Watson}},\ }\href {\doibase https://doi.org/10.1112/jlms/s1-8.3.194}
  {\bibfield  {journal} {\bibinfo  {journal} {Journal of the London
  Mathematical Society}\ }\textbf {\bibinfo {volume} {s1-8}},\ \bibinfo {pages}
  {194} (\bibinfo {year} {1933})}\BibitemShut {NoStop}%
\bibitem [{\citenamefont {Weiss}(2008)}]{qds_ohm}%
  \BibitemOpen
  \bibfield  {author} {\bibinfo {author} {\bibfnamefont {U.}~\bibnamefont
  {Weiss}},\ }\href {\doibase 10.1142/6738} {\emph {\bibinfo {title} {Quantum
  Dissipative Systems}}},\ \bibinfo {edition} {3rd}\ ed.\ (\bibinfo
  {publisher} {World Scientific},\ \bibinfo {address} {Singapore},\ \bibinfo
  {year} {2008})\BibitemShut {NoStop}%
\bibitem [{\citenamefont {Chakraborty}\ and\ \citenamefont
  {Sensarma}(2018)}]{nonmarkov_oqs}%
  \BibitemOpen
  \bibfield  {author} {\bibinfo {author} {\bibfnamefont {A.}~\bibnamefont
  {Chakraborty}}\ and\ \bibinfo {author} {\bibfnamefont {R.}~\bibnamefont
  {Sensarma}},\ }\href {\doibase 10.1103/PhysRevB.97.104306} {\bibfield
  {journal} {\bibinfo  {journal} {Phys. Rev. B}\ }\textbf {\bibinfo {volume}
  {97}},\ \bibinfo {pages} {104306} (\bibinfo {year} {2018})}\BibitemShut
  {NoStop}%
\bibitem [{\citenamefont {Dawson}(1897)}]{dawson}%
  \BibitemOpen
  \bibfield  {author} {\bibinfo {author} {\bibfnamefont {H.~G.}\ \bibnamefont
  {Dawson}},\ }\href {\doibase https://doi.org/10.1112/plms/s1-29.1.519}
  {\bibfield  {journal} {\bibinfo  {journal} {Proceedings of the London
  Mathematical Society}\ }\textbf {\bibinfo {volume} {s1-29}},\ \bibinfo
  {pages} {519} (\bibinfo {year} {1897})}\BibitemShut {NoStop}%
\bibitem [{\citenamefont {Abrikosov}\ \emph {et~al.}(1975)\citenamefont
  {Abrikosov}, \citenamefont {Gorkov}, \citenamefont {Dzyaloshinski},\ and\
  \citenamefont {Silverman}}]{agd}%
  \BibitemOpen
  \bibfield  {author} {\bibinfo {author} {\bibfnamefont {A.~A.}\ \bibnamefont
  {Abrikosov}}, \bibinfo {author} {\bibfnamefont {L.~P.}\ \bibnamefont
  {Gorkov}}, \bibinfo {author} {\bibfnamefont {I.~E.}\ \bibnamefont
  {Dzyaloshinski}}, \ and\ \bibinfo {author} {\bibfnamefont {R.~A.}\
  \bibnamefont {Silverman}},\ }\href@noop {} {\emph {\bibinfo {title} {Methods
  of Quantum Field Theory in Statistical Physics}}}\ (\bibinfo  {publisher}
  {Dover},\ \bibinfo {address} {New York},\ \bibinfo {year} {1975})\BibitemShut
  {NoStop}%
\bibitem [{\citenamefont {Giustino}(2017)}]{ephint}%
  \BibitemOpen
  \bibfield  {author} {\bibinfo {author} {\bibfnamefont {F.}~\bibnamefont
  {Giustino}},\ }\href {\doibase 10.1103/RevModPhys.89.015003} {\bibfield
  {journal} {\bibinfo  {journal} {Rev. Mod. Phys.}\ }\textbf {\bibinfo {volume}
  {89}},\ \bibinfo {pages} {015003} (\bibinfo {year} {2017})}\BibitemShut
  {NoStop}%
\bibitem [{\citenamefont {Lai}\ \emph {et~al.}(2021)\citenamefont {Lai},
  \citenamefont {Xie},\ and\ \citenamefont {Zhang}}]{ephintobs}%
  \BibitemOpen
  \bibfield  {author} {\bibinfo {author} {\bibfnamefont {J.-M.}\ \bibnamefont
  {Lai}}, \bibinfo {author} {\bibfnamefont {Y.-R.}\ \bibnamefont {Xie}}, \ and\
  \bibinfo {author} {\bibfnamefont {J.}~\bibnamefont {Zhang}},\ }\href
  {\doibase 10.1007/s12274-020-2943-1} {\bibfield  {journal} {\bibinfo
  {journal} {Nano Research}\ }\textbf {\bibinfo {volume} {14}},\ \bibinfo
  {pages} {1711} (\bibinfo {year} {2021})}\BibitemShut {NoStop}%
\bibitem [{\citenamefont {Huang}\ \emph {et~al.}(2016)\citenamefont {Huang},
  \citenamefont {Zhou}, \citenamefont {Huang}, \citenamefont {Wu},
  \citenamefont {Gao}, \citenamefont {Qu},\ and\ \citenamefont
  {Chu}}]{phtelint1}%
  \BibitemOpen
  \bibfield  {author} {\bibinfo {author} {\bibfnamefont {Z.}~\bibnamefont
  {Huang}}, \bibinfo {author} {\bibfnamefont {W.}~\bibnamefont {Zhou}},
  \bibinfo {author} {\bibfnamefont {J.}~\bibnamefont {Huang}}, \bibinfo
  {author} {\bibfnamefont {J.}~\bibnamefont {Wu}}, \bibinfo {author}
  {\bibfnamefont {Y.}~\bibnamefont {Gao}}, \bibinfo {author} {\bibfnamefont
  {Y.}~\bibnamefont {Qu}}, \ and\ \bibinfo {author} {\bibfnamefont
  {J.}~\bibnamefont {Chu}},\ }\href {\doibase 10.1038/srep22938} {\bibfield
  {journal} {\bibinfo  {journal} {Scientific Reports}\ }\textbf {\bibinfo
  {volume} {6}},\ \bibinfo {pages} {22938} (\bibinfo {year}
  {2016})}\BibitemShut {NoStop}%
\bibitem [{\citenamefont {Yu}\ \emph {et~al.}(2019)\citenamefont {Yu},
  \citenamefont {Peng}, \citenamefont {Yang},\ and\ \citenamefont
  {Li}}]{phtelint2}%
  \BibitemOpen
  \bibfield  {author} {\bibinfo {author} {\bibfnamefont {H.}~\bibnamefont
  {Yu}}, \bibinfo {author} {\bibfnamefont {Y.}~\bibnamefont {Peng}}, \bibinfo
  {author} {\bibfnamefont {Y.}~\bibnamefont {Yang}}, \ and\ \bibinfo {author}
  {\bibfnamefont {Z.-Y.}\ \bibnamefont {Li}},\ }\href {\doibase
  10.1038/s41524-019-0184-1} {\bibfield  {journal} {\bibinfo  {journal} {npj
  Computational Materials}\ }\textbf {\bibinfo {volume} {5}},\ \bibinfo {pages}
  {45} (\bibinfo {year} {2019})}\BibitemShut {NoStop}%
\bibitem [{\citenamefont {Ritsch}\ \emph {et~al.}(2013)\citenamefont {Ritsch},
  \citenamefont {Domokos}, \citenamefont {Brennecke},\ and\ \citenamefont
  {Esslinger}}]{coldatmcvtrev}%
  \BibitemOpen
  \bibfield  {author} {\bibinfo {author} {\bibfnamefont {H.}~\bibnamefont
  {Ritsch}}, \bibinfo {author} {\bibfnamefont {P.}~\bibnamefont {Domokos}},
  \bibinfo {author} {\bibfnamefont {F.}~\bibnamefont {Brennecke}}, \ and\
  \bibinfo {author} {\bibfnamefont {T.}~\bibnamefont {Esslinger}},\ }\href
  {\doibase 10.1103/RevModPhys.85.553} {\bibfield  {journal} {\bibinfo
  {journal} {Rev. Mod. Phys.}\ }\textbf {\bibinfo {volume} {85}},\ \bibinfo
  {pages} {553} (\bibinfo {year} {2013})}\BibitemShut {NoStop}%
\bibitem [{\citenamefont {Mass\'o}\ \emph {et~al.}(2002)\citenamefont
  {Mass\'o}, \citenamefont {Rota},\ and\ \citenamefont
  {Zsembinszki}}]{axiontherm}%
  \BibitemOpen
  \bibfield  {author} {\bibinfo {author} {\bibfnamefont {E.}~\bibnamefont
  {Mass\'o}}, \bibinfo {author} {\bibfnamefont {F.}~\bibnamefont {Rota}}, \
  and\ \bibinfo {author} {\bibfnamefont {G.}~\bibnamefont {Zsembinszki}},\
  }\href {\doibase 10.1103/PhysRevD.66.023004} {\bibfield  {journal} {\bibinfo
  {journal} {Phys. Rev. D}\ }\textbf {\bibinfo {volume} {66}},\ \bibinfo
  {pages} {023004} (\bibinfo {year} {2002})}\BibitemShut {NoStop}%
\bibitem [{\citenamefont {Mahan}(2000)}]{mahan}%
  \BibitemOpen
  \bibfield  {author} {\bibinfo {author} {\bibfnamefont {G.~D.}\ \bibnamefont
  {Mahan}},\ }\href {\doibase https://doi.org/10.1007/978-1-4757-5714-9} {\emph
  {\bibinfo {title} {Many-Particle Physics}}}\ (\bibinfo  {publisher} {Springer
  New York},\ \bibinfo {year} {2000})\BibitemShut {NoStop}%
\bibitem [{\citenamefont {Giamarchi}(2003)}]{giamarchi}%
  \BibitemOpen
  \bibfield  {author} {\bibinfo {author} {\bibfnamefont {T.}~\bibnamefont
  {Giamarchi}},\ }\href
  {https://oxford.universitypressscholarship.com/view/10.1093/acprof:oso/9780198525004.001.0001/acprof-9780198525004}
  {\emph {\bibinfo {title} {Quantum physics in one dimension}}}\ (\bibinfo
  {publisher} {Oxford Unversity Press},\ \bibinfo {address} {New York},\
  \bibinfo {year} {2003})\BibitemShut {NoStop}%
\bibitem [{\citenamefont {Yang}\ \emph {et~al.}(2019)\citenamefont {Yang},
  \citenamefont {Tao}, \citenamefont {Liu}, \citenamefont {Liu}, \citenamefont
  {Zhang}, \citenamefont {Akter}, \citenamefont {Zhao}, \citenamefont {Xu},
  \citenamefont {Xu}, \citenamefont {Mao}, \citenamefont {Chen},\ and\
  \citenamefont {Li}}]{phnNbSe3nano}%
  \BibitemOpen
  \bibfield  {author} {\bibinfo {author} {\bibfnamefont {L.}~\bibnamefont
  {Yang}}, \bibinfo {author} {\bibfnamefont {Y.}~\bibnamefont {Tao}}, \bibinfo
  {author} {\bibfnamefont {J.}~\bibnamefont {Liu}}, \bibinfo {author}
  {\bibfnamefont {C.}~\bibnamefont {Liu}}, \bibinfo {author} {\bibfnamefont
  {Q.}~\bibnamefont {Zhang}}, \bibinfo {author} {\bibfnamefont
  {M.}~\bibnamefont {Akter}}, \bibinfo {author} {\bibfnamefont
  {Y.}~\bibnamefont {Zhao}}, \bibinfo {author} {\bibfnamefont {T.~T.}\
  \bibnamefont {Xu}}, \bibinfo {author} {\bibfnamefont {Y.}~\bibnamefont {Xu}},
  \bibinfo {author} {\bibfnamefont {Z.}~\bibnamefont {Mao}}, \bibinfo {author}
  {\bibfnamefont {Y.}~\bibnamefont {Chen}}, \ and\ \bibinfo {author}
  {\bibfnamefont {D.}~\bibnamefont {Li}},\ }\href {\doibase
  10.1021/acs.nanolett.8b04206} {\bibfield  {journal} {\bibinfo  {journal}
  {Nano Letters}\ }\textbf {\bibinfo {volume} {19}},\ \bibinfo {pages} {415}
  (\bibinfo {year} {2019})}\BibitemShut {NoStop}%
\bibitem [{\citenamefont {Hoffmann}\ \emph {et~al.}(1980)\citenamefont
  {Hoffmann}, \citenamefont {Shaik}, \citenamefont {Scott}, \citenamefont
  {Whangbo},\ and\ \citenamefont {Foshee}}]{bandNbSe3}%
  \BibitemOpen
  \bibfield  {author} {\bibinfo {author} {\bibfnamefont {R.}~\bibnamefont
  {Hoffmann}}, \bibinfo {author} {\bibfnamefont {S.}~\bibnamefont {Shaik}},
  \bibinfo {author} {\bibfnamefont {J.}~\bibnamefont {Scott}}, \bibinfo
  {author} {\bibfnamefont {M.-H.}\ \bibnamefont {Whangbo}}, \ and\ \bibinfo
  {author} {\bibfnamefont {M.~J.}\ \bibnamefont {Foshee}},\ }\href {\doibase
  https://doi.org/10.1016/0022-4596(80)90230-3} {\bibfield  {journal} {\bibinfo
   {journal} {Journal of Solid State Chemistry}\ }\textbf {\bibinfo {volume}
  {34}},\ \bibinfo {pages} {263} (\bibinfo {year} {1980})}\BibitemShut
  {NoStop}%
\bibitem [{\citenamefont {Arfken}\ \emph {et~al.}(2013)\citenamefont {Arfken},
  \citenamefont {Weber},\ and\ \citenamefont {Harris}}]{arfken}%
  \BibitemOpen
  \bibfield  {author} {\bibinfo {author} {\bibfnamefont {G.~B.}\ \bibnamefont
  {Arfken}}, \bibinfo {author} {\bibfnamefont {H.~J.}\ \bibnamefont {Weber}}, \
  and\ \bibinfo {author} {\bibfnamefont {F.~E.}\ \bibnamefont {Harris}},\ }in\
  \href {\doibase https://doi.org/10.1016/B978-0-12-384654-9.00012-8} {\emph
  {\bibinfo {booktitle} {Mathematical Methods for Physicists (Seventh
  Edition)}}},\ \bibinfo {editor} {edited by\ \bibinfo {editor} {\bibfnamefont
  {G.~B.}\ \bibnamefont {Arfken}}, \bibinfo {editor} {\bibfnamefont {H.~J.}\
  \bibnamefont {Weber}}, \ and\ \bibinfo {editor} {\bibfnamefont {F.~E.}\
  \bibnamefont {Harris}}}\ (\bibinfo  {publisher} {Academic Press},\ \bibinfo
  {address} {Boston},\ \bibinfo {year} {2013})\ \bibinfo {edition} {seventh
  edition}\ ed.,\ pp.\ \bibinfo {pages} {551--598}\BibitemShut {NoStop}%
\bibitem [{\citenamefont {Pletikosi\ifmmode~\acute{c}\else \'{c}\fi{}}\ \emph
  {et~al.}(2014)\citenamefont {Pletikosi\ifmmode~\acute{c}\else \'{c}\fi{}},
  \citenamefont {Ali}, \citenamefont {Fedorov}, \citenamefont {Cava},\ and\
  \citenamefont {Valla}}]{semimetal1}%
  \BibitemOpen
  \bibfield  {author} {\bibinfo {author} {\bibfnamefont {I.}~\bibnamefont
  {Pletikosi\ifmmode~\acute{c}\else \'{c}\fi{}}}, \bibinfo {author}
  {\bibfnamefont {M.~N.}\ \bibnamefont {Ali}}, \bibinfo {author} {\bibfnamefont
  {A.~V.}\ \bibnamefont {Fedorov}}, \bibinfo {author} {\bibfnamefont {R.~J.}\
  \bibnamefont {Cava}}, \ and\ \bibinfo {author} {\bibfnamefont
  {T.}~\bibnamefont {Valla}},\ }\href {\doibase 10.1103/PhysRevLett.113.216601}
  {\bibfield  {journal} {\bibinfo  {journal} {Phys. Rev. Lett.}\ }\textbf
  {\bibinfo {volume} {113}},\ \bibinfo {pages} {216601} (\bibinfo {year}
  {2014})}\BibitemShut {NoStop}%
\bibitem [{\citenamefont {Jiang}\ \emph {et~al.}(2015)\citenamefont {Jiang},
  \citenamefont {Tang}, \citenamefont {Pan}, \citenamefont {Liu}, \citenamefont
  {Niu}, \citenamefont {Wang}, \citenamefont {Xu}, \citenamefont {Yang},
  \citenamefont {Xie}, \citenamefont {Song}, \citenamefont {Dudin},
  \citenamefont {Kim}, \citenamefont {Hoesch}, \citenamefont {Das},
  \citenamefont {Vobornik}, \citenamefont {Wan},\ and\ \citenamefont
  {Feng}}]{semimetal2}%
  \BibitemOpen
  \bibfield  {author} {\bibinfo {author} {\bibfnamefont {J.}~\bibnamefont
  {Jiang}}, \bibinfo {author} {\bibfnamefont {F.}~\bibnamefont {Tang}},
  \bibinfo {author} {\bibfnamefont {X.~C.}\ \bibnamefont {Pan}}, \bibinfo
  {author} {\bibfnamefont {H.~M.}\ \bibnamefont {Liu}}, \bibinfo {author}
  {\bibfnamefont {X.~H.}\ \bibnamefont {Niu}}, \bibinfo {author} {\bibfnamefont
  {Y.~X.}\ \bibnamefont {Wang}}, \bibinfo {author} {\bibfnamefont {D.~F.}\
  \bibnamefont {Xu}}, \bibinfo {author} {\bibfnamefont {H.~F.}\ \bibnamefont
  {Yang}}, \bibinfo {author} {\bibfnamefont {B.~P.}\ \bibnamefont {Xie}},
  \bibinfo {author} {\bibfnamefont {F.~Q.}\ \bibnamefont {Song}}, \bibinfo
  {author} {\bibfnamefont {P.}~\bibnamefont {Dudin}}, \bibinfo {author}
  {\bibfnamefont {T.~K.}\ \bibnamefont {Kim}}, \bibinfo {author} {\bibfnamefont
  {M.}~\bibnamefont {Hoesch}}, \bibinfo {author} {\bibfnamefont {P.~K.}\
  \bibnamefont {Das}}, \bibinfo {author} {\bibfnamefont {I.}~\bibnamefont
  {Vobornik}}, \bibinfo {author} {\bibfnamefont {X.~G.}\ \bibnamefont {Wan}}, \
  and\ \bibinfo {author} {\bibfnamefont {D.~L.}\ \bibnamefont {Feng}},\ }\href
  {\doibase 10.1103/PhysRevLett.115.166601} {\bibfield  {journal} {\bibinfo
  {journal} {Phys. Rev. Lett.}\ }\textbf {\bibinfo {volume} {115}},\ \bibinfo
  {pages} {166601} (\bibinfo {year} {2015})}\BibitemShut {NoStop}%
\bibitem [{\citenamefont {Zeng}\ \emph {et~al.}(2016)\citenamefont {Zeng},
  \citenamefont {Lou}, \citenamefont {Wu}, \citenamefont {Xu}, \citenamefont
  {Guo}, \citenamefont {Kong}, \citenamefont {Zhong}, \citenamefont {Ma},
  \citenamefont {Fu}, \citenamefont {Richard}, \citenamefont {Wang},
  \citenamefont {Liu}, \citenamefont {Lu}, \citenamefont {Huang}, \citenamefont
  {Fang}, \citenamefont {Sun}, \citenamefont {Wang}, \citenamefont {Wang},
  \citenamefont {Shi}, \citenamefont {Weng}, \citenamefont {Lei}, \citenamefont
  {Liu}, \citenamefont {Wang}, \citenamefont {Qian}, \citenamefont {Luo},\ and\
  \citenamefont {Ding}}]{semimetal3}%
  \BibitemOpen
  \bibfield  {author} {\bibinfo {author} {\bibfnamefont {L.-K.}\ \bibnamefont
  {Zeng}}, \bibinfo {author} {\bibfnamefont {R.}~\bibnamefont {Lou}}, \bibinfo
  {author} {\bibfnamefont {D.-S.}\ \bibnamefont {Wu}}, \bibinfo {author}
  {\bibfnamefont {Q.~N.}\ \bibnamefont {Xu}}, \bibinfo {author} {\bibfnamefont
  {P.-J.}\ \bibnamefont {Guo}}, \bibinfo {author} {\bibfnamefont {L.-Y.}\
  \bibnamefont {Kong}}, \bibinfo {author} {\bibfnamefont {Y.-G.}\ \bibnamefont
  {Zhong}}, \bibinfo {author} {\bibfnamefont {J.-Z.}\ \bibnamefont {Ma}},
  \bibinfo {author} {\bibfnamefont {B.-B.}\ \bibnamefont {Fu}}, \bibinfo
  {author} {\bibfnamefont {P.}~\bibnamefont {Richard}}, \bibinfo {author}
  {\bibfnamefont {P.}~\bibnamefont {Wang}}, \bibinfo {author} {\bibfnamefont
  {G.~T.}\ \bibnamefont {Liu}}, \bibinfo {author} {\bibfnamefont
  {L.}~\bibnamefont {Lu}}, \bibinfo {author} {\bibfnamefont {Y.-B.}\
  \bibnamefont {Huang}}, \bibinfo {author} {\bibfnamefont {C.}~\bibnamefont
  {Fang}}, \bibinfo {author} {\bibfnamefont {S.-S.}\ \bibnamefont {Sun}},
  \bibinfo {author} {\bibfnamefont {Q.}~\bibnamefont {Wang}}, \bibinfo {author}
  {\bibfnamefont {L.}~\bibnamefont {Wang}}, \bibinfo {author} {\bibfnamefont
  {Y.-G.}\ \bibnamefont {Shi}}, \bibinfo {author} {\bibfnamefont {H.~M.}\
  \bibnamefont {Weng}}, \bibinfo {author} {\bibfnamefont {H.-C.}\ \bibnamefont
  {Lei}}, \bibinfo {author} {\bibfnamefont {K.}~\bibnamefont {Liu}}, \bibinfo
  {author} {\bibfnamefont {S.-C.}\ \bibnamefont {Wang}}, \bibinfo {author}
  {\bibfnamefont {T.}~\bibnamefont {Qian}}, \bibinfo {author} {\bibfnamefont
  {J.-L.}\ \bibnamefont {Luo}}, \ and\ \bibinfo {author} {\bibfnamefont
  {H.}~\bibnamefont {Ding}},\ }\href {\doibase 10.1103/PhysRevLett.117.127204}
  {\bibfield  {journal} {\bibinfo  {journal} {Phys. Rev. Lett.}\ }\textbf
  {\bibinfo {volume} {117}},\ \bibinfo {pages} {127204} (\bibinfo {year}
  {2016})}\BibitemShut {NoStop}%
\bibitem [{\citenamefont {McCann}\ and\ \citenamefont
  {Koshino}(2013)}]{graphene_band}%
  \BibitemOpen
  \bibfield  {author} {\bibinfo {author} {\bibfnamefont {E.}~\bibnamefont
  {McCann}}\ and\ \bibinfo {author} {\bibfnamefont {M.}~\bibnamefont
  {Koshino}},\ }\href {\doibase 10.1088/0034-4885/76/5/056503} {\bibfield
  {journal} {\bibinfo  {journal} {Reports on Progress in Physics}\ }\textbf
  {\bibinfo {volume} {76}},\ \bibinfo {pages} {056503} (\bibinfo {year}
  {2013})}\BibitemShut {NoStop}%
\bibitem [{\citenamefont {Das~Sarma}\ \emph {et~al.}(2010)\citenamefont
  {Das~Sarma}, \citenamefont {Hwang},\ and\ \citenamefont
  {Rossi}}]{graphene_transport}%
  \BibitemOpen
  \bibfield  {author} {\bibinfo {author} {\bibfnamefont {S.}~\bibnamefont
  {Das~Sarma}}, \bibinfo {author} {\bibfnamefont {E.~H.}\ \bibnamefont
  {Hwang}}, \ and\ \bibinfo {author} {\bibfnamefont {E.}~\bibnamefont
  {Rossi}},\ }\href {\doibase 10.1103/PhysRevB.81.161407} {\bibfield  {journal}
  {\bibinfo  {journal} {Phys. Rev. B}\ }\textbf {\bibinfo {volume} {81}},\
  \bibinfo {pages} {161407(R)} (\bibinfo {year} {2010})}\BibitemShut {NoStop}%
\bibitem [{\citenamefont {Vafek}\ and\ \citenamefont
  {Yang}(2010)}]{graphene_int}%
  \BibitemOpen
  \bibfield  {author} {\bibinfo {author} {\bibfnamefont {O.}~\bibnamefont
  {Vafek}}\ and\ \bibinfo {author} {\bibfnamefont {K.}~\bibnamefont {Yang}},\
  }\href {\doibase 10.1103/PhysRevB.81.041401} {\bibfield  {journal} {\bibinfo
  {journal} {Phys. Rev. B}\ }\textbf {\bibinfo {volume} {81}},\ \bibinfo
  {pages} {041401(R)} (\bibinfo {year} {2010})}\BibitemShut {NoStop}%
\bibitem [{\citenamefont {Ashcroft}\ and\ \citenamefont
  {Mermin}(1976)}]{ashcroft}%
  \BibitemOpen
  \bibfield  {author} {\bibinfo {author} {\bibfnamefont {N.~W.}\ \bibnamefont
  {Ashcroft}}\ and\ \bibinfo {author} {\bibfnamefont {N.~D.}\ \bibnamefont
  {Mermin}},\ }\href@noop {} {\emph {\bibinfo {title} {Solid State
  {P}hysics}}}\ (\bibinfo  {publisher} {Holt-Saunders},\ \bibinfo {address}
  {Philadelphia},\ \bibinfo {year} {1976})\BibitemShut {NoStop}%
\end{thebibliography}%
\end{document}